\documentclass[reprint]{revtex4-1}
\pdfoutput=1
\usepackage{amsmath,nicefrac,units,multirow,dcolumn,float}
\usepackage[american]{babel}
\usepackage[T1]{fontenc}
\usepackage[utf8]{inputenc}
\usepackage{graphicx}
\DeclareGraphicsExtensions{.pdf}
\graphicspath{{./graphics/}}

\usepackage{xcolor}
\usepackage[normalem]{ulem}

\newcommand{\gpaw}{{GPAW}}

\DeclareMathOperator\erf{erf}
\DeclareMathOperator\erfc{erfc}

\begin{document}

\title{Charge transfer excitations with range separated functionals using improved virtual orbitals} 
\date{\today}

\author{Rolf \surname{Würdemann}}
\affiliation{Freiburger Materialforschungszentrum, Universität Freiburg, Stefan-Meier-Straße 21, D-79104 Freiburg, Germany}
\email{Rolf.Wuerdemann@fmf.uni-freiburg.de}
\author{Michael \surname{Walter}}
\affiliation{Freiburger Zentrum für interaktive Werkstoffe und bioinspirierte Technologien, Universität Freiburg, Georges-Köhler-Allee 105, D-79110 Freiburg, Germany}
\affiliation{Fraunhofer IWM, Wöhlerstrasse 11, D-79108 Freiburg, Germany}

\begin{abstract}
We present an implementation of range separated functionals 
utilizing the Slater-function on grids in real space 
in the projector augmented waves method.
The screened Poisson equation is solved to 
evaluate the necessary screened exchange integrals on Cartesian grids.
The implementation is verified against existing literature and
applied to the description of charge transfer excitations.
We find very slow convergence for calculations within linear response 
time-dependent density functional theory and unoccupied
orbitals of the canonical Fock operator.
Convergence can be severely improved by using 
Huzinaga's virtual orbitals instead. 
This combination furthermore enables an accurate determination
of long-range charge transfer excitations by means of 
ground-state calculations.
\end{abstract}


\maketitle
\section{Introduction}

The study of intra-molecular charge transfer excitations (CTE)
is of interest in photovoltaics\cite{sariciftci_photoinduced_1992},
organic electronics\cite{samori_epitaxial_2002} and molecular
and organic magnetism\cite{blundell_organic_2004}.
Within a single particle picture, the simplest CTE is the excitation of an
electron from the highest molecular orbital (HOMO) of a
donor to the lowest unoccupied orbital (LUMO) of a distant acceptor.
Denoting the distance between donor and acceptor by $R$,
Mulliken derived the energetics of this process in the
asymptote of large distances $R$ to (atomic units are used throughout)
\begin{align}
  E_\text{CTE} \approx \text{IP}_D - \text{EA}_A -
  \frac{1}{R},\label{eq:MullCTE}
\end{align}
where $\text{IP}_D$ is the ionization potential of the donor,
$\text{EA}_A$ the electron affinity of the acceptor and
$\frac{1}{R}$ approximates the Coulomb energy
between the excited electron and the hole it left
behind\cite{mulliken_molecular_1969}.

Density functional theory (DFT)\cite{hohenberg_inhomogeneous_1964}
in the formulation of
\citeauthor{kohn_self-consistent_1965}\cite{kohn_self-consistent_1965}
is the method of choice for \emph{ab initio}
calculations of electronic properties of condensed matter
due to its advantageous cost to accuracy ratio.
In contrast to wave-function based methods, DFT expresses
the total energy of a given system as
functional of the electron density $n$.
Although DFT is exact in principle, the exact form of this
functional itself is unknown and
has to be approximated in practice\cite{livshits_well-tempered_2007,chai_systematic_2008,cohen_insights_2008}.
In Kohn-Sham (KS) DFT the density is build from occupied
non-interacting single-particle orbitals $\psi_i$ via
$n = \sum_i f_i |\psi_i|^2$, where $f_i$ denotes the occupation number.
The total energy is expressed as a sum of
density functionals for the different contributions
\begin{align}
  E_\text{tot} [n] = T_\text{S} [n]  + V_\text{ext} [n] + U_\text{H}[n] + 
  E_\text{xc} [n],\label{eq:E_KS_DFT}
\end{align}
where the $T_\text{S}$ denotes the kinetic energy of the
non-interacting system, $V_\text{ext}$ the energy  of the
density in the external potential and
$U_\text{H}$ the classical Coulomb energy of the
density with itself.
These quantities can be calculated exactly.
All other energy contributions are collected in the
exchange-correlation energy $E_\text{xc}$
which is approximated.
Within a generalized KS scheme\cite{baer_tuned_2010}
$E_\text{xc}$ can be further split into the contributions
from exchange as in Hartree-Fock theory (HFT)
\begin{align}
  E_\text{x} = -\frac{1}{2}   \int \int 
  \frac{\left|\sum_i f_i\psi_i^\ast (\vec{r}_1) \psi_i (\vec{r}_2)\right|^2}
  {|\vec{r}_1 - \vec{r}_2|}
  \text{d}\vec{r}_1 \text{d}\vec{r}_2,
  \label{eq:Ex}
\end{align}
and correlation $E_\text{c}$ that contains all energy contributions
missing in the other terms\cite{kohn_self-consistent_1965}.

Several types of approximations for $E_\text{xc}$ are in use.
The
local (LDA)\cite{vosko_accurate_1980,perdew_self-interaction_1981,perdew_accurate_1992}
functional approximates $E_\text{xc}$ by the local values of the
density, while semi-local GGA\cite{gill_new_1996,perdew_generalized_1996}
take local density gradients
and MGGA\cite{adamo_meta-gga_2000,tao_climbing_2003} local
values of the kinetic electron density into account
(we will call local and semi-local functionals as local functionals for
brevity in the following).
The accuracy of local functionals is sometimes improved
by  hybrid functionals that
combine the exchange from local functionals
with the ``exact'' exchange integrals from HFT eq. (\ref{eq:Ex})
by a fixed ratio\cite{becke_new_1993}.
While these approximations work fairly well for
equilibrium properties\cite{baer_density_2005,cohen_insights_2008},
local functionals as well as hybrids fail badly in the description of
CTEs because of their missing
ability to describe non-local interactions correctly
\cite{dreuw_long-range_2003,baerends_kohnsham_2013,kummel_charge-transfer_2017,Maitra17}.

Range separated functionals (RSF), that combine the exchange from the 
local functionals with the non-local exchange from HFT based 
on the spatial distance between two points $\vec{r}_1$ and $\vec{r}_2$
are able to predict the energetics and oscillator strengths of 
CTEs by linear response time-dependent DFT \mbox{(lrTDDFT)}\cite{tawada_long-range-corrected_2004,livshits_well-tempered_2007,chai_systematic_2008,akinaga_intramolecular_2009,rohrdanz_long-range-corrected_2009,stein_reliable_2009,stein_prediction_2009,baer_tuned_2010,kronik_excitation_2012,zhang_non-self-consistent_2012}.

RSF use a separation function
$\omega_\text{RSF}$ to split the Coulomb interaction kernel
of the exchange integral (\ref{eq:Ex}) into two parts\cite{yanai_new_2004}
\begin{align}
  \frac{1}{r_{12}} =& \underbrace{\frac{1 - \left [ \alpha + \beta  \left ( 1 - \omega_\text{RSF}(\gamma, r_{12})\right )\right]}{r_{12}}}_\text{SR, DFT}\notag \\
   &+ \underbrace{\frac{\alpha + \beta \left (1- \omega_\text{RSF}(\gamma, r_{12})\right )}{r_{12}},}_\text{LR, HFT} \label{eq:CAM}
\end{align}
where $\gamma$ is a separation parameter, and $\alpha$ and
$\beta$ are mixing parameters for spatially fixed and
range separated mixing, respectively.
The exchange from the local functionals is usually 
used in the short-range (SR) part, while
non-local exchange integrals from HFT are used for long-range
(LR) exchange. 
There also exists a class of RSF that uses exact exchange at
short-range and the exchange from a local functional for 
the long-range part and is very popular for the description
of periodic systems\cite{heyd_hybrid_2003,heyd_erratum:_2006}, which is
not the topic of our investigations, however.
The correlation energy $E_\text{c}$ is approximated by a local functional 
globally\cite{iikura_long-range_2001} .

The unoccupied states entering lrTDDFT using the canonical
Fock operator of HFT approximate (exited) 
EAs\cite{stein_fundamental_2010,kronik_excitation_2012}.
Coulomb-repulsion and exchange interaction 
cancel for occupied states making them subject to the interaction 
with $N-1$ electrons in an $N$ electron system. 
This cancellation is absent for unoccupied (virtual) states, 
making them subject to the interaction with all $N$ electrons.
Therefore these unoccupied states are 
rather inappropriate for neutral excited state
calculations.

The canonical Fock operator of HFT is not the only possible choice
for the calculation of unoccupied states, as any Hermitian rotation 
within the space of virtual orbitals is allowed\cite{huzinaga_virtual_1970}.
Therefore it is possible to create so called 
improved virtual orbitals (IVOs) that are able to approximate
certain excitations already at the single particle 
level rather accurately\cite{kelly_correlation_1963,kelly_many-body_1964,huzinaga_virtual_1970,huzinaga_virtual_1971} and we will show
that these orbitals are also well suited for the
calculation of charge transfer excitations within RSF.

This work is organized as follows: 
The following section describes the numerical methods applied and
section \ref{sec:RSF} details the implementation of RSF
in real space grids within the projector agmented wave method.
Sec. \ref{sec:verify} presents the verification 
of our implementation against existing literature and
sec. \ref{sec:CTE} applies RSF and its combination with IVOs 
to obtain CTE energies.
The manuscript finally ends with conclusions.

\section{Methods}

DFT using 
the real space grid implementation of the projector augmented waves
(PAW)\cite{blochl_projector_1994} in the \gpaw{}
package\cite{mortensen_real-space_2005,enkovaara_electronic_2010}
was used for all calculations performed in this work.
PAW is an all-electron method, which has shown to provide 
very similar results as converged basis sets for a test set of 
small molecules\cite{kresse_ultrasoft_1999} and 
transition metals\cite{valiev_calculations_2003,Wurdemann2015}.
With the exception of transition metals,
where $3s$, $3p$, $3d$ and $4s$ shells were treated as valence electrons,
all closed shells were subject to the frozen core approximation
and (half-)open shells were treated as valence electrons.
Relativistic effects were applied to the closed shells in the frozen
cores in the scalar-relativistic approximation
of \citeauthor{koelling_technique_1977}\cite{koelling_technique_1977}.
If not stated otherwise, a grid-spacing of
$h=\unit[0.18]{\text{\AA}}$ was used for the smooth KS
wave function and a simulation box which contains
at least $\unit[6]{\text{\AA}}$ space around each atom was applied.
Non-periodic boundary conditions were applied in all three directions
and all calculations were done spin-polarized.
Only collinear spin alignments were considered.
The exchange correlation functionals PBE\cite{perdew_generalized_1996}, 
the hybrid PBE0\cite{adamo_toward_1998} and the 
RSFs LCY-BLYP\cite{akinaga_range-separation_2008}, 
LCY-PBE\cite{seth_range-separated_2012} and 
CAMY-B3LYP\cite{akinaga_range-separation_2008} were used.
Linear response time-dependent density functional theory 
(lrTDDFT)\cite{casida_time-dependent_2009,walter_time-dependent_2008} 
for RSF was implemented along the 
work of \citeauthor{tawada_long-range-corrected_2004}\cite{tawada_long-range-corrected_2004} and
\citeauthor{akinaga_intramolecular_2009}\cite{akinaga_intramolecular_2009}.
Twelve unoccupied bands were used in the lrTDDFT calculations unless
stated otherwise.

\section{Implementation of RSF}
\label{sec:RSF}

Two functions are frequently used as separation function
(\ref{eq:CAM}) in literature.
One is the complementary error-function\cite{iikura_long-range_2001,tawada_long-range-corrected_2004,yanai_new_2004,baer_avoiding_2006,peach_assessment_2006,vydrov_tests_2007,chai_systematic_2008,rohrdanz_simultaneous_2008,wong_absorption_2009}
$\omega_\text{RSF}(\gamma, r_{12}) = \erfc (\gamma r_{12})$
that enables efficient evaluation when the KS orbitals are
represented by Gaussian functions, and the other is
the Slater-function\cite{baer_avoiding_2006,akinaga_range-separation_2008,akinaga_intramolecular_2009,seth_range-separated_2012}
\begin{align}
  \omega_\text{RSF}(\gamma, r_{12}) &= e^{ - \gamma r_{12}}. \label{eq:Yukawa}
\end{align}
which we will apply in our work. 
The Slater-function is the natural choice for a screened Coulomb potential,
as it leads to the Yukawa potential\cite{yukawa_interaction_1935}
that can be derived to be the effective one-electron potential
in many-electron systems\cite{seth_range-separated_2012}.
Calculations utilizing the Slater-function were found to give superior
results for the calculation of charge transfer and
Rydberg excitations compared to the use of the
complementary error-function \cite{akinaga_range-separation_2008,akinaga_intramolecular_2009}.

The calculation of the exact exchange in a RSF is straightforward 
in principle.
Regarding only the long-range part in eq. (\ref{eq:CAM}) and 
setting $\alpha = 0$ and $\beta =1$ for brevity,
we have to evaluate the exchange integral
\begin{align}
  K^\text{RSF}_{ij} &= \left(ij | 1 - \omega_{\text{RSF}} \left(\gamma, r_{12}\right) | ji \right) \label{eq:RSF_EXX_om}
\end{align}
where a Mulliken-like notation
\begin{align}
  \left( ij| \hat{x} | ji \right) &= \notag \\
  \int \int& \frac{\psi_i^\ast (\vec{r}_1) \psi_j (\vec{r}_1) \, \hat{x}\, 
    \psi_j^\ast (\vec{r}_2) \psi_i (\vec{r}_2)}{|\vec{r}_1 - \vec{r}_2|}
  \, \text{d}\vec{r}_1 \text{d}\vec{r}_2
  \label{eq:Iex}
\end{align}
with the one-particle functions $\psi_{i,j}$ and the operator
$\hat{x}$ is used.
The first part of eq. (\ref{eq:RSF_EXX_om}) is the standard
exchange integral $K_{ij}=\left(ij | ji \right)$ from
HFT\cite{rostgaard_exact_2006,enkovaara_electronic_2010} and the
additional
term is the screened exchange integral
\begin{align}
  K^\gamma_{ij} &=  \left(ij |\omega_\text{RSF} \left(r_{12}\right)| ji \right)
  \; .
  \label{eq:IRSF}
\end{align}
In PAW the KS wave-function (WF)
$\psi_i$ is represented as a combination of a soft pseudo
WF, $\tilde{\psi}_i$ and atom-centered (local)
corrections\cite{blochl_projector_1994,mortensen_real-space_2005,enkovaara_electronic_2010}
\begin{align}
  \psi_i  &= \tilde{\psi}_i + \sum_\alpha \sum_k
  \left ( | \phi_k^\alpha \rangle - | \tilde{\phi}_k^\alpha
  \rangle \right ) \mathcal{P}_{ik}^\alpha, \label{eq:PAW_Trans}
\end{align}
where $\phi_k^\alpha$ and $\tilde{\phi}_k^\alpha$ denote atom centered
all-electron and soft partial WFs, respectively, and $\mathcal{P}_{ik}^\alpha$ is
a projection operator which maps the pseudo WF on the partial WF.
The all-electron and soft partial WF match outside of the
atom centered augmentation sphere.
The band-indices $i$ and $k$ contain the main quantum numbers
and $\alpha$ is the atomic index which runs over all
atoms in the calculation.

In our implementation, the pseudo WFs are evaluated on
three dimensional Cartesian grids in real space
while the partial WFs are evaluated on radial grids\cite{mortensen_real-space_2005,enkovaara_electronic_2010}.
Using the exchange density $n_{ij} = \psi_i^\ast \psi_j$ and
the local exchange density
$n^\alpha_{ij} = \sum_{k_1 k_2} \phi_{k_1}^\alpha \phi_{k_2}^\alpha \mathcal{P}_{ik_1}^{\alpha\ast}\mathcal{P}_{jk_2}^{\alpha}$ the integral (\ref{eq:IRSF}) can be written as
\begin{align}
  \left(\left(n_{ij}\right)\right)^\gamma &= \left(\left(\tilde{n}_{ij}\right)\right)^\gamma +
  \sum_\alpha \left [ \left(\left(n^\alpha_{ij}\right)\right)^\gamma - \left(\left(\tilde{n}^\alpha_{ij}\right)\right)^\gamma \right],\label{eq:ScrExPawNoComp}
\end{align}
where $$\left(\left(n_{ij}\right)\right)^\gamma = \left(n_{ij}|n_{ji}\right)^\gamma  = \left(n_{ij} | \exp \left( -\gamma r_{12}\right)| n_{ji} \right)$$
is used as shortcut for the screened exchange interaction of the
exchange density with itself.
Due to the non-locality of the screened exchange operator,
a straight application of eq. (\ref{eq:ScrExPawNoComp}) would
lead to cross-terms between local functions located on different
atomic sites and the need to integrate on 
incompatible grids\cite{blochl_projector_1994,mortensen_real-space_2005}.
To avoid these, compensation charges $\tilde{Z}_{ij}^{\alpha}$ are
introduced\cite{rostgaard_exact_2006,enkovaara_electronic_2010} such that using
$\tilde{\varrho}_{ij} = \tilde{n}_{ij} + \sum_\alpha \tilde{Z}_{ij}^\alpha$
allows to write
\begin{align}
  \left(\left(n_{ij}\right)\right)^\gamma &=\left(\left(\tilde{\varrho}_{ij}\right)\right)^\gamma + \sum_\alpha \Delta K_{ij}^{\alpha\gamma},
  \label{eq:ScrExPawComp}
\end{align}
where $\left(\left(\tilde{\varrho}_{ij}\right)\right)^\gamma$
is free of local contributions. The evaluation of
the $\Delta K_{ij}^{\alpha\gamma}$ using the integration kernel
from \citeauthor{rico_repulsion_2012}\cite{rico_repulsion_2012}
is detailed in supporting information (SI).

Direct evaluation of the double-integral
\begin{align}
I &= \int \int \tilde{\varrho}_{ij}(1)  \frac{e^{-\gamma r_{12}}}{r_{12}} \tilde{\varrho}_{ji}(2)  \text{d}\vec{r}_1 \text{d}\vec{r}_2 \label{eq:IYukawa}
\end{align}
on a
three dimensional Cartesian grid is not only time-consuming,
but also suffers from the singularity at
$\vec{r}_1 = \vec{r}_2$. To circumvent this, the integral
is solved using the method of the Green's functions
 \begin{align}
   I &= \int \tilde{\varrho}_{ij}(\vec{r}_1) \tilde{v}_{ji}(\vec{r_1})
   d\vec{r}_1 \; .\label{eq:IntGreen}
\end{align}
The potential $\tilde{v}_{ij}$ is calculated by
solving the screened Poisson or modified Helmholtz
equation\cite{jackson_classical_1998,greengard_new_2002}
\begin{align}
  \left ( \nabla^2 - \gamma^2 \right ) \tilde{v}_{ij}(\vec{r_1}) &= - 4 \pi \tilde{\varrho}_{ij}(\vec{r_1}) \; . \label{eq:ScreenedPoisson2}
\end{align}
A finite difference scheme together with a root
finder\cite{mortensen_real-space_2005,enkovaara_electronic_2010}
is chosen, where a constant representing $\gamma$ is added to the
central point of the finite difference stencil\cite{wajid_modified_2014}.
The potential $\tilde{v}_{ij}$ of a charge system decays very slowly and
the (screened) Poisson equation (\ref{eq:ScreenedPoisson2}) 
is therefore applied for neutral charge distributions 
only\cite{enkovaara_electronic_2010}.
Charged systems are neutralized by subtracting a Gaussian density 
for which the solution is known analytically (see SI for details).

Range separated functionals also contain a contribution of the
density functional exchange.
\citeauthor{akinaga_range-separation_2008} derived an analytic expression 
for the exchange contribution of a GGA in the case of a 
Slater-function based RSF, 
$E_\text{x}^\text{GGA} \left (\gamma \right)$\cite{akinaga_range-separation_2008}.
\citeauthor{seth_range-separated_2012} discovered, that this expression 
leads to numerical instabilities for very small densities and derived 
a superior expression for small densities based on a power series 
expansion\cite{seth_range-separated_2012}.
Both expressions along with analytic expressions for the first,
second, and third derivatives of $E_\text{x}^\text{GGA}(\gamma)$
based on the analytic expression derived by
\citeauthor{akinaga_range-separation_2008} were
implemented in \emph{libxc}\cite{marques_libxc:_2012,LehtolaRecentdevelopmentslibxc2018}.

\section{Verification of the implementation}
\label{sec:verify}

\begin{table*}
  \begin{tabular}{lD{.}{.}{2}|D{.}{.}{2}|D{.}{.}{3}|D{.}{.}{2}|D{.}{.}{2}|D{.}{.}{3}|D{.}{.}{2}|D{.}{.}{2}|D{.}{.}{3}|D{.}{.}{2}}
    &
    &  \multicolumn{3}{c|}{TiO\textsubscript{2}} 
    &  \multicolumn{3}{c|}{CuCl} 
    &  \multicolumn{3}{c}{CrO\textsubscript{3}} 
    \\
    Functional 
    &\multicolumn{1}{c|}{$\gamma$}
    &  \multicolumn{1}{c|}{ours} 
    &  \multicolumn{1}{c|}{lit.} 
    &  \multicolumn{1}{c|}{$\text{d}E$} 
    &  \multicolumn{1}{c|}{ours} 
    &  \multicolumn{1}{c|}{lit.} 
    &  \multicolumn{1}{c|}{$\text{d}E$} 
    &  \multicolumn{1}{c|}{ours} 
    &  \multicolumn{1}{c|}{lit.} 
    &  \multicolumn{1}{c}{$\text{d}E$}  \\    
    \hline
    PBE              &          & 7.56                                        & 7.53^a & 0.03 & 3.74                                        & 3.73^a & 0.01& 5.67                                        & 5.66^a & 0.01\\
    PBE              &          & \multicolumn{1}{c|}{--"--} & 7.56^b & 0.00 & \multicolumn{1}{c|}{--"--} & 3.67^b & 0.07& \multicolumn{1}{c|}{--"--} & 6.07^b & -0.4\\
    PBE0		     &          & 6.45                                        & 6.44^a & 0.01 & 3.54                                        & 3.62^a & -0.08& 4.24                                        & 4.28^a & -0.04\\
    PBE0            &          & \multicolumn{1}{c|}{--"--} & 6.47^b & -0.02 & \multicolumn{1}{c|}{--"--} & 3.50^b & 0.04& \multicolumn{1}{c|}{--"--} & 4.63^b & -0.39\\
    LCY-BLYP      &  0.70 & 6.24                                       & 6.33^b & -0.09 & 3.44                                       & 3.69^b & -0.25& 4.16                                       & 4.75^b & -0.59\\
                         &  0.75 & 6.12                                       & 6.21^b & -0.09 & 3.43                                       & 3.68^b & -0.25& 4.01                                       & 4.59^b & -0.58\\
    LCY-PBE        & 0.75 & 6.38                                       & 6.47^b & -0.09 & 3.63                                       & 3.88^b & -0.25& 3.93                                       & 4.74^b & -0.81\\
                          & 0.90 & 6.04                                       & 6.14^b & -0.10& 3.60                                       & 3.84^b & -0.24& 3.49                                       & 4.29^b & -0.8\\
    CAMY-B3LYP & 0.34 & 6.32                                       & 6.38^b & -0.06& 3.32                                       & 3.43^b & -0.11& 4.34                                       & 4.89^b & -0.55\\
    $\overline{\text{d}E}_\text{RSF}$& & & & -0.09& & & -0.22& & & -0.67\\
    exp.    &          && 6.62^c &&& 3.81^c & && 4.97^c &\\
    \hline
  \end{tabular}
  \caption{Mean ligand removal enthalpies (see text) for TiO$_2$, CuCl, and CrO$_3$
    from our calculations compared to the literature.
    $dE$ denotes the difference between our results and the
    literature values.
    Geometries were taken from
    \citeauthor{johnson_tests_2009}\cite{johnson_tests_2009} and
    energies are given in $\unit{eV}$.  References and basis-sets: $^a$6–311++G(3df,3pd) ref. \citenum{johnson_tests_2009},
  $^b$TZ2P ref. \citenum{seth_range-separated_2012},
  $^c$experiment ref. \citenum{linstrom_neutral_2016}.
  } 
  \label{tab:YukSeth}
\end{table*}
A large amount of work was devoted to RSF in the 
literature\cite{iikura_long-range_2001,tawada_long-range-corrected_2004,
toulouse_long-rangechar21short-range_2004,yanai_new_2004,baer_density_2005,baer_avoiding_2006,
peach_assessment_2006,vydrov_assessment_2006,vydrov_importance_2006,cohen_development_2007,
livshits_well-tempered_2007,gerber_range_2007,vydrov_tests_2007,akinaga_range-separation_2008,
chai_systematic_2008,henderson_range_2008,livshits_density_2008,rohrdanz_simultaneous_2008,
akinaga_intramolecular_2009,livshits_deleterious_2009,rohrdanz_long-range-corrected_2009,stein_prediction_2009,
stein_reliable_2009,wong_absorption_2009,baer_tuned_2010,stein_fundamental_2010,refaely-abramson_fundamental_2011,kronik_excitation_2012,seth_range-separated_2012,zhang_non-self-consistent_2012,
seth_modeling_2013,autschbach_delocalization_2014,cabral_do_couto_performance_2015}.
In order to verify our implementation, we have re-calculated some
of the published properties in our grid-based approach.
\citeauthor{seth_range-separated_2012}\cite{seth_range-separated_2012} 
calculated the mean 
ligand removal enthalpies defined 
as\cite{johnson_tests_2009,seth_range-separated_2012} 
\begin{align}
  \bar{E}_{L} &= \frac{n E(L) + m E(M) - E(M_m L_n)}{n + m - 1}
\end{align}
for a group of molecules including 
transition metals.
We have calculated this quantity for the molecules TiO\textsubscript{2}, 
CuCl, and CrO\textsubscript{3} and compared our results against the values 
published by \citeauthor{johnson_tests_2009}\cite{johnson_tests_2009} 
for PBE and PBE0, as well as
\citeauthor{seth_range-separated_2012}\cite{seth_range-separated_2012} 
for PBE, PBE0 and a group of RSF in tab. \ref{tab:YukSeth}. 
As the literature values were calculated 
without relativistic corrections, we have neglected 
relativistic effects in the reported values also.
A grid spacing of $h = \unit[0.16]{\text{\AA}}$ was necessary to correctly
describe $3d$-splitting\cite{Wurdemann2015}
(see SI for details).

Generally, our PBE and PBE0 values are in good agreement to the 
results obtained from both groups for TiO\textsubscript{2} and CuCl.
There is a difference of about $\unit[0.4]{eV}$ to the
work of \citeauthor{seth_range-separated_2012} for CrO\textsubscript{3}, 
while our values are a in good agreement to the 
work of \citeauthor{johnson_tests_2009}.
These differences can attributed the different basis sets used.
While \citeauthor{johnson_tests_2009} used the rather large 
6–311++G(3df,3pd) basis set, \citeauthor{seth_range-separated_2012} 
use a smaller TZ2P basis set that is apparently not large enough. 
We have observed similar strong basis set effects
in particular if chromium is involved already in
prior studies for chromium\cite{Wurdemann2015}.

RSF results using Slater functions are unfortunately only 
available from the TZ2P basis set.
While our results are in good agreement 
to \citeauthor{seth_range-separated_2012} 
for TiO$_2$ ($\le \unit[-0.09]{eV}$ deviation),
they already differ by up to $\unit[0.25]{eV}$ for CuCl.
CAMY-B3LYP, which includes only a fraction of the screened exchange, 
generally leads to the smallest deviations.
RSFs are obviously very sensitive to basis set limitations due to
the long range of the effective single particle 
potential\cite{baer_tuned_2010}.
We therefore trust the values obtained by our method which 
represent the large basis set limit.

In comparison to experiment, PBE ligand removal energies are
rather accurate for CuCl, but this functional over-binds TiO$_2$ and CrO$_3$.
In the latter molecules $d$-orbitals contribute to binding and
might be responsible for this overbinding. 
In contrast, the hybrid PBE0 as well as the RSFs tend to underbind
in all three molecules, in particular if $d$-orbitals are involved.
Interestingly, this trend is similar to the overestimation
of $d$-binding in PBE and the lack of proper $d$-binding by 
hybrids we have observed in the Cr-dimer before\cite{Wurdemann2015}.

The value of the separation parameter $\gamma$ can not be defined
rigorously and its
optimal choice is under discussion. 
A system dependence is to
be expected\cite{iikura_long-range_2001,baer_avoiding_2006,livshits_well-tempered_2007,chai_systematic_2008,livshits_density_2008,baer_tuned_2010,stein_fundamental_2010,refaely-abramson_fundamental_2011,kronik_excitation_2012}.
The group around Roi Baer devised schemes to optimize
$\gamma$ without the use of
empirical parameters\cite{livshits_well-tempered_2007,livshits_density_2008,stein_reliable_2009} by forcing the difference between the 
ionization potential (IP) calculated from
total energy differences and
the negative eigenvalue of the HOMO $-\epsilon_\text{HOMO}$
of an $N$ electron system to vanish\cite{livshits_well-tempered_2007}
\begin{align}
   \underbrace{\left[ E_\text{gs}\left(N, \gamma\right)  - E_\text{gs}\left(N-1, \gamma\right) \right]}_{\text{IP}(N)} \notag \\
  \stackrel{\gamma = \gamma_\text{opt}}{\equiv} - \epsilon_\text{HOMO}(N, \gamma) \; . \label{eq:fitgammaIP}
\end{align}
This condition is fulfilled for the exact 
functional\cite{almbladh_exact_1985},
but is usually violated by local and 
hybrid approximations\cite{livshits_well-tempered_2007}.
\citeauthor{livshits_density_2008} also devised an approach to
determine $\gamma$ for the calculation of binding energy curves of
symmetric bi-radical cations which imposes a 
match of the slopes of the energy curves for the
charged and neutral molecule\cite{livshits_density_2008}
\begin{align}
  E_\text{gs} \left(N-\frac{1}{2}, \gamma\right) - E_\text{gs} \left(N-1, \gamma\right) \notag\\
  \quad \stackrel{\gamma = \gamma_\text{opt}}{\equiv} 
  E_\text{gs} \left(N, \gamma\right) - E_\text{gs} \left(N-\frac{1}{2}, \gamma\right) \; .\label{eq:fitgammaFrac}
\end{align}
Their group found that both approaches 
give  almost identical values of $\gamma_\text{opt}$ 
for the same system\cite{baer_tuned_2010}, which we confirm
in the case of Cr$_2$.
We will denote an RSF using an optimized value of
$\gamma$ obtained by eqs. (\ref{eq:fitgammaIP}, \ref{eq:fitgammaFrac})
by appending an asterisk, e.g. LCY-PBE${}^*$.

\begin{table*}
  \begin{tabular}{l|D{.}{.}{2}|D{.}{.}{2}|D{.}{.}{2}|D{.}{.}{2}|D{.}{.}{2}|D{.}{.}{2}|D{.}{.}{2}|D{.}{.}{2}}
      \multicolumn{1}{c|}{}
    &  \multicolumn{2}{c|}{Cr} 
    &  \multicolumn{2}{c|}{Cr\textsubscript{2}} 
    &  \multicolumn{2}{c|}{CO} 
    &  \multicolumn{2}{c}{N$_2$} \\
      \multicolumn{1}{c|}{}
    &  \multicolumn{1}{c|}{$-\epsilon_\text{H}$}
    &  \multicolumn{1}{c|}{IP}
    &  \multicolumn{1}{c|}{$-\epsilon_\text{H}$}
    &  \multicolumn{1}{c|}{IP}
    &  \multicolumn{1}{c|}{$-\epsilon_\text{H}$}
    &  \multicolumn{1}{c|}{IP}
    &  \multicolumn{1}{c|}{$-\epsilon_\text{H}$}
    &  \multicolumn{1}{c}{IP} \\            
    \hline
    exp.\cite{linstrom_ion_2016}                    & & 6.77 & & 6.4 & & 14.01 &  & 15.58   \\ 
    PBE              & 3.70 & & 4.29 &  &  9.09 & &  10.24  &\\
    PBE0           & 4.95 & &   &&  10.79 && 12.16 &\\
    BNL${}^*$\cite{livshits_well-tempered_2007}    &   & &  && 14.3&& 16.6 &\\    
    LCY-PBE${}^*$                                  & 6.79 && 6.85 && 14.32 &14.31 & 16.33 & 16.32  \\
    $\gamma^\text{BNL}_\text{opt} (a_0^{-1})$\cite{livshits_well-tempered_2007} &\multicolumn{2}{c|}{}  & \multicolumn{2}{c|}{}          &  \multicolumn{2}{c|}{0.6}  & \multicolumn{2}{c}{0.6}  \\    
    $\gamma^\text{LCY}_\text{opt} (a_0^{-1})$  & \multicolumn{2}{c|}{0.72} &  \multicolumn{2}{c|}{0.45}  &   \multicolumn{2}{c|}{0.81}  & \multicolumn{2}{c}{0.99} 
  \end{tabular}
  \caption{Negative eigenvalue of the HOMO $-\epsilon_\text{H}$, IP and values of the screening parameter $\gamma$ for Cr, Cr\textsubscript{2},  CO and N\textsubscript{2} calculated using different functionals. All energies given in $\unit{eV}$.
  } 
  \label{tab:TuneGammaMol}
\end{table*}

We used eq. (\ref{eq:fitgammaIP}) to obtain $\gamma_\text{opt}$
for Cr, Cr\textsubscript{2}, CO and N\textsubscript{2}
and verified that the eigenvalues for the HOMO as well as the 
experimental value of the ionization potential match in this case.
The resulting screening parameters for the RSF BNL\textsuperscript{*} 
and LCY-PBE\textsuperscript{*} are listed in tab. \ref{tab:TuneGammaMol}.
BNL\textsuperscript{*} is a LDA based RSF used by \citeauthor{livshits_well-tempered_2007} which utilizes the error-function instead of the Slater function\cite{livshits_well-tempered_2007}. For the gradient corrected PBE and the hybrid-functional PBE0 the eigenvalues of the HOMO doesn't match the experimental ionization potential. This is different for the RSF: The eigenvalues for the HOMO are in quite good, for N\textsubscript{2}, to, in the case of Cr, perfect agreement to the experimental ionization potential. For the cases of CO and N\textsubscript{2} also a good agreement between the values from \citeauthor{livshits_well-tempered_2007} and this work is achieved. For the values of the screening parameter a $\gamma^\text{LCY}_\text{opt} \approx \frac{3}{2} \gamma^\text{BNL}_\text{opt}$ dependency between the used screening functions, Slater vs. error-function, was stated\cite{shimazaki_band_2008}. The comparison between the values listed in tab. \ref{tab:TuneGammaMol} supports this dependency.

\begin{table}[h]
  \begin{tabular}{l|D{.}{.}{2}|r |r|r|r}
  	 \multicolumn{2}{c}{}& \multicolumn{2}{c|}{LCY-PBE\textsuperscript{*}} & \multicolumn{2}{c}{BNL\textsuperscript{*}\cite{baer_tuned_2010}} \\
      \multicolumn{1}{c|}{state}
    &  \multicolumn{1}{c|}{IP\cite{potts_photoelectron_1972} (eV)} 
    &  \multicolumn{1}{c|}{Koop.} 
    &  \multicolumn{1}{c|}{lr.} 
    &  \multicolumn{1}{c|}{Koop.} 
    &  \multicolumn{1}{c}{lr.}\\
    \hline
       1b\textsubscript{1} & 12.62 & 1\% & 1\% & -1\% & -1\%      \\
       3a\textsubscript{1} & 14.74 & 0\% & 4\% &  -3\% & -3\%  \\
       1b\textsubscript{2} & 18.51 & 0\% & 3\% &  -1\% & -1\% \\
       2a\textsubscript{1}${}^a$ & 32.20 & 0\%  & 4\%  & -1\% &  0\%
  \end{tabular}
  \caption{Deviation of the calculated IPs of H${}_2$O calculated with LCY-PBE${}^*$ and using Koopmans theorem or lrTDDFT along with the experimental values and values calculated by \citeauthor{baer_tuned_2010}\cite{baer_tuned_2010}. In this work, $\gamma$ was tuned to $\unit[0.831]{a_0^{-1}}$. $a$: Ref. \citenum{baer_tuned_2010} gave $\unit[30.9]{eV}$ as ionization potential for the 2a\textsubscript{1} state. Koop.: prediction by use of Koopman's theorem, lr.: prediction by use of $\Delta$-SCF and lrTDDT.}
  \label{tab:TuneGammaH2O}
\end{table}

\citeauthor{baer_tuned_2010} also discussed the impact of the tuning of $\gamma$ on the inner ionization energies (ionization into an excited state of the cation). They stated, that by tuning $\gamma$ one is able to predict the inner ionization energies not only by the combination of a $\Delta$SCF and lrTDDFT calculation but also directly by the density of states of the neutral molecule in the sense of Koopmans theorem\cite{koopmans_uber_1934} from HFT\cite{baer_tuned_2010}. In this work their example, H\textsubscript{2}O was also verified. The deviations between the calculated IPs and the experimental values are shown in tab. \ref{tab:TuneGammaH2O}. The calculated values are in a very good agreement to each other and to experiment, despite the issue, that \citeauthor{baer_tuned_2010} gave a ionization potential for the 2a\textsubscript{1} state which differs from the value used as reference in both works.

\section{Charge transfer excitations}
\label{sec:CTE}

In this section we investigate the description of
charge transfer excitations (CTE) within RSF.
One of the frequently used model systems to study CTE is the
ethylene-tetrafluoroethylene dimer\cite{dreuw_long-range_2003,tawada_long-range-corrected_2004,zhao_density_2006,peach_assessment_2006,livshits_well-tempered_2007,chai_systematic_2008,rohrdanz_long-range-corrected_2009,zhang_non-self-consistent_2012}.
This choice is not fortunate, as 
both constituents exhibit a negative EA\cite{chiu_temporary_1979},
which leads to CTE that overlap with the continuum at least for infinite
separation.
Therefore we use the alternative Na\textsubscript{2}--NaCl
complex, where Na\textsubscript{2} is the donor and NaCl the acceptor with
a positive EA (experimental adiabatic EA of 
0.73 eV\cite{miller_electron_1986}).
In order to catch the largely delocalized excited states we increased
the amount of space within our simulation box to
$x_\text{vac} = \unit[11]{\text{\AA}}$ around each atom and
decreased the grid spacing to $h=\unit[0.2]{\text{\AA}}$
due to the higher computational effort.

\begin{table}[h]
  \begin{tabular}{l|D{.}{.}{9}|D{.}{.}{2}|D{.}{.}{2}|D{.}{.}{2}}
  	 \multicolumn{1}{c|}{Mol.} &
  	 \multicolumn{1}{c|}{IP\textsubscript{exp.}\cite{linstrom_ion_2016}} &
  	 \multicolumn{1}{c|}{IP\textsubscript{calc}} &
  	 \multicolumn{1}{c|}{$-\epsilon_\text{H}$}&
  	 \multicolumn{1}{c}{$\text{d}E$}\\
     \hline
  	Na\textsubscript{2}        & 4.892\pm0.003 & 4.93 & 4.94 & -0.01\\
  	NaCl\textsuperscript{-}  & 0.727\pm0.010 & 0.79 & 0.78 & 0.01
  \end{tabular}
  \caption{Experimental, IP\textsubscript{exp.}, and calculated, 
    IP\textsubscript{calc}, ionization potential for the molecules 
    Na\textsubscript{2} and NaCl\textsuperscript{-} along with the 
    negative eigenvalue of the HOMO $-\epsilon_\text{H}$ and their 
    difference  
    $\text{d}E = {\rm IP}\textsubscript{calc} + \epsilon_\text{H}$. 
    All values are in eV.
  }
  \label{tab:FitNa2NaCl}
\end{table}

Using eq. (\ref{eq:fitgammaIP}) to calculate $\gamma_\text{opt}$ 
for the individual 
molecules leads to $\gamma_\text{opt} = \unit[0.38]{a_0^{-1}}$ for 
Na\textsubscript{2} and $\gamma_\text{opt} = \unit[0.40]{a_0^{-1}}$ for 
NaCl\textsuperscript{-}.
In order to obtain the optimal range separation parameter for the 
combined system, the Na\textsubscript{2}--NaCl
complex, we use minimization of the function \cite{stein_reliable_2009}
\begin{align}
  \label{eq:comp1}
  J(\gamma) = &\sum_{i=D^0,A^-} \Big| \epsilon_\text{HOMO}^{i} (\gamma) + 
              \text{IP}_i (N, \gamma)\Big| 
\end{align}
with $\text{IP}_i (N, \gamma) = E_\text{gs}^i (N_i-1, \gamma) - E_\text{gs}^i (N_i, \gamma)$,
where $D^0$ denotes the neutral donor, $A^-$ the acceptor anion 
and $N$ the number of electrons.  
The two molecules were considered separately 
where the experimental geometries of the neutral molecules 
from ref. \citenum{linstrom_constants_2016} were used.
This treatment results in a value of 
$\gamma_\text{opt} = \unit[0.39]{a_0^{-1}}$, which leads to the energies
in table \ref{tab:FitNa2NaCl} that exhibit good agreement to experiment. 
The eigenvalue of the NaCl LUMO ($\epsilon_\text{LUMO} =\unit[-0.57]{eV}$) 
differs from the eigenvalue of the  NaCl$^-$ HOMO which equals 
the NaCl EA through (\ref{eq:comp1}) by $\approx \unit[0.2]{eV}$. 
This effect is known 
and can be attributed to the derivative discontinuity\cite{stein_fundamental_2010,kronik_excitation_2012}.

\begin{figure}
  \centering
  \includegraphics{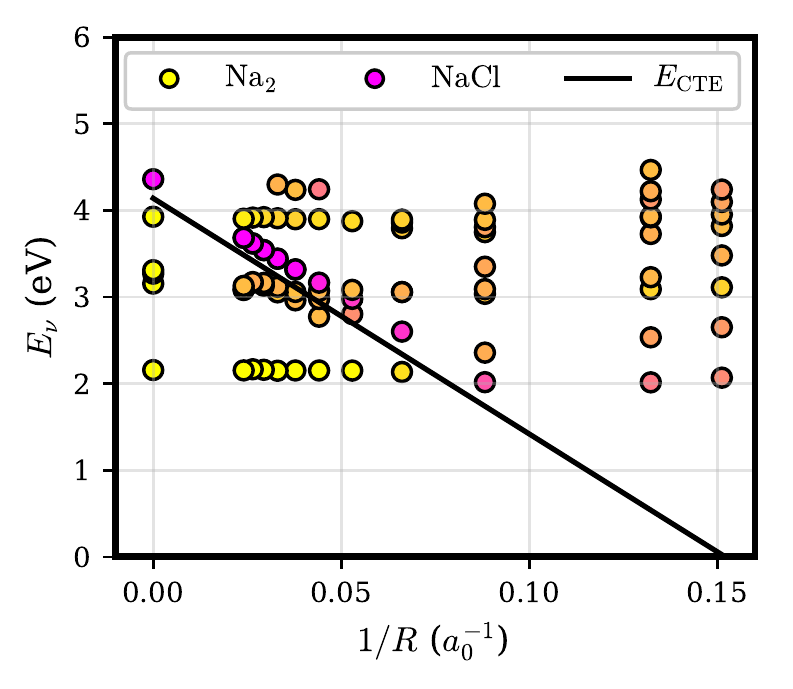}
  \caption{Linear response TDDFT excitations of the 
    Na\textsubscript{2}--NaCl-system depending on the
    inverse distance between the two molecules $\nicefrac{1}{R}$.
    The colors indicate localization of the unoccupied states involved in
    the excitations on Na\textsubscript{2} (yellow) or NaCl (purple). 
    Solid line: excitation energy after eq. (\ref{eq:MullCTE}).
  }
\label{fig:CTE_Na2NaCl_lr}
\end{figure}
In order to study CT excitations,
the molecules were placed with their axes parallel to each other.
We first consider the singlet excited state 
spectrum of the donor acceptor pair
calculated by linear response TDDFT
depending on the molecular separation as depicted in 
fig. \ref{fig:CTE_Na2NaCl_lr}.
The separation $R$ is given by the separation of the two parallel
molecular axes.
The excitations are colored by the weights of the 
involved unoccupied orbitals on the individual molecules. 
Only excitations with more than $\unit[50]{\%}$ contribution
from the Na$_2$ HOMO and either an oscillator strength $\ge 10^{-2}$
or $\unit[90]{\%}$ weight on NaCl are considered.
The excitation spectrum shows a clear CTE where the involved unoccupied
states are clearly located on NaCl and its energy follows the
expectation of Mullikens law eq. (\ref{eq:MullCTE}) for larger separations 
($1/R\le 0.08~a_0^{-1}$)
as expected.
There is a small constant deviation from the expectations of 
eq. (\ref{eq:MullCTE}) that
arises from the difference between the eigenvalue of the 
NaCl LUMO 
and the NaCl EA [entering in eq. (\ref{eq:MullCTE})].
This is validated by the NaCl point at $1/R=0$ which is placed 
at the energy of IP$_D$ + $\epsilon_\text{LUMO, NaCl}$.

There is more interaction between the two molecules 
for smaller distances. Therefore the excited states start to 
mix and their nature is harder to identify. 
The excitation energies exhibit a noticeable 
blur even for large $R$, where little
interaction
between the molecules would be expected. 
This is particularly visible in excitations localized on the 
Na\textsubscript{2} molecule, which exhibit a large degree of 
delocalization, 
viewable by the coloring of excitations around $\unit[3]{eV}$.
\begin{figure}
  \centering
  \includegraphics{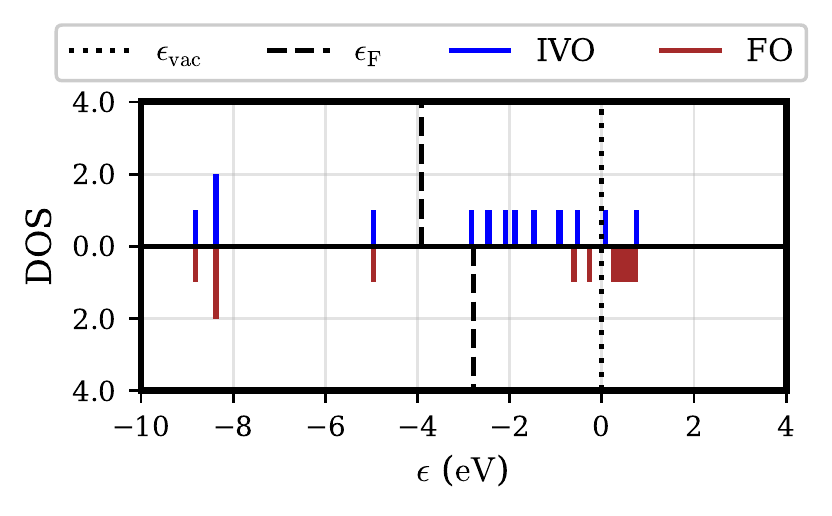}
  \caption{Density of electronic states (DOS)
    of the
    Na\textsubscript{2}--NaCl-system at distance $\unit[8]{\text{\AA}}$
    for the canonical Fock operator (FO) and for
    improved virtual orbitals (IVO).
    The vacuum level $\epsilon_\text{vac}$ is indicated 
    and used as energy reference. 
    The Fermi level is set to $\epsilon_\text{F} = 
    \frac{\epsilon_\text{HOMO} + \epsilon_\text{LUMO}}{2}$ 
     (all states below $\epsilon_\text{F}$ are 
    occupied, all states above are unoccupied).  
  }
  \label{fig:DOS}
\end{figure}
This numerical noise can be attributed to the fact 
that already the third unoccupied state in the Na$_2$ ground-state 
calculation is above the vacuum level and thus this state 
and all higher ones are influenced by eigenstates of the 
simulation box (fig. \ref{fig:DOS}).
This hinders the convergence of these excitations
in the number of unoccupied states involved as shown below.

In HFT the Coulomb- and exchange interaction cancel for an
occupied state with itself,  
such that these states are subject to the
interaction with $N-1$ electrons in an $N$ electron system.
This cancellation is absent
for
unoccupied (virtual) states 
making them subject to the interaction with all $N$ electrons.
Unoccupied states therefore ``see'' a neutralized core 
in a neutral system and thus approximate (exited) EAs 
in HFT
and not excitations of the neutral system\cite{koopmans_uber_1934,baerends_kohnsham_2013}.
As many neutral closed shell molecules exhibit a
negative EA \cite{baerends_kohnsham_2013},
the eigenvalues of unoccupied states in HFT type calculations
like RSF become positive, i.e. they reside in the 
continuum (see fig. \ref{fig:DOS}).
In calculations utilizing basis-sets, these states are stabilized
by the finite size of the basis-set\cite{rosch_comment_1997,tozer_computation_2005,baerends_kohnsham_2013} and their properties 
thus get strongly basis set dependent.
Similarly, these states strongly couple to eigenstates of the
simulation box in grid-based calculations used here.
Therefore RSF DFT calculations become numerically very demanding on
grids due to numerical
instabilities\cite{wurdemann_berechnung_2016}.

\section{Improved virtual orbitals}

Improved virtual orbitals are a remedy for the difficulties
mentioned above. 
The canonical Hartree-Fock operator is not unique in case of unoccupied
orbitals in HFT.
These states do not contribute to the Slater determinant built
from occupied states exclusively, any set of orbitals
orthogonal to the occupied states could serve as valid
set of unoccupied states.
Following the work of \citeauthor{kelly_correlation_1963}\cite{kelly_correlation_1963,kelly_many-body_1964},  \citeauthor{huzinaga_virtual_1970}
devised a scheme to use this freedom to better approximate 
excitations in HFT\cite{huzinaga_virtual_1970,huzinaga_virtual_1971}.
In this scheme Coulomb and exchange interactions between
a virtual hole in an initial occupied orbital $k$ and
the virtual orbitals are described by a
modified Fock-Operator \cite{huzinaga_virtual_1970}
\begin{align}
  \hat{F}^\text{IVO}_i &= \hat{F}_i + \hat{V} \\
  \hat{V} &= \left( 1 - \hat{P} \right)
  \label{eq:HuzProV}
  \hat{\Omega}_k\left( 1 - \hat{P} \right) \\
  \hat {P} &= \sum_i^N \left | \psi_i \rangle \langle \psi_i \right |.
  \label{eq:HuzPro}
\end{align}
$\hat{F}_i$ denotes the canonical Fock-Operator and
$\psi_i$ the non-interacting single particle Hartree-Fock or,
in the case of the present study, KS orbitals.
$\hat{P}$ separates the space of the unoccupied orbitals
from the occupied ones, circumventing slight changes in the
eigenstates of the occupied states which occur
otherwise\cite{kelly_many-body_1964,huzinaga_virtual_1970}.
The rotation operator $\hat{\Omega}_k$ can be chosen
arbitrarily as long as it is Hermitian\cite{huzinaga_virtual_1970}.
\citeauthor{huzinaga_virtual_1970} suggested to use\cite{huzinaga_virtual_1971}
\begin{align}
  \hat{\Omega}_{k} &= - \hat{J}_k +  \hat{K}_k \pm \hat{K}_k
  \label{eq:Huzi_Omega}
\end{align}
for closed shell systems (as is the case discussed below)
following the work of \citeauthor{hunt_excited_1969}\cite{hunt_excited_1969}.
$\hat{J}_k$ denotes the Coulomb-, $\hat{K}_k$ the exchange
operator and $k$ the band-index of the orbital
to excite from.
The second exchange term can be used to approximate either
singlet (``+'', $\hat{\Omega}_k^\text{S}$)  and
triplet (``-'', $\hat{\Omega}_k^\text{T}$) excitations,
or can be omitted ($\hat{\Omega}_k^\text{A}$) to approximate their average.
The initial orbital $k$ can be chosen
arbitrarily \cite{huzinaga_virtual_1970,huzinaga_virtual_1971}
and determines the nature of the excitations to be described.
Virtual orbitals subject to this scheme are called
improved virtual orbitals (IVO)
\cite{huzinaga_virtual_1970,huzinaga_virtual_1971}.

The IVO scheme (\ref{eq:HuzPro}) can also be applied within RSF 
setting $\hat{K}_k = \hat{K}_k^\text{RSF}$ as in eq. 
(\ref{eq:RSF_EXX_om})
and $\hat{J}_k = \hat{J}_k^\text{RSF}$ 
as the corresponding screened Coulomb counterpart.
We have disregarded the orthogonalization through operator $\hat{P}$
due to numerical instabilities and $\hat{\Omega}_k$ was
applied to the unoccupied states only.
The matrix elements in the exact exchange part of the 
Hamiltonian mixing occupied and unoccupied states were 
set to zero consistently.
We have verified that both approaches 
lead to virtually identical eigenvalues (see SI).
As we will investigate the possibility of the combination of RSF and IVOs for the prediction of  excitation energies by means of ground-state calculations we have used the singlet form: $\hat{\Omega}_k^\text{S} = - \hat{J}_k + 2 \hat{K}_k$,
where $k$ is the quantum number of the HOMO located on Na$_2$.
The resulting density of states is depicted in 
fig. \ref{fig:DOS}. While the occupied state energies of the 
canonical Fock operator and the IVO operator agree by definition, 
the IVO operator leads to many more states below the 
vacuum level $\epsilon_\text{vac}$.
There are infinitely many Rydberg states below 
$\epsilon_\text{vac}$ in principle,
but these do not appear due to the
finite box size in our calculations.

It was shown, that 
lrTDHFT excitation energies in the Tamm-Dancoff approximation and
IVO Eigenenergies agree in a two-orbital two 
electron model\cite{casida_progress_2012}.
Therefore the IVOs can be expected to
represent a good basis for lrTDDFT calculations.
Their use involves slight modifications of the usual
formalism in the formulation of lrTDDFT as
generalized eigenvalue problem\cite{dreuw_long-range_2003,casida_time-dependent_2009,akinaga_intramolecular_2009}
\begin{align}
  \begin{pmatrix}
    \mathbf{A} & \mathbf{B} \\
    \mathbf{B^\ast} & \mathbf{A^\ast}
  \end{pmatrix} \begin{pmatrix} \vec{X} \\ 
    \vec{Y} 
  \end{pmatrix} &= \omega 
  \begin{pmatrix}
    \mathbf{1} & \mathbf{0} \\
    \mathbf{0} & -\mathbf{1}
  \end{pmatrix} 
  \begin{pmatrix} \vec{X} \\ 
    \vec{Y} 
  \end{pmatrix},
\end{align}
where $\vec{X}/\vec{Y}$ denote the 
excitations/de-excitations, and $\mathbf{1}$ the unity-matrix.
The matrix elements of $\mathbf{A}$ and $\mathbf{B}$ are given 
by
\begin{align}  
  A_{ia\sigma,jb\tau} &= \delta_{ij}\delta_{ab} \delta_{\sigma\tau} \left( \epsilon_a - \epsilon_i \right) + K_{ia\sigma,jb\tau} \label{eq:lrA} \\
B_{ia\sigma,jb\tau} &=  K_{ia\sigma,bj\tau}\label{eq:lrB}
\end{align}
with the $K_{ia\sigma,jb\tau}$ written in the most general form (see SI)
\begin{align}
K_{ia\sigma,jb\tau} &= \left( i_\sigma a_\sigma|\nicefrac{1}{r_{12}}|j_\tau b_\tau \right) \notag \\
&\quad + \left( 1 - \alpha - \beta \right) \left(i_\sigma a_\sigma |f_\text{xc}|j_\tau b_\tau\right) \notag \\
  & \quad + \beta \left(i_\sigma a_\sigma |f_\text{xc}^\text{RSF}|j_\tau b_\tau\right) \notag \\
  & \quad - \delta_{\sigma\tau} \left(i j\left |\frac{\alpha + \beta \left ( 1 - \omega_\text{RSF}\right) }{r_{12}}\right |a b\right) \; .\label{eq:lrK}
\end{align}
Mulliken notation 
\begin{align}
\left( ab | \hat{x} | cd \right) =&\notag \\
 \int \int a^\ast(\vec{r_1}) b(\vec{r_1})& \hat{x} (\vec{r_1}, \vec{r_2}, ...)  c^\ast(\vec{r_2}) d(\vec{r_2}) \text{d}\vec{r_1} \text{d}\vec{r_2}\notag 
\end{align} was used and
$f_\text{xc} =  \frac{\delta^2E_\text{xc}}{\delta\varrho\left(\vec{r_1}\right)\delta\varrho\left(\vec{r_2}\right)}$ is exchange-correlation kernel of the local
functional and $f_\text{xc}^\text{RSF}$ the damped exchange-correlation kernel
derived from $E_\text{x}^\text{GGA} (\gamma) + E_\text{c}^\text{GGA} = E_\text{xc}^\text{GGA} (\gamma)$.
Occupied orbital indices are denoted by $i$ and $j$,
unoccupied orbitals by $a$ and $b$, and $\sigma$ and $\tau$ are the 
spin-indices, while
$\alpha$ and $\beta$ are the mixing parameters from the CAM scheme
eq. (\ref{eq:CAM}).  
The use of IVOs requires to modify the matrix 
$\mathbf{A}$ to\cite{berman_fast_1979}
\begin{align}
A_{ia\sigma,jb\tau}^\text{IVO} &= A_{ia\sigma,jb\tau} \notag \\
& \quad + \delta_{ab} \delta_{\sigma\tau}\left[ \left(aa|kk\right) - \left(ak|ka\right) \right . \notag \\
& \left .\quad \quad \mp \left(ak|ka\right)\right] \notag\\
&= A_{ia\sigma,jb\tau} - \delta_{ab} \delta_{\sigma\tau} \left\langle a \left| \hat{\Omega}_k \right| a \right\rangle\label{eq:LRIVO} 
\end{align}
where $k$ denotes the excitation orbital and $\hat{\Omega}_k$ 
is defined in (\ref{eq:Huzi_Omega}). The matrix $\mathbf{B}$ remains the same.

\begin{figure}
  \centering
  \includegraphics{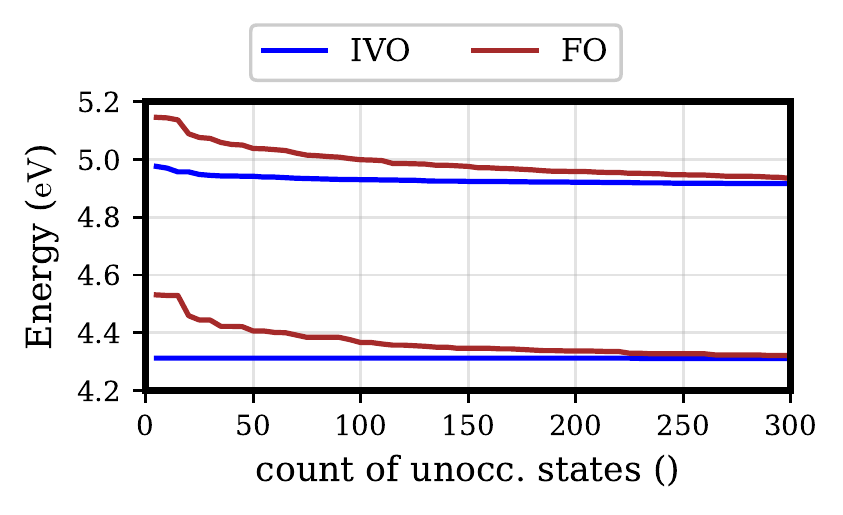}
  \caption{
    Convergence of the two lowest singlet excitation energies
    of the isolated NaCl molecule using
    the canonical Fock operator (FO) and
    improved virtual orbitals (IVO).
  }
  \label{fig:nacl_tddft_j}
\end{figure} 
The use of IVO indeed facilitates the calculation of excitations using lrTDDFT 
as shown in fig. \ref{fig:nacl_tddft_j} for the two
first singlet excitations of the isolated NaCl molecule.
These are excitations mainly from the degenerated HOMO and HOMO-1 to the 
LUMO and converge rapidly in the IVO basis.
In contrast, more than 200 unoccupied orbitals of the 
canonical HF operator are needed
to arrive at converged energies, which shows that
this basis is not very 
appropriate for the description of excitations.
It is known, that excitations calculated
by linear response time dependent HFT need a
large linear combination of single particle excitations
(i.e. a high number of unoccupied states),
while excitations calculated by lrTDDFT using the kernels of
local functionals can often be described 
by an individual excitation\cite{vanMeerPhysicalMeaningVirtual2014,vanMeerNaturalexcitationorbitals2017}.
The twelve unoccupied states that are used in the lrTDDFT calculations
for the  Na$_2$--NaCl-system presented in fig. \ref{fig:CTE_Na2NaCl_lr}
utilizing the canonical FO are therefore far from converged
for the neutral excitations.
This is seen by the CTE state and the first excitation of Na$_2$
in fig. \ref{fig:CTE_Na2NaCl_lr}, which are rather clear 
as these only involve states with negative eigenvalues
which do not couple with the eigenstates of the simulation box. 

\begin{figure}
  \centering
  \includegraphics{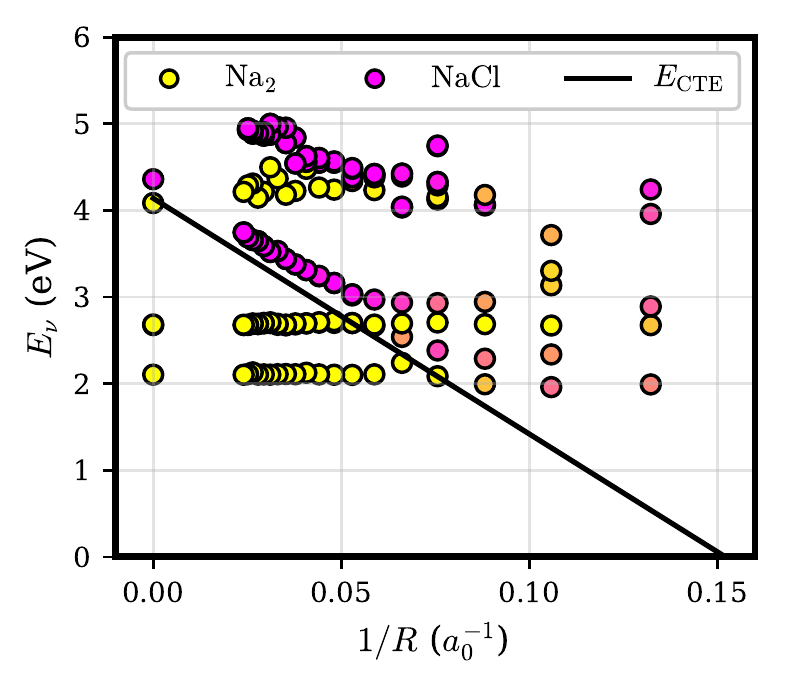}
  \caption{
  	Linear response TDDFT excitations of the 
    Na\textsubscript{2}--NaCl-system depending on the
    inverse distance between the two molecules $\nicefrac{1}{R}$ as in fig. \ref{fig:CTE_Na2NaCl_lr} using IVOs with $\hat{\Omega}_k^S$ and $k$ was chosen as the HOMO of the Na$_2$ molecule. Besides the use of the IVOs, everything like fig. \ref{fig:CTE_Na2NaCl_lr}.
  }
  \label{fig:CTE_Na2NaCl}
\end{figure}
The IVOs provide a better basis for the calculation of excitations
as can be seen in the lrTDDFT spectrum for the Na$_2$--NaCl system
utilizing IVOs depicted in fig. \ref{fig:CTE_Na2NaCl}.
The depicted states are selected and colored as in fig. \ref{fig:CTE_Na2NaCl_lr}.
The ``hole'' is in the HOMO of the Na$_2$ molecule.
The spectrum gets much clearer than in fig. \ref{fig:CTE_Na2NaCl_lr}
and  the energy of the second excitation on Na$_2$ is clearly lower
and it is strongly localized on Na$_2$
due to better convergence. 
Again, the calculated CTE energies are in perfect agreement to 
the behavior predicted by Mullikens law eq. (\ref{eq:MullCTE}),
except the offset due to the difference between the 
eigenvalue of the NaCl LUMO
and its calculated EA.
A second CTE state approximately $\unit[1]{eV}$ above the first 
can be identified. which is not the case in 
fig. \ref{fig:CTE_Na2NaCl_lr}. 
This is an effect of the stronger stabilization due to the artificial hole.

This suggests that RSFs might be also used to calculate the energetics
of CTEs by means of ground-state calculations 
within the IVO formalism.
This conjecture can be further rationalized
in particular for CTEs, where HOMO and LUMO are 
spatially strongly separated
such that most of the weight in eq. (\ref{eq:CAM}) resides on the
exchange integrals from HFT.
By adding $\hat{V}$ from eq. (\ref{eq:HuzProV}) to the 
Kohn-Sham-Hamiltonian
and taking the HOMO as the hole-state $k$,
the eigenvalue of the LUMO in a RSF becomes
$\epsilon_\text{LUMO} \approx -\text{EA}_A^\text{calc} - \hat{J}_k$,
where it was used that the HOMO and LUMO orbitals do not overlap
which leads to vanishing exchange ($\hat{K}_k \to 0$).
With $\epsilon_\text{HOMO}=-\text{IP}_D^\text{calc}$ the
energetic difference between HOMO and LUMO results in
\begin{align}
  \epsilon_\text{LUMO} - \epsilon_\text{HOMO} &\approx
  \text{IP}_D^\text{calc} -\text{EA}_A^\text{calc} - \frac{1}{R},
\end{align}
which is equal to the desired result of eq. (\ref{eq:MullCTE}). 

\begin{figure}
  \centering
  \includegraphics{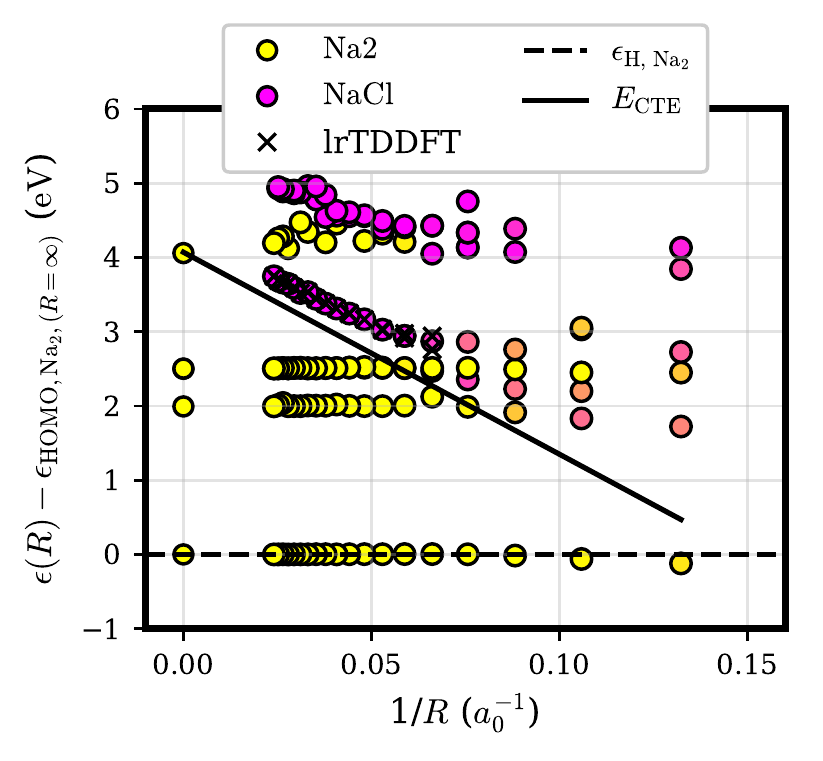}
  \caption{
    Eigenvalues of the Na\textsubscript{2}--NaCl-system 
    utilizing IVOs with $\hat{\Omega}_k^S$, where $k$ is the HOMO of the
    system located on Na$_2$. The Eigenvalues are given relative
    to the HOMO for $R\to\infty$ (dashed line) and their dependence
    on the inverse distance between the two molecules 
    $\nicefrac{1}{R}$ is shown. 
    The states are colored by their weight on the individual molecules. 
    Solid line: eigenvalue of the LUMO of the NaCl molecule according 
    to Mullikens law. 
    The stars mark the CTE
    energies from TDDFT utilizing IVOs (fig. \ref{fig:CTE_Na2NaCl}).
  }
  \label{fig:CTE_Na2NaCl_IVO}
\end{figure}
The possibility to calculate CTE  by the combination of RSF and IVO by means of ground-state calculations is confirmed by fig. \ref{fig:CTE_Na2NaCl_IVO}.
It shows the 
eigenvalues of the HOMO as well as the eigenvalues of the
unoccupied states of the Na$_{2}$--NaCl--System 
calculated by the utilization of the IVOs with 
$\hat{\Omega}_k^\text{S}$ as rotation operator in dependence 
of the intermolecular
distance $R$ similar to figs. \ref{fig:CTE_Na2NaCl_lr}
and \ref{fig:CTE_Na2NaCl}. 
In the asymptote $R\to\infty$ ($\frac{1}{R} \to 0$) the eigenvalues from the eigenstates of the isolated Na\textsubscript{2} molecule, which were subject to the modified Fock operator $\hat{F}^\text{IVO}$ with $\hat{\Omega}_k^\text{S}$, are shown. 
The eigenstates were colored by their weight on the individual molecules, where the projected local density of states was used. 
Similar to figs. \ref{fig:CTE_Na2NaCl_lr} and \ref{fig:CTE_Na2NaCl} only states with an oscillator strength $\ge 10^{-2}$ for excitations from the Na$_2$-HOMO or $\unit[90]{\%}$ weight on NaCl are considered.
For $R > \approx \unit[10]{\text{\AA}}$  ($\frac{1}{R} \le \unit[0.05]{a_0^{-1}}$) the eigenvalue of the LUMO located on the NaCl molecule, which corresponds to the CTE state, is  clearly visible and follows the straight line defined by Mullikens law with a constant slight offset. In this region the calculated eigenvalues of the LUMO are in perfect agreement with the energies predicted by lrTDDFT utilizing IVOs.
The offset to Mullikens law is based on the difference between the 
eigenvalue of the NaCl LUMO and its calculated EA, see above. 
As in fig. \ref{fig:CTE_Na2NaCl} a second CTE state 
approximately one $\unit{eV}$ above the first CTE state can be identified.

While RSF open a way to calculate the energetics
and oscillator strengths of Rydberg- and  charge transfer
excitations by lrTDDFT, the combination of RSF with IVOs opens
a way to calculate the energetics of these excitations by
means of ground-state calculations. 
Generalized lrTDDFT calculations using empty orbitals of the canonical FO 
need a much larger number of unoccupied states and thus are very hard to 
converge. This problem can be circumvented by utilizing IVOs that provide
usable energies based on a rather small basis. 
But for lrTDDFT one needs to calculate Coulomb and exchange for every pair of excitation and de-excitation. Therefore lrTDDFT need a lot of computational resources compared
to ground-state calculations. Thus utilizing the combination of RSF and IVOs one can calculate the 
energetics of CTEs on a light computational footprint.

\section{Conclusions}

In this article the implementation of RSFs utilizing the Slater function 
using the method of PAW on real space grids 
was presented.
The screened Poisson equation was used to calculate the 
RSF exchange integrals on Cartesian grids, while integrations on
radial grids were performed using the integration kernel devised by \citeauthor{rico_repulsion_2012}.

The implementation was verified against literature, where excellent agreement
has been found. 
Slight differences had been traced to the use of 
non-converged basis sets used in the literature, which underlines
the importance of the basis set choice.
Comparison between calculated and experimental mean 
ligand removal energies unveiled a poor description of binding situations 
including $d$-electrons by RSF and hybrid functionals that 
might be attributed of the poor treatment of $d$-binding in HFT.

lrTDDFT within the generalized Kohn-Sham scheme using canonical unoccupied 
Fock-orbitals  is hard to converge as the empty states approximate 
(excited) electron affinities. 
These orbitals are thus inappropriate to describe neutral excitations. 
As a remedy, we combined Huzinaga's IVOs
with RSF and extended the linear response TDDFT coupling matrix accordingly. 
The RSF IVO orbitals are a superior basis for the calculation of excitations 
with much better convergence properties than these of the canonical FO.
These orbitals and their energies not only improve the calculation
of lrTDDFT. Due to their construction, they also open a way to 
calculate the energetics of CTEs 
by means of ground-state calculations. 
The much smaller numerical footprint of a ground state calculation 
as compared to lrTDDFT might enable calculation of these energies 
for systems unreachable by lrTDDFT.

\section*{Acknowledgements}
R. W. acknowledges funding by the Freiburger Materialforschungszentrum
and also thanks Miguel Marques from the \emph{libxc} project 
for a bug fix and further improvements on the code implemented 
in \emph{libxc}.
Computational resources of FZ-Jülich\cite{krause_jureca:_2016} 
(project HFR08) are thankfully acknowledged and
the authors acknowledge support by the state of 
Baden-Württemberg through bwHPC
and the German Research Foundation (DFG) 
through grant no INST 39/963-1 FUGG. 

\section*{Supporting information available}
The evaluation of the local terms for the screened exchange, 
the influence of the grid-spacing and the amount of vacuum 
around each atom on the eigenvalues and derived energies, as 
well as the analytic expressions for the first, second and third 
derivative of the exchange term for an RSF with the use of the 
Slater function, the analytic solution of the screened 
Poisson equation for a Gaussian shaped density along with 
it's derivation, the effects of dropping the projection 
operator $\hat{P}$ on the IVOs, excitation energies for the disodium 
molecule calculated by lrTDDFT utilizing RSF and IVOs with the three 
different forms of $\Omega_k$ and the 
rationale for the exchange terms in lrTDDFT are given in the 
supporting information.
 
\bibliography{./Yukawa_GPAW}

\begin{thebibliography}{102}%
\makeatletter
\providecommand \@ifxundefined [1]{%
 \@ifx{#1\undefined}
}%
\providecommand \@ifnum [1]{%
 \ifnum #1\expandafter \@firstoftwo
 \else \expandafter \@secondoftwo
 \fi
}%
\providecommand \@ifx [1]{%
 \ifx #1\expandafter \@firstoftwo
 \else \expandafter \@secondoftwo
 \fi
}%
\providecommand \natexlab [1]{#1}%
\providecommand \enquote  [1]{``#1''}%
\providecommand \bibnamefont  [1]{#1}%
\providecommand \bibfnamefont [1]{#1}%
\providecommand \citenamefont [1]{#1}%
\providecommand \href@noop [0]{\@secondoftwo}%
\providecommand \href [0]{\begingroup \@sanitize@url \@href}%
\providecommand \@href[1]{\@@startlink{#1}\@@href}%
\providecommand \@@href[1]{\endgroup#1\@@endlink}%
\providecommand \@sanitize@url [0]{\catcode `\\12\catcode `\$12\catcode
  `\&12\catcode `\#12\catcode `\^12\catcode `\_12\catcode `\%12\relax}%
\providecommand \@@startlink[1]{}%
\providecommand \@@endlink[0]{}%
\providecommand \url  [0]{\begingroup\@sanitize@url \@url }%
\providecommand \@url [1]{\endgroup\@href {#1}{\urlprefix }}%
\providecommand \urlprefix  [0]{URL }%
\providecommand \Eprint [0]{\href }%
\providecommand \doibase [0]{http://dx.doi.org/}%
\providecommand \selectlanguage [0]{\@gobble}%
\providecommand \bibinfo  [0]{\@secondoftwo}%
\providecommand \bibfield  [0]{\@secondoftwo}%
\providecommand \translation [1]{[#1]}%
\providecommand \BibitemOpen [0]{}%
\providecommand \bibitemStop [0]{}%
\providecommand \bibitemNoStop [0]{.\EOS\space}%
\providecommand \EOS [0]{\spacefactor3000\relax}%
\providecommand \BibitemShut  [1]{\csname bibitem#1\endcsname}%
\let\auto@bib@innerbib\@empty
\bibitem [{\citenamefont {Sariciftci}\ \emph {et~al.}(1992)\citenamefont
  {Sariciftci}, \citenamefont {Smilowitz}, \citenamefont {Heeger},\ and\
  \citenamefont {Wudl}}]{sariciftci_photoinduced_1992}%
  \BibitemOpen
  \bibfield  {author} {\bibinfo {author} {\bibfnamefont {N.~S.}\ \bibnamefont
  {Sariciftci}}, \bibinfo {author} {\bibfnamefont {L.}~\bibnamefont
  {Smilowitz}}, \bibinfo {author} {\bibfnamefont {A.~J.}\ \bibnamefont
  {Heeger}}, \ and\ \bibinfo {author} {\bibfnamefont {F.}~\bibnamefont
  {Wudl}},\ }\href {\doibase 10.1126/science.258.5087.1474} {\bibfield
  {journal} {\bibinfo  {journal} {Science}\ }\textbf {\bibinfo {volume}
  {258}},\ \bibinfo {pages} {1474} (\bibinfo {year} {1992})}\BibitemShut
  {NoStop}%
\bibitem [{\citenamefont {Samor{\'\i}}\ \emph {et~al.}(2002)\citenamefont
  {Samor{\'\i}}, \citenamefont {Severin}, \citenamefont {Simpson},
  \citenamefont {M{\"u}llen},\ and\ \citenamefont
  {Rabe}}]{samori_epitaxial_2002}%
  \BibitemOpen
  \bibfield  {author} {\bibinfo {author} {\bibfnamefont {P.}~\bibnamefont
  {Samor{\'\i}}}, \bibinfo {author} {\bibfnamefont {N.}~\bibnamefont
  {Severin}}, \bibinfo {author} {\bibfnamefont {C.~D.}\ \bibnamefont
  {Simpson}}, \bibinfo {author} {\bibfnamefont {K.}~\bibnamefont {M{\"u}llen}},
  \ and\ \bibinfo {author} {\bibfnamefont {J.~P.}\ \bibnamefont {Rabe}},\
  }\href {\doibase 10.1021/ja020323q} {\bibfield  {journal} {\bibinfo
  {journal} {J. Am. Chem. Soc.}\ }\textbf {\bibinfo {volume} {124}},\ \bibinfo
  {pages} {9454} (\bibinfo {year} {2002})}\BibitemShut {NoStop}%
\bibitem [{\citenamefont {Blundell}\ and\ \citenamefont
  {Pratt}(2004)}]{blundell_organic_2004}%
  \BibitemOpen
  \bibfield  {author} {\bibinfo {author} {\bibfnamefont {S.~J.}\ \bibnamefont
  {Blundell}}\ and\ \bibinfo {author} {\bibfnamefont {F.~L.}\ \bibnamefont
  {Pratt}},\ }\href {\doibase 10.1088/0953-8984/16/24/R03} {\bibfield
  {journal} {\bibinfo  {journal} {J. Phys.: Condens. Matter}\ }\textbf
  {\bibinfo {volume} {16}},\ \bibinfo {pages} {R771} (\bibinfo {year}
  {2004})}\BibitemShut {NoStop}%
\bibitem [{\citenamefont {Mulliken}\ and\ \citenamefont
  {Person}(1969)}]{mulliken_molecular_1969}%
  \BibitemOpen
  \bibfield  {author} {\bibinfo {author} {\bibfnamefont {R.~S.}\ \bibnamefont
  {Mulliken}}\ and\ \bibinfo {author} {\bibfnamefont {W.~B.}\ \bibnamefont
  {Person}},\ }\href@noop {} {\emph {\bibinfo {title} {Molecular Complexes: A
  Lecture and Reprint Volume}}}\ (\bibinfo  {publisher}
  {{Wiley-Interscience}},\ \bibinfo {address} {New York; London},\ \bibinfo
  {year} {1969})\ \bibinfo {note} {oCLC: 499972894}\BibitemShut {NoStop}%
\bibitem [{\citenamefont {Hohenberg}\ and\ \citenamefont
  {Kohn}(1964)}]{hohenberg_inhomogeneous_1964}%
  \BibitemOpen
  \bibfield  {author} {\bibinfo {author} {\bibfnamefont {P.}~\bibnamefont
  {Hohenberg}}\ and\ \bibinfo {author} {\bibfnamefont {W.}~\bibnamefont
  {Kohn}},\ }\href {\doibase 10.1103/PhysRev.136.B864} {\bibfield  {journal}
  {\bibinfo  {journal} {Phys. Rev.}\ }\textbf {\bibinfo {volume} {136}},\
  \bibinfo {pages} {B864} (\bibinfo {year} {1964})}\BibitemShut {NoStop}%
\bibitem [{\citenamefont {Kohn}\ and\ \citenamefont
  {Sham}(1965)}]{kohn_self-consistent_1965}%
  \BibitemOpen
  \bibfield  {author} {\bibinfo {author} {\bibfnamefont {W.}~\bibnamefont
  {Kohn}}\ and\ \bibinfo {author} {\bibfnamefont {L.~J.}\ \bibnamefont
  {Sham}},\ }\href {\doibase 10.1103/PhysRev.140.A1133} {\bibfield  {journal}
  {\bibinfo  {journal} {Phys. Rev.}\ }\textbf {\bibinfo {volume} {140}},\
  \bibinfo {pages} {A1133} (\bibinfo {year} {1965})}\BibitemShut {NoStop}%
\bibitem [{\citenamefont {Livshits}\ and\ \citenamefont
  {Baer}(2007)}]{livshits_well-tempered_2007}%
  \BibitemOpen
  \bibfield  {author} {\bibinfo {author} {\bibfnamefont {E.}~\bibnamefont
  {Livshits}}\ and\ \bibinfo {author} {\bibfnamefont {R.}~\bibnamefont
  {Baer}},\ }\href {\doibase 10.1039/B617919C} {\bibfield  {journal} {\bibinfo
  {journal} {Phys. Chem. Chem. Phys.}\ }\textbf {\bibinfo {volume} {9}},\
  \bibinfo {pages} {2932} (\bibinfo {year} {2007})},\ \Eprint
  {http://arxiv.org/abs/0701493} {arXiv:0701493 [cond-mat]} \BibitemShut
  {NoStop}%
\bibitem [{\citenamefont {Chai}\ and\ \citenamefont
  {Head-Gordon}(2008)}]{chai_systematic_2008}%
  \BibitemOpen
  \bibfield  {author} {\bibinfo {author} {\bibfnamefont {J.-D.}\ \bibnamefont
  {Chai}}\ and\ \bibinfo {author} {\bibfnamefont {M.}~\bibnamefont
  {Head-Gordon}},\ }\href {\doibase 10.1063/1.2834918} {\bibfield  {journal}
  {\bibinfo  {journal} {J. Chem. Phys.}\ }\textbf {\bibinfo {volume} {128}},\
  \bibinfo {pages} {084106} (\bibinfo {year} {2008})}\BibitemShut {NoStop}%
\bibitem [{\citenamefont {Cohen}\ \emph {et~al.}(2008)\citenamefont {Cohen},
  \citenamefont {Mori-S{\'a}nchez},\ and\ \citenamefont
  {Yang}}]{cohen_insights_2008}%
  \BibitemOpen
  \bibfield  {author} {\bibinfo {author} {\bibfnamefont {A.~J.}\ \bibnamefont
  {Cohen}}, \bibinfo {author} {\bibfnamefont {P.}~\bibnamefont
  {Mori-S{\'a}nchez}}, \ and\ \bibinfo {author} {\bibfnamefont
  {W.}~\bibnamefont {Yang}},\ }\href {\doibase 10.1126/science.1158722}
  {\bibfield  {journal} {\bibinfo  {journal} {Science}\ }\textbf {\bibinfo
  {volume} {321}},\ \bibinfo {pages} {792} (\bibinfo {year}
  {2008})}\BibitemShut {NoStop}%
\bibitem [{\citenamefont {Baer}\ \emph {et~al.}(2010)\citenamefont {Baer},
  \citenamefont {Livshits},\ and\ \citenamefont {Salzner}}]{baer_tuned_2010}%
  \BibitemOpen
  \bibfield  {author} {\bibinfo {author} {\bibfnamefont {R.}~\bibnamefont
  {Baer}}, \bibinfo {author} {\bibfnamefont {E.}~\bibnamefont {Livshits}}, \
  and\ \bibinfo {author} {\bibfnamefont {U.}~\bibnamefont {Salzner}},\ }\href
  {\doibase 10.1146/annurev.physchem.012809.103321} {\bibfield  {journal}
  {\bibinfo  {journal} {Annu. Rev. Phys. Chem.}\ }\textbf {\bibinfo {volume}
  {61}},\ \bibinfo {pages} {85} (\bibinfo {year} {2010})}\BibitemShut {NoStop}%
\bibitem [{\citenamefont {Vosko}\ \emph {et~al.}(1980)\citenamefont {Vosko},
  \citenamefont {Wilk},\ and\ \citenamefont {Nusair}}]{vosko_accurate_1980}%
  \BibitemOpen
  \bibfield  {author} {\bibinfo {author} {\bibfnamefont {S.~H.}\ \bibnamefont
  {Vosko}}, \bibinfo {author} {\bibfnamefont {L.}~\bibnamefont {Wilk}}, \ and\
  \bibinfo {author} {\bibfnamefont {M.}~\bibnamefont {Nusair}},\ }\href
  {\doibase 10.1139/p80-159} {\bibfield  {journal} {\bibinfo  {journal} {Can.
  J. Phys.}\ }\textbf {\bibinfo {volume} {58}},\ \bibinfo {pages} {1200}
  (\bibinfo {year} {1980})}\BibitemShut {NoStop}%
\bibitem [{\citenamefont {Perdew}\ and\ \citenamefont
  {Zunger}(1981)}]{perdew_self-interaction_1981}%
  \BibitemOpen
  \bibfield  {author} {\bibinfo {author} {\bibfnamefont {J.~P.}\ \bibnamefont
  {Perdew}}\ and\ \bibinfo {author} {\bibfnamefont {A.}~\bibnamefont
  {Zunger}},\ }\href {\doibase 10.1103/PhysRevB.23.5048} {\bibfield  {journal}
  {\bibinfo  {journal} {Phys. Rev. B}\ }\textbf {\bibinfo {volume} {23}},\
  \bibinfo {pages} {5048} (\bibinfo {year} {1981})}\BibitemShut {NoStop}%
\bibitem [{\citenamefont {Perdew}\ and\ \citenamefont
  {Wang}(1992)}]{perdew_accurate_1992}%
  \BibitemOpen
  \bibfield  {author} {\bibinfo {author} {\bibfnamefont {J.~P.}\ \bibnamefont
  {Perdew}}\ and\ \bibinfo {author} {\bibfnamefont {Y.}~\bibnamefont {Wang}},\
  }\href {\doibase 10.1103/PhysRevB.45.13244} {\bibfield  {journal} {\bibinfo
  {journal} {Phys. Rev. B}\ }\textbf {\bibinfo {volume} {45}},\ \bibinfo
  {pages} {13244} (\bibinfo {year} {1992})}\BibitemShut {NoStop}%
\bibitem [{\citenamefont {Gill}(1996)}]{gill_new_1996}%
  \BibitemOpen
  \bibfield  {author} {\bibinfo {author} {\bibfnamefont {P.~M.~W.}\
  \bibnamefont {Gill}},\ }\href {\doibase 10.1080/002689796173813} {\bibfield
  {journal} {\bibinfo  {journal} {Mol. Phys.}\ }\textbf {\bibinfo {volume}
  {89}},\ \bibinfo {pages} {433} (\bibinfo {year} {1996})}\BibitemShut
  {NoStop}%
\bibitem [{\citenamefont {Perdew}\ \emph {et~al.}(1996)\citenamefont {Perdew},
  \citenamefont {Burke},\ and\ \citenamefont
  {Ernzerhof}}]{perdew_generalized_1996}%
  \BibitemOpen
  \bibfield  {author} {\bibinfo {author} {\bibfnamefont {J.~P.}\ \bibnamefont
  {Perdew}}, \bibinfo {author} {\bibfnamefont {K.}~\bibnamefont {Burke}}, \
  and\ \bibinfo {author} {\bibfnamefont {M.}~\bibnamefont {Ernzerhof}},\ }\href
  {\doibase 10.1103/PhysRevLett.77.3865} {\bibfield  {journal} {\bibinfo
  {journal} {Phys. Rev. Lett.}\ }\textbf {\bibinfo {volume} {77}},\ \bibinfo
  {pages} {3865} (\bibinfo {year} {1996})}\BibitemShut {NoStop}%
\bibitem [{\citenamefont {Adamo}\ \emph {et~al.}(2000)\citenamefont {Adamo},
  \citenamefont {Ernzerhof},\ and\ \citenamefont
  {Scuseria}}]{adamo_meta-gga_2000}%
  \BibitemOpen
  \bibfield  {author} {\bibinfo {author} {\bibfnamefont {C.}~\bibnamefont
  {Adamo}}, \bibinfo {author} {\bibfnamefont {M.}~\bibnamefont {Ernzerhof}}, \
  and\ \bibinfo {author} {\bibfnamefont {G.~E.}\ \bibnamefont {Scuseria}},\
  }\href {\doibase 10.1063/1.480838} {\bibfield  {journal} {\bibinfo  {journal}
  {J. Chem. Phys.}\ }\textbf {\bibinfo {volume} {112}},\ \bibinfo {pages}
  {2643} (\bibinfo {year} {2000})}\BibitemShut {NoStop}%
\bibitem [{\citenamefont {Tao}\ \emph {et~al.}(2003)\citenamefont {Tao},
  \citenamefont {Perdew}, \citenamefont {Staroverov},\ and\ \citenamefont
  {Scuseria}}]{tao_climbing_2003}%
  \BibitemOpen
  \bibfield  {author} {\bibinfo {author} {\bibfnamefont {J.}~\bibnamefont
  {Tao}}, \bibinfo {author} {\bibfnamefont {J.~P.}\ \bibnamefont {Perdew}},
  \bibinfo {author} {\bibfnamefont {V.~N.}\ \bibnamefont {Staroverov}}, \ and\
  \bibinfo {author} {\bibfnamefont {G.~E.}\ \bibnamefont {Scuseria}},\ }\href
  {\doibase 10.1103/PhysRevLett.91.146401} {\bibfield  {journal} {\bibinfo
  {journal} {Phys. Rev. Lett.}\ }\textbf {\bibinfo {volume} {91}},\ \bibinfo
  {pages} {146401} (\bibinfo {year} {2003})},\ \Eprint
  {http://arxiv.org/abs/0306203} {arXiv:0306203 [cond-mat]} \BibitemShut
  {NoStop}%
\bibitem [{\citenamefont {Becke}(1993)}]{becke_new_1993}%
  \BibitemOpen
  \bibfield  {author} {\bibinfo {author} {\bibfnamefont {A.~D.}\ \bibnamefont
  {Becke}},\ }\href {\doibase 10.1063/1.464304} {\bibfield  {journal} {\bibinfo
   {journal} {J. Chem. Phys.}\ }\textbf {\bibinfo {volume} {98}},\ \bibinfo
  {pages} {1372} (\bibinfo {year} {1993})}\BibitemShut {NoStop}%
\bibitem [{\citenamefont {Baer}\ and\ \citenamefont
  {Neuhauser}(2005)}]{baer_density_2005}%
  \BibitemOpen
  \bibfield  {author} {\bibinfo {author} {\bibfnamefont {R.}~\bibnamefont
  {Baer}}\ and\ \bibinfo {author} {\bibfnamefont {D.}~\bibnamefont
  {Neuhauser}},\ }\href {\doibase 10.1103/PhysRevLett.94.043002} {\bibfield
  {journal} {\bibinfo  {journal} {Phys. Rev. Lett.}\ }\textbf {\bibinfo
  {volume} {94}},\ \bibinfo {pages} {043002} (\bibinfo {year} {2005})},\
  \Eprint {http://arxiv.org/abs/0408664} {arXiv:0408664 [cond-mat]}
  \BibitemShut {NoStop}%
\bibitem [{\citenamefont {Dreuw}\ \emph {et~al.}(2003)\citenamefont {Dreuw},
  \citenamefont {Weisman},\ and\ \citenamefont
  {Head-Gordon}}]{dreuw_long-range_2003}%
  \BibitemOpen
  \bibfield  {author} {\bibinfo {author} {\bibfnamefont {A.}~\bibnamefont
  {Dreuw}}, \bibinfo {author} {\bibfnamefont {J.~L.}\ \bibnamefont {Weisman}},
  \ and\ \bibinfo {author} {\bibfnamefont {M.}~\bibnamefont {Head-Gordon}},\
  }\href {\doibase 10.1063/1.1590951} {\bibfield  {journal} {\bibinfo
  {journal} {J. Chem. Phys.}\ }\textbf {\bibinfo {volume} {119}},\ \bibinfo
  {pages} {2943} (\bibinfo {year} {2003})}\BibitemShut {NoStop}%
\bibitem [{\citenamefont {Baerends}\ \emph {et~al.}(2013)\citenamefont
  {Baerends}, \citenamefont {Gritsenko},\ and\ \citenamefont {van
  Meer}}]{baerends_kohnsham_2013}%
  \BibitemOpen
  \bibfield  {author} {\bibinfo {author} {\bibfnamefont {E.~J.}\ \bibnamefont
  {Baerends}}, \bibinfo {author} {\bibfnamefont {O.~V.}\ \bibnamefont
  {Gritsenko}}, \ and\ \bibinfo {author} {\bibfnamefont {R.}~\bibnamefont {van
  Meer}},\ }\href {\doibase 10.1039/C3CP52547C} {\bibfield  {journal} {\bibinfo
   {journal} {Phys. Chem. Chem. Phys.}\ }\textbf {\bibinfo {volume} {15}},\
  \bibinfo {pages} {16408} (\bibinfo {year} {2013})}\BibitemShut {NoStop}%
\bibitem [{\citenamefont {K{\"u}mmel}(2017)}]{kummel_charge-transfer_2017}%
  \BibitemOpen
  \bibfield  {author} {\bibinfo {author} {\bibfnamefont {S.}~\bibnamefont
  {K{\"u}mmel}},\ }\href {\doibase 10.1002/aenm.201700440} {\bibfield
  {journal} {\bibinfo  {journal} {Adv. Energy Mater.}\ }\textbf {\bibinfo
  {volume} {7}},\ \bibinfo {pages} {1700440} (\bibinfo {year}
  {2017})}\BibitemShut {NoStop}%
\bibitem [{\citenamefont {Maitra}(2017)}]{Maitra17}%
  \BibitemOpen
  \bibfield  {author} {\bibinfo {author} {\bibfnamefont {N.~T.}\ \bibnamefont
  {Maitra}},\ }\href {\doibase 10.1088/1361-648X/aa836e} {\bibfield  {journal}
  {\bibinfo  {journal} {Journal of Physics: Condensed Matter}\ }\textbf
  {\bibinfo {volume} {29}},\ \bibinfo {pages} {423001} (\bibinfo {year}
  {2017})},\ \Eprint {http://arxiv.org/abs/1707.08054} {arXiv:1707.08054
  [physics.chem-ph]} \BibitemShut {NoStop}%
\bibitem [{\citenamefont {Tawada}\ \emph {et~al.}(2004)\citenamefont {Tawada},
  \citenamefont {Tsuneda}, \citenamefont {Yanagisawa}, \citenamefont {Yanai},\
  and\ \citenamefont {Hirao}}]{tawada_long-range-corrected_2004}%
  \BibitemOpen
  \bibfield  {author} {\bibinfo {author} {\bibfnamefont {Y.}~\bibnamefont
  {Tawada}}, \bibinfo {author} {\bibfnamefont {T.}~\bibnamefont {Tsuneda}},
  \bibinfo {author} {\bibfnamefont {S.}~\bibnamefont {Yanagisawa}}, \bibinfo
  {author} {\bibfnamefont {T.}~\bibnamefont {Yanai}}, \ and\ \bibinfo {author}
  {\bibfnamefont {K.}~\bibnamefont {Hirao}},\ }\href {\doibase
  10.1063/1.1688752} {\bibfield  {journal} {\bibinfo  {journal} {J. Chem.
  Phys.}\ }\textbf {\bibinfo {volume} {120}},\ \bibinfo {pages} {8425}
  (\bibinfo {year} {2004})}\BibitemShut {NoStop}%
\bibitem [{\citenamefont {Akinaga}\ and\ \citenamefont
  {Ten-No}(2009)}]{akinaga_intramolecular_2009}%
  \BibitemOpen
  \bibfield  {author} {\bibinfo {author} {\bibfnamefont {Y.}~\bibnamefont
  {Akinaga}}\ and\ \bibinfo {author} {\bibfnamefont {S.}~\bibnamefont
  {Ten-No}},\ }\href {\doibase 10.1002/qua.22012} {\bibfield  {journal}
  {\bibinfo  {journal} {Int. J. Quantum Chem.}\ }\textbf {\bibinfo {volume}
  {109}},\ \bibinfo {pages} {1905} (\bibinfo {year} {2009})}\BibitemShut
  {NoStop}%
\bibitem [{\citenamefont {Rohrdanz}\ \emph {et~al.}(2009)\citenamefont
  {Rohrdanz}, \citenamefont {Martins},\ and\ \citenamefont
  {Herbert}}]{rohrdanz_long-range-corrected_2009}%
  \BibitemOpen
  \bibfield  {author} {\bibinfo {author} {\bibfnamefont {M.~A.}\ \bibnamefont
  {Rohrdanz}}, \bibinfo {author} {\bibfnamefont {K.~M.}\ \bibnamefont
  {Martins}}, \ and\ \bibinfo {author} {\bibfnamefont {J.~M.}\ \bibnamefont
  {Herbert}},\ }\href {\doibase 10.1063/1.3073302} {\bibfield  {journal}
  {\bibinfo  {journal} {J. Chem. Phys.}\ }\textbf {\bibinfo {volume} {130}},\
  \bibinfo {pages} {054112} (\bibinfo {year} {2009})}\BibitemShut {NoStop}%
\bibitem [{\citenamefont {Stein}\ \emph
  {et~al.}(2009{\natexlab{a}})\citenamefont {Stein}, \citenamefont {Kronik},\
  and\ \citenamefont {Baer}}]{stein_reliable_2009}%
  \BibitemOpen
  \bibfield  {author} {\bibinfo {author} {\bibfnamefont {T.}~\bibnamefont
  {Stein}}, \bibinfo {author} {\bibfnamefont {L.}~\bibnamefont {Kronik}}, \
  and\ \bibinfo {author} {\bibfnamefont {R.}~\bibnamefont {Baer}},\ }\href
  {\doibase 10.1021/ja8087482} {\bibfield  {journal} {\bibinfo  {journal} {J.
  Am. Chem. Soc.}\ }\textbf {\bibinfo {volume} {131}},\ \bibinfo {pages} {2818}
  (\bibinfo {year} {2009}{\natexlab{a}})}\BibitemShut {NoStop}%
\bibitem [{\citenamefont {Stein}\ \emph
  {et~al.}(2009{\natexlab{b}})\citenamefont {Stein}, \citenamefont {Kronik},\
  and\ \citenamefont {Baer}}]{stein_prediction_2009}%
  \BibitemOpen
  \bibfield  {author} {\bibinfo {author} {\bibfnamefont {T.}~\bibnamefont
  {Stein}}, \bibinfo {author} {\bibfnamefont {L.}~\bibnamefont {Kronik}}, \
  and\ \bibinfo {author} {\bibfnamefont {R.}~\bibnamefont {Baer}},\ }\href
  {\doibase 10.1063/1.3269029} {\bibfield  {journal} {\bibinfo  {journal} {J.
  Chem. Phys.}\ }\textbf {\bibinfo {volume} {131}},\ \bibinfo {pages} {244119}
  (\bibinfo {year} {2009}{\natexlab{b}})}\BibitemShut {NoStop}%
\bibitem [{\citenamefont {Kronik}\ \emph {et~al.}(2012)\citenamefont {Kronik},
  \citenamefont {Stein}, \citenamefont {Refaely-Abramson},\ and\ \citenamefont
  {Baer}}]{kronik_excitation_2012}%
  \BibitemOpen
  \bibfield  {author} {\bibinfo {author} {\bibfnamefont {L.}~\bibnamefont
  {Kronik}}, \bibinfo {author} {\bibfnamefont {T.}~\bibnamefont {Stein}},
  \bibinfo {author} {\bibfnamefont {S.}~\bibnamefont {Refaely-Abramson}}, \
  and\ \bibinfo {author} {\bibfnamefont {R.}~\bibnamefont {Baer}},\ }\href
  {\doibase 10.1021/ct2009363} {\bibfield  {journal} {\bibinfo  {journal} {J.
  Chem. Theory Comput.}\ }\textbf {\bibinfo {volume} {8}},\ \bibinfo {pages}
  {1515} (\bibinfo {year} {2012})}\BibitemShut {NoStop}%
\bibitem [{\citenamefont {Zhang}\ \emph {et~al.}(2012)\citenamefont {Zhang},
  \citenamefont {Li},\ and\ \citenamefont
  {Lu}}]{zhang_non-self-consistent_2012}%
  \BibitemOpen
  \bibfield  {author} {\bibinfo {author} {\bibfnamefont {X.}~\bibnamefont
  {Zhang}}, \bibinfo {author} {\bibfnamefont {Z.}~\bibnamefont {Li}}, \ and\
  \bibinfo {author} {\bibfnamefont {G.}~\bibnamefont {Lu}},\ }\href {\doibase
  10.1088/0953-8984/24/20/205801} {\bibfield  {journal} {\bibinfo  {journal}
  {J. Phys.: Condens. Matter}\ }\textbf {\bibinfo {volume} {24}},\ \bibinfo
  {pages} {205801} (\bibinfo {year} {2012})}\BibitemShut {NoStop}%
\bibitem [{\citenamefont {Yanai}\ \emph {et~al.}(2004)\citenamefont {Yanai},
  \citenamefont {Tew},\ and\ \citenamefont {Handy}}]{yanai_new_2004}%
  \BibitemOpen
  \bibfield  {author} {\bibinfo {author} {\bibfnamefont {T.}~\bibnamefont
  {Yanai}}, \bibinfo {author} {\bibfnamefont {D.~P.}\ \bibnamefont {Tew}}, \
  and\ \bibinfo {author} {\bibfnamefont {N.~C.}\ \bibnamefont {Handy}},\ }\href
  {\doibase 10.1016/j.cplett.2004.06.011} {\bibfield  {journal} {\bibinfo
  {journal} {Chemical Physics Letters}\ }\textbf {\bibinfo {volume} {393}},\
  \bibinfo {pages} {51} (\bibinfo {year} {2004})}\BibitemShut {NoStop}%
\bibitem [{\citenamefont {Heyd}\ \emph {et~al.}(2003)\citenamefont {Heyd},
  \citenamefont {Scuseria},\ and\ \citenamefont
  {Ernzerhof}}]{heyd_hybrid_2003}%
  \BibitemOpen
  \bibfield  {author} {\bibinfo {author} {\bibfnamefont {J.}~\bibnamefont
  {Heyd}}, \bibinfo {author} {\bibfnamefont {G.~E.}\ \bibnamefont {Scuseria}},
  \ and\ \bibinfo {author} {\bibfnamefont {M.}~\bibnamefont {Ernzerhof}},\
  }\href {\doibase 10.1063/1.1564060} {\bibfield  {journal} {\bibinfo
  {journal} {The Journal of Chemical Physics}\ }\textbf {\bibinfo {volume}
  {118}},\ \bibinfo {pages} {8207} (\bibinfo {year} {2003})}\BibitemShut
  {NoStop}%
\bibitem [{\citenamefont {Heyd}\ \emph {et~al.}(2006)\citenamefont {Heyd},
  \citenamefont {Scuseria},\ and\ \citenamefont
  {Ernzerhof}}]{heyd_erratum:_2006}%
  \BibitemOpen
  \bibfield  {author} {\bibinfo {author} {\bibfnamefont {J.}~\bibnamefont
  {Heyd}}, \bibinfo {author} {\bibfnamefont {G.~E.}\ \bibnamefont {Scuseria}},
  \ and\ \bibinfo {author} {\bibfnamefont {M.}~\bibnamefont {Ernzerhof}},\
  }\href {\doibase 10.1063/1.2204597} {\bibfield  {journal} {\bibinfo
  {journal} {The Journal of Chemical Physics}\ }\textbf {\bibinfo {volume}
  {124}},\ \bibinfo {pages} {219906} (\bibinfo {year} {2006})}\BibitemShut
  {NoStop}%
\bibitem [{\citenamefont {Iikura}\ \emph {et~al.}(2001)\citenamefont {Iikura},
  \citenamefont {Tsuneda}, \citenamefont {Yanai},\ and\ \citenamefont
  {Hirao}}]{iikura_long-range_2001}%
  \BibitemOpen
  \bibfield  {author} {\bibinfo {author} {\bibfnamefont {H.}~\bibnamefont
  {Iikura}}, \bibinfo {author} {\bibfnamefont {T.}~\bibnamefont {Tsuneda}},
  \bibinfo {author} {\bibfnamefont {T.}~\bibnamefont {Yanai}}, \ and\ \bibinfo
  {author} {\bibfnamefont {K.}~\bibnamefont {Hirao}},\ }\href {\doibase
  10.1063/1.1383587} {\bibfield  {journal} {\bibinfo  {journal} {J. Chem.
  Phys.}\ }\textbf {\bibinfo {volume} {115}},\ \bibinfo {pages} {3540}
  (\bibinfo {year} {2001})}\BibitemShut {NoStop}%
\bibitem [{\citenamefont {Stein}\ \emph {et~al.}(2010)\citenamefont {Stein},
  \citenamefont {Eisenberg}, \citenamefont {Kronik},\ and\ \citenamefont
  {Baer}}]{stein_fundamental_2010}%
  \BibitemOpen
  \bibfield  {author} {\bibinfo {author} {\bibfnamefont {T.}~\bibnamefont
  {Stein}}, \bibinfo {author} {\bibfnamefont {H.}~\bibnamefont {Eisenberg}},
  \bibinfo {author} {\bibfnamefont {L.}~\bibnamefont {Kronik}}, \ and\ \bibinfo
  {author} {\bibfnamefont {R.}~\bibnamefont {Baer}},\ }\href {\doibase
  10.1103/PhysRevLett.105.266802} {\bibfield  {journal} {\bibinfo  {journal}
  {Phys. Rev. Lett.}\ }\textbf {\bibinfo {volume} {105}},\ \bibinfo {pages}
  {266802} (\bibinfo {year} {2010})},\ \Eprint
  {http://arxiv.org/abs/1006.5420v1} {arXiv:1006.5420v1 [cond-mat.mtrl-sci]}
  \BibitemShut {NoStop}%
\bibitem [{\citenamefont {Huzinaga}\ and\ \citenamefont
  {Arnau}(1970)}]{huzinaga_virtual_1970}%
  \BibitemOpen
  \bibfield  {author} {\bibinfo {author} {\bibfnamefont {S.}~\bibnamefont
  {Huzinaga}}\ and\ \bibinfo {author} {\bibfnamefont {C.}~\bibnamefont
  {Arnau}},\ }\href {\doibase 10.1103/PhysRevA.1.1285} {\bibfield  {journal}
  {\bibinfo  {journal} {Phys. Rev. A}\ }\textbf {\bibinfo {volume} {1}},\
  \bibinfo {pages} {1285} (\bibinfo {year} {1970})}\BibitemShut {NoStop}%
\bibitem [{\citenamefont {Kelly}(1963)}]{kelly_correlation_1963}%
  \BibitemOpen
  \bibfield  {author} {\bibinfo {author} {\bibfnamefont {H.~P.}\ \bibnamefont
  {Kelly}},\ }\href {\doibase 10.1103/PhysRev.131.684} {\bibfield  {journal}
  {\bibinfo  {journal} {Phys. Rev.}\ }\textbf {\bibinfo {volume} {131}},\
  \bibinfo {pages} {684} (\bibinfo {year} {1963})}\BibitemShut {NoStop}%
\bibitem [{\citenamefont {Kelly}(1964)}]{kelly_many-body_1964}%
  \BibitemOpen
  \bibfield  {author} {\bibinfo {author} {\bibfnamefont {H.~P.}\ \bibnamefont
  {Kelly}},\ }\href {\doibase 10.1103/PhysRev.136.B896} {\bibfield  {journal}
  {\bibinfo  {journal} {Phys. Rev.}\ }\textbf {\bibinfo {volume} {136}},\
  \bibinfo {pages} {B896} (\bibinfo {year} {1964})}\BibitemShut {NoStop}%
\bibitem [{\citenamefont {Huzinaga}\ and\ \citenamefont
  {Arnau}(1971)}]{huzinaga_virtual_1971}%
  \BibitemOpen
  \bibfield  {author} {\bibinfo {author} {\bibfnamefont {S.}~\bibnamefont
  {Huzinaga}}\ and\ \bibinfo {author} {\bibfnamefont {C.}~\bibnamefont
  {Arnau}},\ }\href {\doibase 10.1063/1.1675123} {\bibfield  {journal}
  {\bibinfo  {journal} {J. Chem. Phys.}\ }\textbf {\bibinfo {volume} {54}},\
  \bibinfo {pages} {1948} (\bibinfo {year} {1971})}\BibitemShut {NoStop}%
\bibitem [{\citenamefont {Bl{\"o}chl}(1994)}]{blochl_projector_1994}%
  \BibitemOpen
  \bibfield  {author} {\bibinfo {author} {\bibfnamefont {P.~E.}\ \bibnamefont
  {Bl{\"o}chl}},\ }\href {\doibase 10.1103/PhysRevB.50.17953} {\bibfield
  {journal} {\bibinfo  {journal} {Phys. Rev. B}\ }\textbf {\bibinfo {volume}
  {50}},\ \bibinfo {pages} {17953} (\bibinfo {year} {1994})}\BibitemShut
  {NoStop}%
\bibitem [{\citenamefont {Mortensen}\ \emph {et~al.}(2005)\citenamefont
  {Mortensen}, \citenamefont {Hansen},\ and\ \citenamefont
  {Jacobsen}}]{mortensen_real-space_2005}%
  \BibitemOpen
  \bibfield  {author} {\bibinfo {author} {\bibfnamefont {J.~J.}\ \bibnamefont
  {Mortensen}}, \bibinfo {author} {\bibfnamefont {L.~B.}\ \bibnamefont
  {Hansen}}, \ and\ \bibinfo {author} {\bibfnamefont {K.~W.}\ \bibnamefont
  {Jacobsen}},\ }\href {\doibase 10.1103/PhysRevB.71.035109} {\bibfield
  {journal} {\bibinfo  {journal} {Phys. Rev. B}\ }\textbf {\bibinfo {volume}
  {71}},\ \bibinfo {pages} {035109} (\bibinfo {year} {2005})},\ \Eprint
  {http://arxiv.org/abs/0411218} {arXiv:0411218 [cond-mat.mtrl-sci]}
  \BibitemShut {NoStop}%
\bibitem [{\citenamefont {Enkovaara}\ \emph {et~al.}(2010)\citenamefont
  {Enkovaara}, \citenamefont {Rostgaard}, \citenamefont {Mortensen},
  \citenamefont {Chen}, \citenamefont {Du{\l}ak}, \citenamefont {Ferrighi},
  \citenamefont {Gavnholt}, \citenamefont {Glinsvad}, \citenamefont {Haikola},
  \citenamefont {Hansen}, \citenamefont {Kristoffersen}, \citenamefont
  {Kuisma}, \citenamefont {Larsen}, \citenamefont {Lehtovaara}, \citenamefont
  {Ljungberg}, \citenamefont {Lopez-Acevedo}, \citenamefont {Moses},
  \citenamefont {Ojanen}, \citenamefont {Olsen}, \citenamefont {Petzold},
  \citenamefont {Romero}, \citenamefont {Stausholm-M{\o}ller}, \citenamefont
  {Strange}, \citenamefont {Tritsaris}, \citenamefont {Vanin}, \citenamefont
  {Walter}, \citenamefont {Hammer}, \citenamefont {H{\"a}kkinen}, \citenamefont
  {Madsen}, \citenamefont {Nieminen}, \citenamefont {N{\o}rskov}, \citenamefont
  {Puska}, \citenamefont {Rantala}, \citenamefont {Schi{\o}tz}, \citenamefont
  {Thygesen},\ and\ \citenamefont {Jacobsen}}]{enkovaara_electronic_2010}%
  \BibitemOpen
  \bibfield  {author} {\bibinfo {author} {\bibfnamefont {J.}~\bibnamefont
  {Enkovaara}}, \bibinfo {author} {\bibfnamefont {C.}~\bibnamefont
  {Rostgaard}}, \bibinfo {author} {\bibfnamefont {J.~J.}\ \bibnamefont
  {Mortensen}}, \bibinfo {author} {\bibfnamefont {J.}~\bibnamefont {Chen}},
  \bibinfo {author} {\bibfnamefont {M.}~\bibnamefont {Du{\l}ak}}, \bibinfo
  {author} {\bibfnamefont {L.}~\bibnamefont {Ferrighi}}, \bibinfo {author}
  {\bibfnamefont {J.}~\bibnamefont {Gavnholt}}, \bibinfo {author}
  {\bibfnamefont {C.}~\bibnamefont {Glinsvad}}, \bibinfo {author}
  {\bibfnamefont {V.}~\bibnamefont {Haikola}}, \bibinfo {author} {\bibfnamefont
  {H.~A.}\ \bibnamefont {Hansen}}, \bibinfo {author} {\bibfnamefont {H.~H.}\
  \bibnamefont {Kristoffersen}}, \bibinfo {author} {\bibfnamefont
  {M.}~\bibnamefont {Kuisma}}, \bibinfo {author} {\bibfnamefont {A.~H.}\
  \bibnamefont {Larsen}}, \bibinfo {author} {\bibfnamefont {L.}~\bibnamefont
  {Lehtovaara}}, \bibinfo {author} {\bibfnamefont {M.}~\bibnamefont
  {Ljungberg}}, \bibinfo {author} {\bibfnamefont {O.}~\bibnamefont
  {Lopez-Acevedo}}, \bibinfo {author} {\bibfnamefont {P.~G.}\ \bibnamefont
  {Moses}}, \bibinfo {author} {\bibfnamefont {J.}~\bibnamefont {Ojanen}},
  \bibinfo {author} {\bibfnamefont {T.}~\bibnamefont {Olsen}}, \bibinfo
  {author} {\bibfnamefont {V.}~\bibnamefont {Petzold}}, \bibinfo {author}
  {\bibfnamefont {N.~A.}\ \bibnamefont {Romero}}, \bibinfo {author}
  {\bibfnamefont {J.}~\bibnamefont {Stausholm-M{\o}ller}}, \bibinfo {author}
  {\bibfnamefont {M.}~\bibnamefont {Strange}}, \bibinfo {author} {\bibfnamefont
  {G.~A.}\ \bibnamefont {Tritsaris}}, \bibinfo {author} {\bibfnamefont
  {M.}~\bibnamefont {Vanin}}, \bibinfo {author} {\bibfnamefont
  {M.}~\bibnamefont {Walter}}, \bibinfo {author} {\bibfnamefont
  {B.}~\bibnamefont {Hammer}}, \bibinfo {author} {\bibfnamefont
  {H.}~\bibnamefont {H{\"a}kkinen}}, \bibinfo {author} {\bibfnamefont
  {G.~K.~H.}\ \bibnamefont {Madsen}}, \bibinfo {author} {\bibfnamefont {R.~M.}\
  \bibnamefont {Nieminen}}, \bibinfo {author} {\bibfnamefont {J.~K.}\
  \bibnamefont {N{\o}rskov}}, \bibinfo {author} {\bibfnamefont
  {M.}~\bibnamefont {Puska}}, \bibinfo {author} {\bibfnamefont {T.~T.}\
  \bibnamefont {Rantala}}, \bibinfo {author} {\bibfnamefont {J.}~\bibnamefont
  {Schi{\o}tz}}, \bibinfo {author} {\bibfnamefont {K.~S.}\ \bibnamefont
  {Thygesen}}, \ and\ \bibinfo {author} {\bibfnamefont {K.~W.}\ \bibnamefont
  {Jacobsen}},\ }\href {\doibase 10.1088/0953-8984/22/25/253202} {\bibfield
  {journal} {\bibinfo  {journal} {J. Phys.: Condens. Matter}\ }\textbf
  {\bibinfo {volume} {22}},\ \bibinfo {pages} {253202} (\bibinfo {year}
  {2010})}\BibitemShut {NoStop}%
\bibitem [{\citenamefont {Kresse}\ and\ \citenamefont
  {Joubert}(1999)}]{kresse_ultrasoft_1999}%
  \BibitemOpen
  \bibfield  {author} {\bibinfo {author} {\bibfnamefont {G.}~\bibnamefont
  {Kresse}}\ and\ \bibinfo {author} {\bibfnamefont {D.}~\bibnamefont
  {Joubert}},\ }\href {\doibase 10.1103/PhysRevB.59.1758} {\bibfield  {journal}
  {\bibinfo  {journal} {Phys. Rev. B}\ }\textbf {\bibinfo {volume} {59}},\
  \bibinfo {pages} {1758} (\bibinfo {year} {1999})}\BibitemShut {NoStop}%
\bibitem [{\citenamefont {Valiev}\ \emph {et~al.}(2003)\citenamefont {Valiev},
  \citenamefont {Bylaska},\ and\ \citenamefont
  {Weare}}]{valiev_calculations_2003}%
  \BibitemOpen
  \bibfield  {author} {\bibinfo {author} {\bibfnamefont {M.}~\bibnamefont
  {Valiev}}, \bibinfo {author} {\bibfnamefont {E.~J.}\ \bibnamefont {Bylaska}},
  \ and\ \bibinfo {author} {\bibfnamefont {J.~H.}\ \bibnamefont {Weare}},\
  }\href {\doibase 10.1063/1.1602694} {\bibfield  {journal} {\bibinfo
  {journal} {J. Chem. Phys.}\ }\textbf {\bibinfo {volume} {119}},\ \bibinfo
  {pages} {5955} (\bibinfo {year} {2003})}\BibitemShut {NoStop}%
\bibitem [{\citenamefont {W{\"u}rdemann}\ \emph {et~al.}(2015)\citenamefont
  {W{\"u}rdemann}, \citenamefont {Kristoffersen}, \citenamefont {Moseler},\
  and\ \citenamefont {Walter}}]{Wurdemann2015}%
  \BibitemOpen
  \bibfield  {author} {\bibinfo {author} {\bibfnamefont {R.}~\bibnamefont
  {W{\"u}rdemann}}, \bibinfo {author} {\bibfnamefont {H.~H.}\ \bibnamefont
  {Kristoffersen}}, \bibinfo {author} {\bibfnamefont {M.}~\bibnamefont
  {Moseler}}, \ and\ \bibinfo {author} {\bibfnamefont {M.}~\bibnamefont
  {Walter}},\ }\href {\doibase 10.1063/1.4915265} {\bibfield  {journal}
  {\bibinfo  {journal} {J. Chem. Phys.}\ }\textbf {\bibinfo {volume} {142}},\
  \bibinfo {pages} {124316} (\bibinfo {year} {2015})}\BibitemShut {NoStop}%
\bibitem [{\citenamefont {Koelling}\ and\ \citenamefont
  {Harmon}(1977)}]{koelling_technique_1977}%
  \BibitemOpen
  \bibfield  {author} {\bibinfo {author} {\bibfnamefont {D.~D.}\ \bibnamefont
  {Koelling}}\ and\ \bibinfo {author} {\bibfnamefont {B.~N.}\ \bibnamefont
  {Harmon}},\ }\href {\doibase 10.1088/0022-3719/10/16/019} {\bibfield
  {journal} {\bibinfo  {journal} {J. Phys. C: Solid State Phys.}\ }\textbf
  {\bibinfo {volume} {10}},\ \bibinfo {pages} {3107} (\bibinfo {year}
  {1977})}\BibitemShut {NoStop}%
\bibitem [{\citenamefont {Adamo}\ and\ \citenamefont
  {Barone}(1998)}]{adamo_toward_1998}%
  \BibitemOpen
  \bibfield  {author} {\bibinfo {author} {\bibfnamefont {C.}~\bibnamefont
  {Adamo}}\ and\ \bibinfo {author} {\bibfnamefont {V.}~\bibnamefont {Barone}},\
  }\href {\doibase 10.1016/S0009-2614(98)01201-9} {\bibfield  {journal}
  {\bibinfo  {journal} {Chemical Physics Letters}\ }\textbf {\bibinfo {volume}
  {298}},\ \bibinfo {pages} {113} (\bibinfo {year} {1998})}\BibitemShut
  {NoStop}%
\bibitem [{\citenamefont {Akinaga}\ and\ \citenamefont
  {Ten-no}(2008)}]{akinaga_range-separation_2008}%
  \BibitemOpen
  \bibfield  {author} {\bibinfo {author} {\bibfnamefont {Y.}~\bibnamefont
  {Akinaga}}\ and\ \bibinfo {author} {\bibfnamefont {S.}~\bibnamefont
  {Ten-no}},\ }\href {\doibase 10.1016/j.cplett.2008.07.103} {\bibfield
  {journal} {\bibinfo  {journal} {Chemical Physics Letters}\ }\textbf {\bibinfo
  {volume} {462}},\ \bibinfo {pages} {348} (\bibinfo {year}
  {2008})}\BibitemShut {NoStop}%
\bibitem [{\citenamefont {Seth}\ and\ \citenamefont
  {Ziegler}(2012)}]{seth_range-separated_2012}%
  \BibitemOpen
  \bibfield  {author} {\bibinfo {author} {\bibfnamefont {M.}~\bibnamefont
  {Seth}}\ and\ \bibinfo {author} {\bibfnamefont {T.}~\bibnamefont {Ziegler}},\
  }\href {\doibase 10.1021/ct300006h} {\bibfield  {journal} {\bibinfo
  {journal} {J. Chem. Theory Comput.}\ }\textbf {\bibinfo {volume} {8}},\
  \bibinfo {pages} {901} (\bibinfo {year} {2012})}\BibitemShut {NoStop}%
\bibitem [{\citenamefont {Casida}(2009)}]{casida_time-dependent_2009}%
  \BibitemOpen
  \bibfield  {author} {\bibinfo {author} {\bibfnamefont {M.~E.}\ \bibnamefont
  {Casida}},\ }\href {\doibase 10.1016/j.theochem.2009.08.018} {\bibfield
  {journal} {\bibinfo  {journal} {Journal of Molecular Structure: THEOCHEM}\
  }\bibinfo {series} {Time-dependent density-functional theory for molecules
  and molecular solids},\ \textbf {\bibinfo {volume} {914}},\ \bibinfo {pages}
  {3} (\bibinfo {year} {2009})}\BibitemShut {NoStop}%
\bibitem [{\citenamefont {Walter}\ \emph {et~al.}(2008)\citenamefont {Walter},
  \citenamefont {H{\"a}kkinen}, \citenamefont {Lehtovaara}, \citenamefont
  {Puska}, \citenamefont {Enkovaara}, \citenamefont {Rostgaard},\ and\
  \citenamefont {Mortensen}}]{walter_time-dependent_2008}%
  \BibitemOpen
  \bibfield  {author} {\bibinfo {author} {\bibfnamefont {M.}~\bibnamefont
  {Walter}}, \bibinfo {author} {\bibfnamefont {H.}~\bibnamefont
  {H{\"a}kkinen}}, \bibinfo {author} {\bibfnamefont {L.}~\bibnamefont
  {Lehtovaara}}, \bibinfo {author} {\bibfnamefont {M.}~\bibnamefont {Puska}},
  \bibinfo {author} {\bibfnamefont {J.}~\bibnamefont {Enkovaara}}, \bibinfo
  {author} {\bibfnamefont {C.}~\bibnamefont {Rostgaard}}, \ and\ \bibinfo
  {author} {\bibfnamefont {J.~J.}\ \bibnamefont {Mortensen}},\ }\href {\doibase
  10.1063/1.2943138} {\bibfield  {journal} {\bibinfo  {journal} {J. Chem.
  Phys.}\ }\textbf {\bibinfo {volume} {128}},\ \bibinfo {pages} {244101}
  (\bibinfo {year} {2008})}\BibitemShut {NoStop}%
\bibitem [{\citenamefont {Baer}\ \emph {et~al.}(2006)\citenamefont {Baer},
  \citenamefont {Livshits},\ and\ \citenamefont
  {Neuhauser}}]{baer_avoiding_2006}%
  \BibitemOpen
  \bibfield  {author} {\bibinfo {author} {\bibfnamefont {R.}~\bibnamefont
  {Baer}}, \bibinfo {author} {\bibfnamefont {E.}~\bibnamefont {Livshits}}, \
  and\ \bibinfo {author} {\bibfnamefont {D.}~\bibnamefont {Neuhauser}},\ }\href
  {\doibase 10.1016/j.chemphys.2006.06.041} {\bibfield  {journal} {\bibinfo
  {journal} {Chemical Physics}\ }\bibinfo {series} {Electron Correlation and
  Multimode Dynamics in Molecules(in honour of Lorenz S. Cederbaum)},\ \textbf
  {\bibinfo {volume} {329}},\ \bibinfo {pages} {266} (\bibinfo {year}
  {2006})}\BibitemShut {NoStop}%
\bibitem [{\citenamefont {Peach}\ \emph {et~al.}(2006)\citenamefont {Peach},
  \citenamefont {Helgaker}, \citenamefont {Sa{\l}ek}, \citenamefont {Keal},
  \citenamefont {Lutn{\ae}s}, \citenamefont {Tozer},\ and\ \citenamefont
  {Handy}}]{peach_assessment_2006}%
  \BibitemOpen
  \bibfield  {author} {\bibinfo {author} {\bibfnamefont {M.~J.~G.}\
  \bibnamefont {Peach}}, \bibinfo {author} {\bibfnamefont {T.}~\bibnamefont
  {Helgaker}}, \bibinfo {author} {\bibfnamefont {P.}~\bibnamefont {Sa{\l}ek}},
  \bibinfo {author} {\bibfnamefont {T.~W.}\ \bibnamefont {Keal}}, \bibinfo
  {author} {\bibfnamefont {O.~B.}\ \bibnamefont {Lutn{\ae}s}}, \bibinfo
  {author} {\bibfnamefont {D.~J.}\ \bibnamefont {Tozer}}, \ and\ \bibinfo
  {author} {\bibfnamefont {N.~C.}\ \bibnamefont {Handy}},\ }\href {\doibase
  10.1039/B511865D} {\bibfield  {journal} {\bibinfo  {journal} {Phys. Chem.
  Chem. Phys.}\ }\textbf {\bibinfo {volume} {8}},\ \bibinfo {pages} {558}
  (\bibinfo {year} {2006})}\BibitemShut {NoStop}%
\bibitem [{\citenamefont {Vydrov}\ \emph {et~al.}(2007)\citenamefont {Vydrov},
  \citenamefont {Scuseria},\ and\ \citenamefont {Perdew}}]{vydrov_tests_2007}%
  \BibitemOpen
  \bibfield  {author} {\bibinfo {author} {\bibfnamefont {O.~A.}\ \bibnamefont
  {Vydrov}}, \bibinfo {author} {\bibfnamefont {G.~E.}\ \bibnamefont
  {Scuseria}}, \ and\ \bibinfo {author} {\bibfnamefont {J.~P.}\ \bibnamefont
  {Perdew}},\ }\href {\doibase 10.1063/1.2723119} {\bibfield  {journal}
  {\bibinfo  {journal} {J. Chem. Phys.}\ }\textbf {\bibinfo {volume} {126}},\
  \bibinfo {pages} {154109} (\bibinfo {year} {2007})}\BibitemShut {NoStop}%
\bibitem [{\citenamefont {Rohrdanz}\ and\ \citenamefont
  {Herbert}(2008)}]{rohrdanz_simultaneous_2008}%
  \BibitemOpen
  \bibfield  {author} {\bibinfo {author} {\bibfnamefont {M.~A.}\ \bibnamefont
  {Rohrdanz}}\ and\ \bibinfo {author} {\bibfnamefont {J.~M.}\ \bibnamefont
  {Herbert}},\ }\href {\doibase 10.1063/1.2954017} {\bibfield  {journal}
  {\bibinfo  {journal} {J. Chem. Phys.}\ }\textbf {\bibinfo {volume} {129}},\
  \bibinfo {pages} {034107} (\bibinfo {year} {2008})}\BibitemShut {NoStop}%
\bibitem [{\citenamefont {Wong}\ \emph {et~al.}(2009)\citenamefont {Wong},
  \citenamefont {Piacenza},\ and\ \citenamefont {Sala}}]{wong_absorption_2009}%
  \BibitemOpen
  \bibfield  {author} {\bibinfo {author} {\bibfnamefont {B.~M.}\ \bibnamefont
  {Wong}}, \bibinfo {author} {\bibfnamefont {M.}~\bibnamefont {Piacenza}}, \
  and\ \bibinfo {author} {\bibfnamefont {F.~D.}\ \bibnamefont {Sala}},\ }\href
  {\doibase 10.1039/B901743G} {\bibfield  {journal} {\bibinfo  {journal} {Phys.
  Chem. Chem. Phys.}\ }\textbf {\bibinfo {volume} {11}},\ \bibinfo {pages}
  {4498} (\bibinfo {year} {2009})},\ \Eprint {http://arxiv.org/abs/0904.3918}
  {arXiv:0904.3918 [physics.chem-ph]} \BibitemShut {NoStop}%
\bibitem [{\citenamefont {Yukawa}(1935)}]{yukawa_interaction_1935}%
  \BibitemOpen
  \bibfield  {author} {\bibinfo {author} {\bibfnamefont {H.}~\bibnamefont
  {Yukawa}},\ }\href@noop {} {\bibfield  {journal} {\bibinfo  {journal} {Proc.
  Phys.-Math. Soc. Jpn. 3rd Ser.}\ }\textbf {\bibinfo {volume} {17}},\ \bibinfo
  {pages} {48} (\bibinfo {year} {1935})}\BibitemShut {NoStop}%
\bibitem [{\citenamefont {Rostgaard}(2006)}]{rostgaard_exact_2006}%
  \BibitemOpen
  \bibfield  {author} {\bibinfo {author} {\bibfnamefont {C.}~\bibnamefont
  {Rostgaard}},\ }\emph {\bibinfo {title} {{Exact exchange in density
  functional calculations}}},\ \href@noop {} {Master's thesis},\ \bibinfo
  {school} {Technical University Denmark}, \bibinfo {address} {Lyngby, Denmark}
  (\bibinfo {year} {2006})\BibitemShut {NoStop}%
\bibitem [{\citenamefont {Rico}\ \emph {et~al.}(2012)\citenamefont {Rico},
  \citenamefont {L{\'o}pez}, \citenamefont {Ram{\'\i}rez},\ and\ \citenamefont
  {Ema}}]{rico_repulsion_2012}%
  \BibitemOpen
  \bibfield  {author} {\bibinfo {author} {\bibfnamefont {J.~F.}\ \bibnamefont
  {Rico}}, \bibinfo {author} {\bibfnamefont {R.}~\bibnamefont {L{\'o}pez}},
  \bibinfo {author} {\bibfnamefont {G.}~\bibnamefont {Ram{\'\i}rez}}, \ and\
  \bibinfo {author} {\bibfnamefont {I.}~\bibnamefont {Ema}},\ }\href {\doibase
  10.1007/s00214-012-1304-x} {\bibfield  {journal} {\bibinfo  {journal} {Theor
  Chem Acc}\ }\textbf {\bibinfo {volume} {132}},\ \bibinfo {pages} {1}
  (\bibinfo {year} {2012})}\BibitemShut {NoStop}%
\bibitem [{\citenamefont {Jackson}(1998)}]{jackson_classical_1998}%
  \BibitemOpen
  \bibfield  {author} {\bibinfo {author} {\bibfnamefont {J.~D.}\ \bibnamefont
  {Jackson}},\ }\href@noop {} {\emph {\bibinfo {title} {Classical
  {{Electrodynamics}}}}},\ \bibinfo {edition} {3rd}\ ed.\ (\bibinfo
  {publisher} {{Wiley-Interscience}},\ \bibinfo {year} {1998})\BibitemShut
  {NoStop}%
\bibitem [{\citenamefont {Greengard}\ and\ \citenamefont
  {Huang}(2002)}]{greengard_new_2002}%
  \BibitemOpen
  \bibfield  {author} {\bibinfo {author} {\bibfnamefont {L.~F.}\ \bibnamefont
  {Greengard}}\ and\ \bibinfo {author} {\bibfnamefont {J.}~\bibnamefont
  {Huang}},\ }\href {\doibase 10.1006/jcph.2002.7110} {\bibfield  {journal}
  {\bibinfo  {journal} {Journal of Computational Physics}\ }\textbf {\bibinfo
  {volume} {180}},\ \bibinfo {pages} {642} (\bibinfo {year}
  {2002})}\BibitemShut {NoStop}%
\bibitem [{\citenamefont {Wajid}\ \emph {et~al.}(2014)\citenamefont {Wajid},
  \citenamefont {Ahmed}, \citenamefont {Iqbal},\ and\ \citenamefont
  {Arshad}}]{wajid_modified_2014}%
  \BibitemOpen
  \bibfield  {author} {\bibinfo {author} {\bibfnamefont {H.~A.}\ \bibnamefont
  {Wajid}}, \bibinfo {author} {\bibfnamefont {N.}~\bibnamefont {Ahmed}},
  \bibinfo {author} {\bibfnamefont {H.}~\bibnamefont {Iqbal}}, \ and\ \bibinfo
  {author} {\bibfnamefont {M.~S.}\ \bibnamefont {Arshad}},\ }\href {\doibase
  10.1155/2014/673106} {\bibfield  {journal} {\bibinfo  {journal} {J. Appl.
  Math.}\ }\textbf {\bibinfo {volume} {2014}},\ \bibinfo {pages} {e673106}
  (\bibinfo {year} {2014})}\BibitemShut {NoStop}%
\bibitem [{\citenamefont {Marques}\ \emph {et~al.}(2012)\citenamefont
  {Marques}, \citenamefont {Oliveira},\ and\ \citenamefont
  {Burnus}}]{marques_libxc:_2012}%
  \BibitemOpen
  \bibfield  {author} {\bibinfo {author} {\bibfnamefont {M.~A.~L.}\
  \bibnamefont {Marques}}, \bibinfo {author} {\bibfnamefont {M.~J.~T.}\
  \bibnamefont {Oliveira}}, \ and\ \bibinfo {author} {\bibfnamefont
  {T.}~\bibnamefont {Burnus}},\ }\href {\doibase 10.1016/j.cpc.2012.05.007}
  {\bibfield  {journal} {\bibinfo  {journal} {Computer Physics Communications}\
  }\textbf {\bibinfo {volume} {183}},\ \bibinfo {pages} {2272} (\bibinfo {year}
  {2012})},\ \Eprint {http://arxiv.org/abs/1203.1739} {arXiv:1203.1739
  [cond-mat.mtrl-sci]} \BibitemShut {NoStop}%
\bibitem [{\citenamefont {Lehtola}\ \emph {et~al.}(2018)\citenamefont
  {Lehtola}, \citenamefont {Steigemann}, \citenamefont {Oliveira},\ and\
  \citenamefont {Marques}}]{LehtolaRecentdevelopmentslibxc2018}%
  \BibitemOpen
  \bibfield  {author} {\bibinfo {author} {\bibfnamefont {S.}~\bibnamefont
  {Lehtola}}, \bibinfo {author} {\bibfnamefont {C.}~\bibnamefont {Steigemann}},
  \bibinfo {author} {\bibfnamefont {M.~J.~T.}\ \bibnamefont {Oliveira}}, \ and\
  \bibinfo {author} {\bibfnamefont {M.~A.~L.}\ \bibnamefont {Marques}},\ }\href
  {\doibase 10.1016/j.softx.2017.11.002} {\bibfield  {journal} {\bibinfo
  {journal} {SoftwareX}\ }\textbf {\bibinfo {volume} {7}},\ \bibinfo {pages}
  {1} (\bibinfo {year} {2018})}\BibitemShut {NoStop}%
\bibitem [{\citenamefont {Johnson}\ and\ \citenamefont
  {Becke}(2009)}]{johnson_tests_2009}%
  \BibitemOpen
  \bibfield  {author} {\bibinfo {author} {\bibfnamefont {E.~R.}\ \bibnamefont
  {Johnson}}\ and\ \bibinfo {author} {\bibfnamefont {A.~D.}\ \bibnamefont
  {Becke}},\ }\href {\doibase 10.1139/V09-102} {\bibfield  {journal} {\bibinfo
  {journal} {Can. J. Chem.}\ }\textbf {\bibinfo {volume} {87}},\ \bibinfo
  {pages} {1369} (\bibinfo {year} {2009})}\BibitemShut {NoStop}%
\bibitem [{\citenamefont {Afeefy}\ \emph {et~al.}(2016)\citenamefont {Afeefy},
  \citenamefont {Liebmann},\ and\ \citenamefont
  {Stein}}]{linstrom_neutral_2016}%
  \BibitemOpen
  \bibfield  {author} {\bibinfo {author} {\bibfnamefont {H.}~\bibnamefont
  {Afeefy}}, \bibinfo {author} {\bibfnamefont {J.}~\bibnamefont {Liebmann}}, \
  and\ \bibinfo {author} {\bibfnamefont {S.}~\bibnamefont {Stein}},\ }in\
  \href@noop {} {\emph {\bibinfo {booktitle} {{{NIST Standard Reference
  Database Number}}}}},\ Vol.~\bibinfo {volume} {69},\ \bibinfo {editor}
  {edited by\ \bibinfo {editor} {\bibfnamefont {P.}~\bibnamefont {Linstrom}}\
  and\ \bibinfo {editor} {\bibfnamefont {W.}~\bibnamefont {Mallard}}}\
  (\bibinfo  {publisher} {{National Institute of Standards and Technology}},\
  \bibinfo {address} {Gaithersburg MD},\ \bibinfo {year} {2016})\BibitemShut
  {NoStop}%
\bibitem [{\citenamefont {Toulouse}\ \emph {et~al.}(2004)\citenamefont
  {Toulouse}, \citenamefont {Colonna},\ and\ \citenamefont
  {Savin}}]{toulouse_long-rangechar21short-range_2004}%
  \BibitemOpen
  \bibfield  {author} {\bibinfo {author} {\bibfnamefont {J.}~\bibnamefont
  {Toulouse}}, \bibinfo {author} {\bibfnamefont {F.}~\bibnamefont {Colonna}}, \
  and\ \bibinfo {author} {\bibfnamefont {A.}~\bibnamefont {Savin}},\ }\href
  {\doibase 10.1103/PhysRevA.70.062505} {\bibfield  {journal} {\bibinfo
  {journal} {Phys. Rev. A}\ }\textbf {\bibinfo {volume} {70}},\ \bibinfo
  {pages} {062505} (\bibinfo {year} {2004})},\ \Eprint
  {http://arxiv.org/abs/0410062} {arXiv:0410062 [physics.chem-ph]} \BibitemShut
  {NoStop}%
\bibitem [{\citenamefont {Vydrov}\ and\ \citenamefont
  {Scuseria}(2006)}]{vydrov_assessment_2006}%
  \BibitemOpen
  \bibfield  {author} {\bibinfo {author} {\bibfnamefont {O.~A.}\ \bibnamefont
  {Vydrov}}\ and\ \bibinfo {author} {\bibfnamefont {G.~E.}\ \bibnamefont
  {Scuseria}},\ }\href {\doibase 10.1063/1.2409292} {\bibfield  {journal}
  {\bibinfo  {journal} {J. Chem. Phys.}\ }\textbf {\bibinfo {volume} {125}},\
  \bibinfo {pages} {234109} (\bibinfo {year} {2006})}\BibitemShut {NoStop}%
\bibitem [{\citenamefont {Vydrov}\ \emph {et~al.}(2006)\citenamefont {Vydrov},
  \citenamefont {Heyd}, \citenamefont {Krukau},\ and\ \citenamefont
  {Scuseria}}]{vydrov_importance_2006}%
  \BibitemOpen
  \bibfield  {author} {\bibinfo {author} {\bibfnamefont {O.~A.}\ \bibnamefont
  {Vydrov}}, \bibinfo {author} {\bibfnamefont {J.}~\bibnamefont {Heyd}},
  \bibinfo {author} {\bibfnamefont {A.~V.}\ \bibnamefont {Krukau}}, \ and\
  \bibinfo {author} {\bibfnamefont {G.~E.}\ \bibnamefont {Scuseria}},\ }\href
  {\doibase 10.1063/1.2244560} {\bibfield  {journal} {\bibinfo  {journal} {J.
  Chem. Phys.}\ }\textbf {\bibinfo {volume} {125}},\ \bibinfo {pages} {074106}
  (\bibinfo {year} {2006})}\BibitemShut {NoStop}%
\bibitem [{\citenamefont {Cohen}\ \emph {et~al.}(2007)\citenamefont {Cohen},
  \citenamefont {Mori-S{\'a}nchez},\ and\ \citenamefont
  {Yang}}]{cohen_development_2007}%
  \BibitemOpen
  \bibfield  {author} {\bibinfo {author} {\bibfnamefont {A.~J.}\ \bibnamefont
  {Cohen}}, \bibinfo {author} {\bibfnamefont {P.}~\bibnamefont
  {Mori-S{\'a}nchez}}, \ and\ \bibinfo {author} {\bibfnamefont
  {W.}~\bibnamefont {Yang}},\ }\href {\doibase 10.1063/1.2741248} {\bibfield
  {journal} {\bibinfo  {journal} {J. Chem. Phys.}\ }\textbf {\bibinfo {volume}
  {126}},\ \bibinfo {pages} {191109} (\bibinfo {year} {2007})}\BibitemShut
  {NoStop}%
\bibitem [{\citenamefont {Gerber}\ \emph {et~al.}(2007)\citenamefont {Gerber},
  \citenamefont {{\'A}ngy{\'a}n}, \citenamefont {Marsman},\ and\ \citenamefont
  {Kresse}}]{gerber_range_2007}%
  \BibitemOpen
  \bibfield  {author} {\bibinfo {author} {\bibfnamefont {I.~C.}\ \bibnamefont
  {Gerber}}, \bibinfo {author} {\bibfnamefont {J.~G.}\ \bibnamefont
  {{\'A}ngy{\'a}n}}, \bibinfo {author} {\bibfnamefont {M.}~\bibnamefont
  {Marsman}}, \ and\ \bibinfo {author} {\bibfnamefont {G.}~\bibnamefont
  {Kresse}},\ }\href {\doibase 10.1063/1.2759209} {\bibfield  {journal}
  {\bibinfo  {journal} {J. Chem. Phys.}\ }\textbf {\bibinfo {volume} {127}},\
  \bibinfo {pages} {054101} (\bibinfo {year} {2007})}\BibitemShut {NoStop}%
\bibitem [{\citenamefont {Henderson}\ \emph {et~al.}(2008)\citenamefont
  {Henderson}, \citenamefont {Janesko},\ and\ \citenamefont
  {Scuseria}}]{henderson_range_2008}%
  \BibitemOpen
  \bibfield  {author} {\bibinfo {author} {\bibfnamefont {T.~M.}\ \bibnamefont
  {Henderson}}, \bibinfo {author} {\bibfnamefont {B.~G.}\ \bibnamefont
  {Janesko}}, \ and\ \bibinfo {author} {\bibfnamefont {G.~E.}\ \bibnamefont
  {Scuseria}},\ }\href {\doibase 10.1021/jp806573k} {\bibfield  {journal}
  {\bibinfo  {journal} {J. Phys. Chem. A}\ }\textbf {\bibinfo {volume} {112}},\
  \bibinfo {pages} {12530} (\bibinfo {year} {2008})}\BibitemShut {NoStop}%
\bibitem [{\citenamefont {Livshits}\ and\ \citenamefont
  {Baer}(2008)}]{livshits_density_2008}%
  \BibitemOpen
  \bibfield  {author} {\bibinfo {author} {\bibfnamefont {E.}~\bibnamefont
  {Livshits}}\ and\ \bibinfo {author} {\bibfnamefont {R.}~\bibnamefont
  {Baer}},\ }\href {\doibase 10.1021/jp803606n} {\bibfield  {journal} {\bibinfo
   {journal} {J. Phys. Chem. A}\ }\textbf {\bibinfo {volume} {112}},\ \bibinfo
  {pages} {12789} (\bibinfo {year} {2008})},\ \Eprint
  {http://arxiv.org/abs/0804.3145} {arXiv:0804.3145 [cond-mat.mtrl-sci]}
  \BibitemShut {NoStop}%
\bibitem [{\citenamefont {Livshits}\ \emph {et~al.}(2009)\citenamefont
  {Livshits}, \citenamefont {Baer},\ and\ \citenamefont
  {Kosloff}}]{livshits_deleterious_2009}%
  \BibitemOpen
  \bibfield  {author} {\bibinfo {author} {\bibfnamefont {E.}~\bibnamefont
  {Livshits}}, \bibinfo {author} {\bibfnamefont {R.}~\bibnamefont {Baer}}, \
  and\ \bibinfo {author} {\bibfnamefont {R.}~\bibnamefont {Kosloff}},\ }\href
  {\doibase 10.1021/jp900892r} {\bibfield  {journal} {\bibinfo  {journal} {J.
  Phys. Chem. A}\ }\textbf {\bibinfo {volume} {113}},\ \bibinfo {pages} {7521}
  (\bibinfo {year} {2009})}\BibitemShut {NoStop}%
\bibitem [{\citenamefont {Refaely-Abramson}\ \emph {et~al.}(2011)\citenamefont
  {Refaely-Abramson}, \citenamefont {Baer},\ and\ \citenamefont
  {Kronik}}]{refaely-abramson_fundamental_2011}%
  \BibitemOpen
  \bibfield  {author} {\bibinfo {author} {\bibfnamefont {S.}~\bibnamefont
  {Refaely-Abramson}}, \bibinfo {author} {\bibfnamefont {R.}~\bibnamefont
  {Baer}}, \ and\ \bibinfo {author} {\bibfnamefont {L.}~\bibnamefont
  {Kronik}},\ }\href {\doibase 10.1103/PhysRevB.84.075144} {\bibfield
  {journal} {\bibinfo  {journal} {Phys. Rev. B}\ }\textbf {\bibinfo {volume}
  {84}},\ \bibinfo {pages} {075144} (\bibinfo {year} {2011})}\BibitemShut
  {NoStop}%
\bibitem [{\citenamefont {Seth}\ \emph {et~al.}(2013)\citenamefont {Seth},
  \citenamefont {Ziegler}, \citenamefont {Steinmetz},\ and\ \citenamefont
  {Grimme}}]{seth_modeling_2013}%
  \BibitemOpen
  \bibfield  {author} {\bibinfo {author} {\bibfnamefont {M.}~\bibnamefont
  {Seth}}, \bibinfo {author} {\bibfnamefont {T.}~\bibnamefont {Ziegler}},
  \bibinfo {author} {\bibfnamefont {M.}~\bibnamefont {Steinmetz}}, \ and\
  \bibinfo {author} {\bibfnamefont {S.}~\bibnamefont {Grimme}},\ }\href
  {\doibase 10.1021/ct301112m} {\bibfield  {journal} {\bibinfo  {journal} {J.
  Chem. Theory Comput.}\ }\textbf {\bibinfo {volume} {9}},\ \bibinfo {pages}
  {2286} (\bibinfo {year} {2013})}\BibitemShut {NoStop}%
\bibitem [{\citenamefont {Autschbach}\ and\ \citenamefont
  {Srebro}(2014)}]{autschbach_delocalization_2014}%
  \BibitemOpen
  \bibfield  {author} {\bibinfo {author} {\bibfnamefont {J.}~\bibnamefont
  {Autschbach}}\ and\ \bibinfo {author} {\bibfnamefont {M.}~\bibnamefont
  {Srebro}},\ }\href {\doibase 10.1021/ar500171t} {\bibfield  {journal}
  {\bibinfo  {journal} {Acc. Chem. Res.}\ }\textbf {\bibinfo {volume} {47}},\
  \bibinfo {pages} {2592} (\bibinfo {year} {2014})}\BibitemShut {NoStop}%
\bibitem [{\citenamefont {{Cabral do Couto}}\ \emph {et~al.}(2015)\citenamefont
  {{Cabral do Couto}}, \citenamefont {Hollas},\ and\ \citenamefont
  {Slav{\'\i}{\v c}ek}}]{cabral_do_couto_performance_2015}%
  \BibitemOpen
  \bibfield  {author} {\bibinfo {author} {\bibfnamefont {P.}~\bibnamefont
  {{Cabral do Couto}}}, \bibinfo {author} {\bibfnamefont {D.}~\bibnamefont
  {Hollas}}, \ and\ \bibinfo {author} {\bibfnamefont {P.}~\bibnamefont
  {Slav{\'\i}{\v c}ek}},\ }\href {\doibase 10.1021/acs.jctc.5b00066} {\bibfield
   {journal} {\bibinfo  {journal} {J. Chem. Theory Comput.}\ }\textbf {\bibinfo
  {volume} {11}},\ \bibinfo {pages} {3234} (\bibinfo {year}
  {2015})}\BibitemShut {NoStop}%
\bibitem [{\citenamefont {Almbladh}\ and\ \citenamefont {{von
  Barth}}(1985)}]{almbladh_exact_1985}%
  \BibitemOpen
  \bibfield  {author} {\bibinfo {author} {\bibfnamefont {C.-O.}\ \bibnamefont
  {Almbladh}}\ and\ \bibinfo {author} {\bibfnamefont {U.}~\bibnamefont {{von
  Barth}}},\ }\href {\doibase 10.1103/PhysRevB.31.3231} {\bibfield  {journal}
  {\bibinfo  {journal} {Phys. Rev. B}\ }\textbf {\bibinfo {volume} {31}},\
  \bibinfo {pages} {3231} (\bibinfo {year} {1985})}\BibitemShut {NoStop}%
\bibitem [{\citenamefont {Lias}\ and\ \citenamefont
  {Liebmann}(2016)}]{linstrom_ion_2016}%
  \BibitemOpen
  \bibfield  {author} {\bibinfo {author} {\bibfnamefont {S.}~\bibnamefont
  {Lias}}\ and\ \bibinfo {author} {\bibfnamefont {J.}~\bibnamefont
  {Liebmann}},\ }in\ \href@noop {} {\emph {\bibinfo {booktitle} {{{NIST
  Standard Reference Database Number}}}}},\ Vol.~\bibinfo {volume} {69},\
  \bibinfo {editor} {edited by\ \bibinfo {editor} {\bibfnamefont
  {P.}~\bibnamefont {Linstrom}}\ and\ \bibinfo {editor} {\bibfnamefont
  {W.}~\bibnamefont {Mallard}}}\ (\bibinfo  {publisher} {{National Institute of
  Standards and Technology}},\ \bibinfo {address} {Gaithersburg MD},\ \bibinfo
  {year} {2016})\BibitemShut {NoStop}%
\bibitem [{\citenamefont {Shimazaki}\ and\ \citenamefont
  {Asai}(2008)}]{shimazaki_band_2008}%
  \BibitemOpen
  \bibfield  {author} {\bibinfo {author} {\bibfnamefont {T.}~\bibnamefont
  {Shimazaki}}\ and\ \bibinfo {author} {\bibfnamefont {Y.}~\bibnamefont
  {Asai}},\ }\href {\doibase 10.1016/j.cplett.2008.10.012} {\bibfield
  {journal} {\bibinfo  {journal} {Chemical Physics Letters}\ }\textbf {\bibinfo
  {volume} {466}},\ \bibinfo {pages} {91} (\bibinfo {year} {2008})}\BibitemShut
  {NoStop}%
\bibitem [{\citenamefont {Potts}\ and\ \citenamefont
  {Price}(1972)}]{potts_photoelectron_1972}%
  \BibitemOpen
  \bibfield  {author} {\bibinfo {author} {\bibfnamefont {A.~W.}\ \bibnamefont
  {Potts}}\ and\ \bibinfo {author} {\bibfnamefont {W.~C.}\ \bibnamefont
  {Price}},\ }\href@noop {} {\bibfield  {journal} {\bibinfo  {journal}
  {Proceedings of the Royal Society of London. Series A, Mathematical and
  Physical Sciences}\ }\textbf {\bibinfo {volume} {326}},\ \bibinfo {pages}
  {181} (\bibinfo {year} {1972})}\BibitemShut {NoStop}%
\bibitem [{\citenamefont {Koopmans}(1934)}]{koopmans_uber_1934}%
  \BibitemOpen
  \bibfield  {author} {\bibinfo {author} {\bibfnamefont {T.}~\bibnamefont
  {Koopmans}},\ }\href {\doibase 10.1016/S0031-8914(34)90011-2} {\bibfield
  {journal} {\bibinfo  {journal} {Physica}\ }\textbf {\bibinfo {volume} {1}},\
  \bibinfo {pages} {104} (\bibinfo {year} {1934})}\BibitemShut {NoStop}%
\bibitem [{\citenamefont {Zhao}\ and\ \citenamefont
  {Truhlar}(2006)}]{zhao_density_2006}%
  \BibitemOpen
  \bibfield  {author} {\bibinfo {author} {\bibfnamefont {Y.}~\bibnamefont
  {Zhao}}\ and\ \bibinfo {author} {\bibfnamefont {D.~G.}\ \bibnamefont
  {Truhlar}},\ }\href {\doibase 10.1021/jp066479k} {\bibfield  {journal}
  {\bibinfo  {journal} {J. Phys. Chem. A}\ }\textbf {\bibinfo {volume} {110}},\
  \bibinfo {pages} {13126} (\bibinfo {year} {2006})}\BibitemShut {NoStop}%
\bibitem [{\citenamefont {Chiu}\ \emph {et~al.}(1979)\citenamefont {Chiu},
  \citenamefont {Burrow},\ and\ \citenamefont {Jordan}}]{chiu_temporary_1979}%
  \BibitemOpen
  \bibfield  {author} {\bibinfo {author} {\bibfnamefont {N.~S.}\ \bibnamefont
  {Chiu}}, \bibinfo {author} {\bibfnamefont {P.~D.}\ \bibnamefont {Burrow}}, \
  and\ \bibinfo {author} {\bibfnamefont {K.~D.}\ \bibnamefont {Jordan}},\
  }\href {\doibase 10.1016/0009-2614(79)80082-2} {\bibfield  {journal}
  {\bibinfo  {journal} {Chemical Physics Letters}\ }\textbf {\bibinfo {volume}
  {68}},\ \bibinfo {pages} {121} (\bibinfo {year} {1979})}\BibitemShut
  {NoStop}%
\bibitem [{\citenamefont {Miller}\ \emph {et~al.}(1986)\citenamefont {Miller},
  \citenamefont {Leopold}, \citenamefont {Murray},\ and\ \citenamefont
  {Lineberger}}]{miller_electron_1986}%
  \BibitemOpen
  \bibfield  {author} {\bibinfo {author} {\bibfnamefont {T.~M.}\ \bibnamefont
  {Miller}}, \bibinfo {author} {\bibfnamefont {D.~G.}\ \bibnamefont {Leopold}},
  \bibinfo {author} {\bibfnamefont {K.~K.}\ \bibnamefont {Murray}}, \ and\
  \bibinfo {author} {\bibfnamefont {W.~C.}\ \bibnamefont {Lineberger}},\ }\href
  {\doibase 10.1063/1.451091} {\bibfield  {journal} {\bibinfo  {journal} {The
  Journal of Chemical Physics}\ }\textbf {\bibinfo {volume} {85}},\ \bibinfo
  {pages} {2368} (\bibinfo {year} {1986})}\BibitemShut {NoStop}%
\bibitem [{\citenamefont {Huber}\ and\ \citenamefont
  {Herzberg}(2016)}]{linstrom_constants_2016}%
  \BibitemOpen
  \bibfield  {author} {\bibinfo {author} {\bibfnamefont {K.~P.}\ \bibnamefont
  {Huber}}\ and\ \bibinfo {author} {\bibfnamefont {G.}~\bibnamefont
  {Herzberg}},\ }in\ \href@noop {} {\emph {\bibinfo {booktitle} {{{NIST
  Standard Reference Database Number}}}}},\ Vol.~\bibinfo {volume} {69},\
  \bibinfo {editor} {edited by\ \bibinfo {editor} {\bibfnamefont
  {P.}~\bibnamefont {Linstrom}}\ and\ \bibinfo {editor} {\bibfnamefont
  {W.}~\bibnamefont {Mallard}}}\ (\bibinfo  {publisher} {{National Institute of
  Standards and Technology}},\ \bibinfo {address} {Gaithersburg MD},\ \bibinfo
  {year} {2016})\BibitemShut {NoStop}%
\bibitem [{\citenamefont {R{\"o}sch}\ and\ \citenamefont
  {Trickey}(1997)}]{rosch_comment_1997}%
  \BibitemOpen
  \bibfield  {author} {\bibinfo {author} {\bibfnamefont {N.}~\bibnamefont
  {R{\"o}sch}}\ and\ \bibinfo {author} {\bibfnamefont {S.~B.}\ \bibnamefont
  {Trickey}},\ }\href {\doibase 10.1063/1.473946} {\bibfield  {journal}
  {\bibinfo  {journal} {J. Chem. Phys.}\ }\textbf {\bibinfo {volume} {106}},\
  \bibinfo {pages} {8940} (\bibinfo {year} {1997})}\BibitemShut {NoStop}%
\bibitem [{\citenamefont {Tozer}\ and\ \citenamefont
  {De~Proft}(2005)}]{tozer_computation_2005}%
  \BibitemOpen
  \bibfield  {author} {\bibinfo {author} {\bibfnamefont {D.~J.}\ \bibnamefont
  {Tozer}}\ and\ \bibinfo {author} {\bibfnamefont {F.}~\bibnamefont
  {De~Proft}},\ }\href {\doibase 10.1021/jp053504y} {\bibfield  {journal}
  {\bibinfo  {journal} {J. Phys. Chem. A}\ }\textbf {\bibinfo {volume} {109}},\
  \bibinfo {pages} {8923} (\bibinfo {year} {2005})}\BibitemShut {NoStop}%
\bibitem [{\citenamefont {W{\"u}rdemann}(2016)}]{wurdemann_berechnung_2016}%
  \BibitemOpen
  \bibfield  {author} {\bibinfo {author} {\bibfnamefont {R.}~\bibnamefont
  {W{\"u}rdemann}},\ }\emph {\bibinfo {title} {{Berechnung optischer Spektren
  und Grundzustandseigenschaften neutraler und geladener Molek{\"u}le mittels
  Dichtefunktionaltheorie}}},\ \href@noop {} {Ph.D. thesis},\ \bibinfo
  {school} {Universit{\"a}t Freiburg}, \bibinfo {address} {Freiburg, Germany}
  (\bibinfo {year} {2016})\BibitemShut {NoStop}%
\bibitem [{\citenamefont {Hunt}\ and\ \citenamefont
  {Goddard~III}(1969)}]{hunt_excited_1969}%
  \BibitemOpen
  \bibfield  {author} {\bibinfo {author} {\bibfnamefont {s.~J.}\ \bibnamefont
  {Hunt}}\ and\ \bibinfo {author} {\bibfnamefont {W.~A.}\ \bibnamefont
  {Goddard~III}},\ }\href {\doibase 10.1016/0009-2614(69)80154-5} {\bibfield
  {journal} {\bibinfo  {journal} {Chemical Physics Letters}\ }\textbf {\bibinfo
  {volume} {3}},\ \bibinfo {pages} {414} (\bibinfo {year} {1969})}\BibitemShut
  {NoStop}%
\bibitem [{\citenamefont {Casida}\ and\ \citenamefont
  {Huix-Rotllant}(2012)}]{casida_progress_2012}%
  \BibitemOpen
  \bibfield  {author} {\bibinfo {author} {\bibfnamefont {M.~E.}\ \bibnamefont
  {Casida}}\ and\ \bibinfo {author} {\bibfnamefont {M.}~\bibnamefont
  {Huix-Rotllant}},\ }\href {\doibase 10.1146/annurev-physchem-032511-143803}
  {\bibfield  {journal} {\bibinfo  {journal} {Annu. Rev. Phys. Chem.}\ }\textbf
  {\bibinfo {volume} {63}},\ \bibinfo {pages} {287} (\bibinfo {year} {2012})},\
  \Eprint {http://arxiv.org/abs/1108.0611} {arXiv:1108.0611 [physics.chem-ph]}
  \BibitemShut {NoStop}%
\bibitem [{\citenamefont {Berman}\ and\ \citenamefont
  {Kaldor}(1979)}]{berman_fast_1979}%
  \BibitemOpen
  \bibfield  {author} {\bibinfo {author} {\bibfnamefont {M.}~\bibnamefont
  {Berman}}\ and\ \bibinfo {author} {\bibfnamefont {U.}~\bibnamefont
  {Kaldor}},\ }\href {\doibase 10.1016/0301-0104(79)85205-2} {\bibfield
  {journal} {\bibinfo  {journal} {Chemical Physics}\ }\textbf {\bibinfo
  {volume} {43}},\ \bibinfo {pages} {375} (\bibinfo {year} {1979})}\BibitemShut
  {NoStop}%
\bibitem [{\citenamefont {{van Meer}}\ \emph {et~al.}(2014)\citenamefont {{van
  Meer}}, \citenamefont {Gritsenko},\ and\ \citenamefont
  {Baerends}}]{vanMeerPhysicalMeaningVirtual2014}%
  \BibitemOpen
  \bibfield  {author} {\bibinfo {author} {\bibfnamefont {R.}~\bibnamefont {{van
  Meer}}}, \bibinfo {author} {\bibfnamefont {O.~V.}\ \bibnamefont {Gritsenko}},
  \ and\ \bibinfo {author} {\bibfnamefont {E.~J.}\ \bibnamefont {Baerends}},\
  }\href {\doibase 10.1021/ct500727c} {\bibfield  {journal} {\bibinfo
  {journal} {J. Chem. Theory Comput.}\ }\textbf {\bibinfo {volume} {10}},\
  \bibinfo {pages} {4432} (\bibinfo {year} {2014})}\BibitemShut {NoStop}%
\bibitem [{\citenamefont {{van Meer}}\ \emph {et~al.}(2017)\citenamefont {{van
  Meer}}, \citenamefont {Gritsenko},\ and\ \citenamefont
  {Baerends}}]{vanMeerNaturalexcitationorbitals2017}%
  \BibitemOpen
  \bibfield  {author} {\bibinfo {author} {\bibfnamefont {R.}~\bibnamefont {{van
  Meer}}}, \bibinfo {author} {\bibfnamefont {O.~V.}\ \bibnamefont {Gritsenko}},
  \ and\ \bibinfo {author} {\bibfnamefont {E.~J.}\ \bibnamefont {Baerends}},\
  }\href {\doibase 10.1063/1.4974327} {\bibfield  {journal} {\bibinfo
  {journal} {The Journal of Chemical Physics}\ }\textbf {\bibinfo {volume}
  {146}},\ \bibinfo {pages} {044119} (\bibinfo {year} {2017})}\BibitemShut
  {NoStop}%
\bibitem [{\citenamefont {Krause}\ and\ \citenamefont
  {Th{\"o}rnig}(2016)}]{krause_jureca:_2016}%
  \BibitemOpen
  \bibfield  {author} {\bibinfo {author} {\bibfnamefont {D.}~\bibnamefont
  {Krause}}\ and\ \bibinfo {author} {\bibfnamefont {P.}~\bibnamefont
  {Th{\"o}rnig}},\ }\href {\doibase 10.17815/jlsrf-2-121} {\bibfield  {journal}
  {\bibinfo  {journal} {J. Large-Scale Res. Facil. JLSRF}\ }\textbf {\bibinfo
  {volume} {2}} (\bibinfo {year} {2016}),\ 10.17815/jlsrf-2-121}\BibitemShut
  {NoStop}%
\bibitem [{{\relax DLMF}()}]{NIST:DLMF}%
  \BibitemOpen
  {\relax DLMF},\ \href {http://dlmf.nist.gov/} {\enquote {\bibinfo {title}
  {{NIST Digital Library of Mathematical Functions}},}\ }\bibinfo
  {howpublished} {http://dlmf.nist.gov/, Release 1.0.10 of 2015-08-07},\
  \bibinfo {note} {online companion to \cite{Olver:2010:NHMF}}\BibitemShut
  {NoStop}%
\bibitem [{\citenamefont {Olver}\ \emph {et~al.}(2010)\citenamefont {Olver},
  \citenamefont {Lozier}, \citenamefont {Boisvert},\ and\ \citenamefont
  {Clark}}]{Olver:2010:NHMF}%
  \BibitemOpen
  \bibinfo {editor} {\bibfnamefont {F.~W.~J.}\ \bibnamefont {Olver}}, \bibinfo
  {editor} {\bibfnamefont {D.~W.}\ \bibnamefont {Lozier}}, \bibinfo {editor}
  {\bibfnamefont {R.~F.}\ \bibnamefont {Boisvert}}, \ and\ \bibinfo {editor}
  {\bibfnamefont {C.~W.}\ \bibnamefont {Clark}},\ eds.,\ \href@noop {} {\emph
  {\bibinfo {title} {{NIST Handbook of Mathematical Functions}}}}\ (\bibinfo
  {publisher} {Cambridge University Press},\ \bibinfo {address} {New York,
  NY},\ \bibinfo {year} {2010})\ \bibinfo {note} {print companion to
  \cite{NIST:DLMF}}\BibitemShut {NoStop}%
\bibitem [{\citenamefont {Gaunt}(1929)}]{gaunt_triplets_1929-1}%
  \BibitemOpen
  \bibfield  {author} {\bibinfo {author} {\bibfnamefont {J.~A.}\ \bibnamefont
  {Gaunt}},\ }\href@noop {} {\bibfield  {journal} {\bibinfo  {journal}
  {Philosophical Transactions of the Royal Society of London. Series A,
  Containing Papers of a Mathematical or Physical Character}\ }\textbf
  {\bibinfo {volume} {228}},\ \bibinfo {pages} {151} (\bibinfo {year}
  {1929})}\BibitemShut {NoStop}%
\bibitem [{\citenamefont {{Maxima}}(2014)}]{maxima_maxima_2014}%
  \BibitemOpen
  \bibfield  {author} {\bibinfo {author} {\bibnamefont {{Maxima}}},\
  }\href@noop {} {\enquote {\bibinfo {title} {Maxima, a {{Computer Algebra
  System}}. {{Version}} 5.34.1},}\ } (\bibinfo {year} {2014})\BibitemShut
  {NoStop}%
\bibitem [{\citenamefont {Kuritz}\ \emph {et~al.}(2011)\citenamefont {Kuritz},
  \citenamefont {Stein}, \citenamefont {Baer},\ and\ \citenamefont
  {Kronik}}]{KuritzChargeTransferLikeExcitationsTimeDependent2011}%
  \BibitemOpen
  \bibfield  {author} {\bibinfo {author} {\bibfnamefont {N.}~\bibnamefont
  {Kuritz}}, \bibinfo {author} {\bibfnamefont {T.}~\bibnamefont {Stein}},
  \bibinfo {author} {\bibfnamefont {R.}~\bibnamefont {Baer}}, \ and\ \bibinfo
  {author} {\bibfnamefont {L.}~\bibnamefont {Kronik}},\ }\href {\doibase
  10.1021/ct2002804} {\bibfield  {journal} {\bibinfo  {journal} {J. Chem.
  Theory Comput.}\ }\textbf {\bibinfo {volume} {7}},\ \bibinfo {pages} {2408}
  (\bibinfo {year} {2011})}\BibitemShut {NoStop}%
\bibitem [{\citenamefont {Li}\ and\ \citenamefont
  {Lyyra}(1999)}]{LiTripletstatesNa21999}%
  \BibitemOpen
  \bibfield  {author} {\bibinfo {author} {\bibfnamefont {L.}~\bibnamefont
  {Li}}\ and\ \bibinfo {author} {\bibfnamefont {A.~M.}\ \bibnamefont {Lyyra}},\
  }\href {\doibase 10.1016/S1386-1425(99)00091-8} {\bibfield  {journal}
  {\bibinfo  {journal} {Spectrochimica Acta Part A: Molecular and Biomolecular
  Spectroscopy}\ }\textbf {\bibinfo {volume} {55}},\ \bibinfo {pages} {2147}
  (\bibinfo {year} {1999})}\BibitemShut {NoStop}%
\end{thebibliography}%
\bibliographystyle{apsrev4-1}

\begin{widetext}

\clearpage
\newpage
\setcounter{page}{1}
\setcounter{section}{0}
\setcounter{equation}{0}
\setcounter{table}{0}
\setcounter{figure}{0}
\renewcommand{\thesection}{SI \arabic{section}}
\renewcommand{\thepage}{SI \arabic{page}}
\renewcommand{\thefigure}{SI \arabic{figure}}
\renewcommand{\thetable}{SI \arabic{table}}

\renewcommand{\theequation}{S\arabic{equation}}

\begin{center}
\textbf{\large{Supporting information for ``Calculation of charge transfer excitations with range separated functionals using improved virtual orbitals on real-space grids''}}
\end{center}

\section*{Introduction}

This supporting information contains details of derivations and results that are too lengthy to be included in
the main paper. 

\section{Evaluation of local terms for screened exchange}

As written in the main text, the evaluation of 
\begin{align}
  \left(\left(n_{ij}\right)\right)^\gamma &= \left(\left(\tilde{n}_{ij}\right)\right)^\gamma +
  \sum_\alpha \left [ \left(\left(n^\alpha_{ij}\right)\right)^\gamma - \left(\left(\tilde{n}^\alpha_{ij}\right)\right)^\gamma \right],\label{eq:SIScrExPawNoComp}
\end{align}
with $$\left(\left(n_{ij}\right)\right)^\gamma = \left(n_{ij}|n_{ji}\right)^\gamma  = \left(n_{ij} |\frac{ \exp \left( -\gamma r_{12}\right)}{r_{12}}| n_{ji} \right)$$
would lead to cross-terms between localized functions located on different atomic sites and the need to integrate on incompatible grids\cite{blochl_projector_1994,mortensen_real-space_2005}. To circumvent this, compensation charges
$\tilde{Z}_{ij}^\alpha = \sum_L Q_{L,ij}^\alpha \hat{g}_L^\alpha$
with the expansion coefficients $Q_{L,ij}^\alpha$ and the
smooth localized function $\hat{g}_L^\alpha$ are
introduced\cite{rostgaard_exact_2006,enkovaara_electronic_2010},
where
$L=(\ell, m)$ denotes the combination of angular $\ell$ and magnetic
quantum numbers $m$.

Using  $\tilde{\varrho}_{ij} = \tilde{n}_{ij} + \sum_\alpha \tilde{Z}_{ij}^\alpha$, 
eq. (\ref{eq:SIScrExPawNoComp}) becomes
\begin{align}
\left(\left(n_{ij}\right)\right)^\gamma &= \left(\left(\tilde{\varrho}_{ij} \right)\right)^\gamma \notag\\
& \quad + \sum_\alpha \left [ \left(\left(n^\alpha_{ij}\right)\right)^\gamma - \left(\left(\tilde{n}^\alpha_{ij} + \tilde{Z}_{ij}^\alpha\right)\right)^\gamma \right] \\
&= \left(\left(\tilde{\varrho}_{ij}\right)\right)^\gamma + \sum_\alpha \Delta K_{ij}^{\alpha\gamma}.\label{eq:ScrExPawCompSI}
\end{align}
The local correction term $\Delta K_{ij}^{\alpha\gamma}$ reads
\begin{align}
\Delta K_{ij}^{\alpha\gamma} &= \sum_{k_1 k_2 k_3 k_4} \mathcal{P}_{ik_1}^{\alpha\ast} \mathcal{P}_{jk_2}^\alpha
 C_{k_1 k_2 k_3 k_4}^{\alpha\gamma} \mathcal{P}_{jk_3}^{\alpha\ast} \mathcal{P}_{ik_4}^\alpha
\end{align}
where the system independent tensor
\begin{align}
C_{k_1 k_2 k_3 k_4}^{\alpha\gamma} &= J_{k_1 k_2 k_3 k_4}^{\alpha\gamma} + \notag \\
& \quad  \sum_L M_{k_1 k_2}^{\alpha L\gamma} \Delta_{L k_3 k_4}^\alpha  + \notag \\ 
& \quad \sum_{LL^\prime} \Delta_{L k_1 k_2}^\alpha N_{LL^\prime}^{\alpha\gamma} \Delta_{L^\prime k_3 k_4}^\alpha ,
\end{align}
with the terms
\begin{align}
J_{k_1 k_2 k_3 k_4}^{\alpha\gamma} &= \frac{1}{2} \Big [ \left(\phi_{k_1}^\alpha \phi_{k_2}^\alpha |\phi_{k_3}^\alpha \phi_{k_4}^\alpha \right)^\gamma  \notag \\
& \left . \quad \quad \quad - \left(\tilde{\phi}_{k_1}^\alpha \tilde{\phi}_{k_2}^\alpha |\tilde{\phi}_{k_3}^\alpha \tilde{\phi}_{k_4}^\alpha \right)^\gamma \right]\\
M_{k_1 k_2}^{\alpha L\gamma} &= \left(\tilde{\phi}_{k_1}^\alpha \tilde{\phi}_{k_2}^\alpha \big |\tilde{g}_L^\alpha\right)^\gamma \\
N_{LL^\prime}^{\alpha\gamma} &= \frac{1}{2} \left(\tilde{g}_L^\alpha |\tilde{g}_{L^\prime}^\alpha\right)^\gamma,
\end{align}
is used. 
The $\Delta_{L k_1 k_2}^\alpha$ are described in refs. \citenum{mortensen_real-space_2005,rostgaard_exact_2006}. Integrations on two types of grids have to be performed to evaluate $K^\gamma_{ij}$ by eq. (\ref{eq:ScrExPawCompSI}): once the integration on a three dimensional Cartesian grid using the soft pseudo-charge, $\tilde{\varrho}_{ij}$ and once integrations on a radial grid using the partial WF, $\phi_k^\alpha$, $\tilde{\phi}_k^\alpha$ and the localized function $\tilde{g}_L^\alpha$. The integration using the screened Poisson equation is discussed in the main text. 

The partial wave functions $\phi_k^\alpha$ and the localized functions $\tilde{g}_L^\alpha$ are implemented as a (real) radial function $R_k (r)$ times a spherical harmonics\cite{mortensen_real-space_2005}
\begin{align}
\phi^\alpha_k &= R_k^\alpha (r)Y_{L_k} (\hat{r}).\notag
\end{align}
Following the work of 
Rico \emph{et al.}\cite{rico_repulsion_2012}, 
the screened exchange integral is evaluated on radial grids as
\begin{align}
I_{k_1 k_2 k_3 k_4}(\gamma) =&  4 \pi \sum_L   G_{L_{k_1} L_{k_2}}^L G_{L_{k_3} L_{k_4}}^L  \notag  \\
& \quad \times \int_0^\infty  R_{k_1} (r_1) R_{k_2} (r_1) r_1^2\notag \\
& \quad \times  \int_0^\infty \frac{I_{\ell + \nicefrac{1}{2}} \left( \gamma r_< \right)K_{\ell + \nicefrac{1}{2}} \left( \gamma r_> \right)}{\sqrt{r_1 r_2}} \notag \\
& \quad \times  R_{k_3} (r_2) R_{k_4} (r_2) r_2^2 dr_1 dr_2 \label{eq:YukRadialIK}, 
\end{align}
where $G_{L_a L_b}^{L}$ denotes the 
Gaunt coefficients\cite{gaunt_triplets_1929-1},  
$I_\ell(x)$ and $K_\ell (x)$ are the modified Bessel functions of the first 
and second type (the latter are also known as MacDonald functions)\cite{NIST:DLMF,Olver:2010:NHMF},
and $r_>$ and $r_<$ denote the smaller/larger of the radial 
coordinates $r_1$ and $r_2$. 
Note, that we have dropped the atomic index $\alpha$ for brevity.

\section{Influence of the grid- and vacuum-spacing on eigenvalues and energies}

The grid-spacing $h$ and the amount of vacuum around each atom $x_\text{vac}$ 
are crucial parameters in grid-based calculations. 
To give a rationale for our chosen parameters, the influence of $h$ on the eigenvalues of the HOMO and the ionization energy for atomic chromium and oxygen is investigated. 

\begin{figure}[h]
  \centering
  \includegraphics{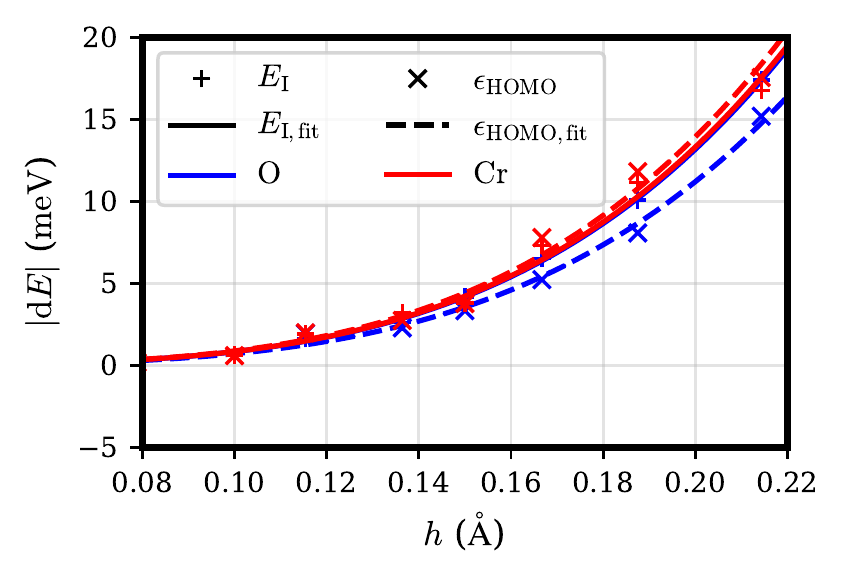}
  \caption{Deviation of $\epsilon_\text{HOMO}$ and $E_\text{I}$ for
    isolated a) oxygen and b) chromium atoms in dependence of 
    the grid-spacing $h$.}
  \label{fig:grid_ion_OCr}
\end{figure} 
Figure \ref{fig:grid_ion_OCr} depicts the absolute deviation 
$|\text{d}E|$ of the ionization energy 
$E_\text{I}$ and the eigenvalue of the HOMO 
$\epsilon_\text{HOMO}$ for atomic chromium and oxygen from the estimated 
value for an infinite dense grid
relative to the grid-spacing. 
The value for the infinitely dense grid
was estimated using an $h^4$ behavior.
This was verified using a fit with $a\cdot h^4 + b$ 
also depicted in fig. \ref{fig:grid_ion_OCr}.
The $h^4$ behavior is to be expected as all 
approximations in \gpaw{} are exact to at least 
$h^3$\cite{mortensen_real-space_2005}.
LCY-PBE with
$\gamma = \unit[0.9]{a_0^{-1}}$ was used in these calculations. 
This high value of the screening parameter leads to a strong decay 
of the screened Gaussian used to neutralize the charge 
introduced by calculating the exchange of a WF with itself. 
Both curves show a nearly identical behavior. 
For $h\le\unit[0.18]{\text{\AA}}$ both 
$E_\text{I}$ and $\epsilon_\text{HOMO}$ are converged within $\unit[10]{meV}$ 
relative to the estimated value for the infinite dense grid. 
Therefore $h=\unit[0.18]{\text{\AA}}$ was chosen for calculations 
not involving transition metals (which suffer from 
further numerical effects, see next section).

\begin{figure}[htb]
  \centering
  \includegraphics{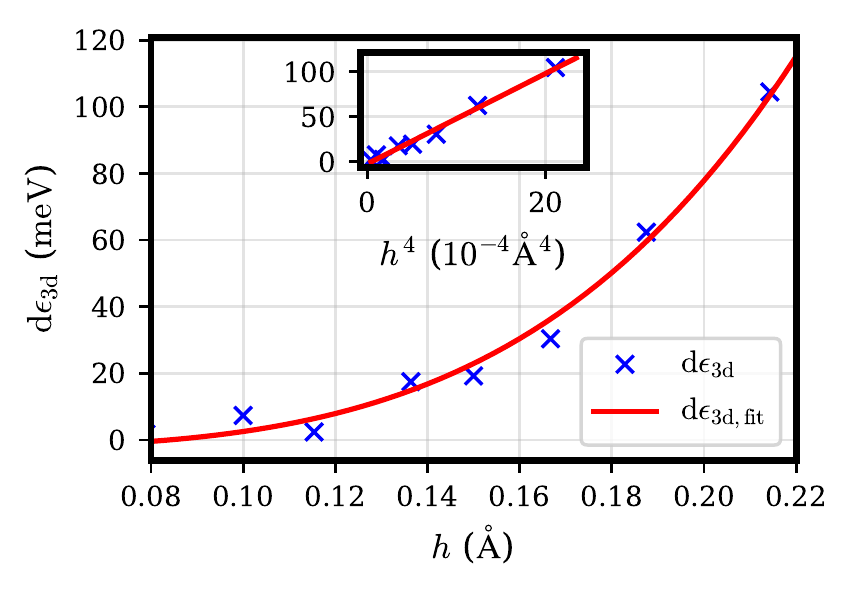}
  \caption{Energetic splitting, $\text{d}\epsilon_{3d}$, for the eigenvalues of the $3d$ shells of the isolated Chromium atom relative to  the grid-spacing $h$.}
  \label{fig:grid_split_Cr}
\end{figure} 
The projection of $d$ type orbitals in transition metals 
on Cartesian grids leads 
to an artificial energetic splitting of the eigenvalues 
of the corresponding states. 
The influence of $h$ on this splitting was investigated 
in order to get parameters accurate for calculations 
involving transition metals. 
Figure \ref{fig:grid_split_Cr} depicts the projection induced 
energetic splitting of the $3d$ shells $\text{d}\epsilon_{3d}$ 
for the example of atomic chromium. 
As indicated by the inset, the splitting also exhibits an $h^4$ behavior.
The error for $h=\unit[0.18]{\text{\AA}}$ stays in the order of 
$\unit[50]{meV}$ and drops 
to $\unit[30]{meV}$ for $h=\unit[0.16]{\text{\AA}}$. 
Thus the latter spacing was chosen when transition metals were considered.

\begin{figure}[htb]
  \centering
  \includegraphics{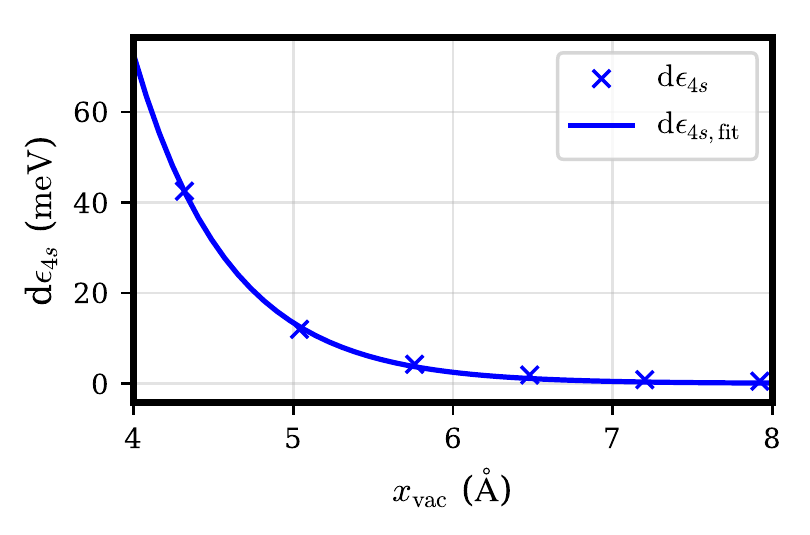}
  \caption{Eigenvalue of the occupied $4s$ orbital $\text{d}\epsilon_{4s}$ 
    for an isolated Chromium atom relative to the distance 
    of the atom from the border of the simulation box $x_\text{vac}$.}
  \label{fig:box_homo}
\end{figure}
In order to study the effect of our finite simulation box, we have used
the occupied $4s$ orbital of atomic chromium which is strongly delocalized.
The behavior of its eigenvalue relative to $x_\text{vac}$ was
taken as a measure to qualify an appropriate box-size.
Figure \ref{fig:box_homo} depicts the deviation of the eigenvalue 
of the occupied $4s$ orbital of the isolated Chromium atom from the 
estimated value for the infinite large simulation box relative to 
$x_\text{vac}$ for the functional LCY-PBE with $\gamma= \unit[0.9]{a_0}$. 
The deviation is below $\unit[30]{meV}$ already for 
$x_\text{vac} > \unit[4.5]{\text{\AA}}$. 
For $x_\text{vac} = \unit[6.0]{\text{\AA}}$ the differences falls 
below $\unit[5]{meV}$ which we regarded as adequate for the 
calculations performed.

\section{Derivatives of the localized range separation function}

The exchange part of a GGA in the case of a Slater-function based RSF 
reads\cite{akinaga_range-separation_2008}
\begin{align}
E_\text{X}^\text{Slater} &= - \frac{1}{2} \sum_\sigma \int \text{d}\vec{r} K_\sigma^\text{GGA} \varrho_\sigma^{\frac{4}{3}} \notag \\
&\quad  \times \underbrace{\left[ 1 - \frac{8}{3} a_\sigma \left \lbrace \arctan \frac{1}{a_\sigma} + \frac{a_\sigma}{4} - \frac{a_\sigma}{4} \left( a_\sigma^2 + 3 \right) \ln \left( 1 + \frac{1}{a_\sigma^2} \right) \right \rbrace  \right]}_{=f(a_\sigma)},\label{eq:ExYuk}
\end{align}
where $\sigma$ denotes the spin, $K_\sigma^\text{GGA}$ the GGA correction, $a_\sigma$ = $\gamma / 2k_\sigma^\text{GGA}$ with $k_\sigma^\text{GGA} = \left( 9 \pi/K_\sigma^\text{GGA}\right)^{\frac{1}{2}} \varrho_\sigma^{\frac{1}{3}}$ as given in refs. \cite{akinaga_range-separation_2008,iikura_long-range_2001}.
To calculate the exchange-correlation potential
\begin{align}
v_\text{XC}[\varrho_\sigma](\vec{r}) = \frac{\delta E_\text{XC}}{\delta \varrho_\sigma(\vec{r})},
\end{align}
the exchange-correlation kernel
\begin{align}
f_\text{XC}[\varrho_\sigma,\varrho_\sigma^\prime](\vec{r}, \vec{r\prime})  = \frac{\delta^2 E_\text{XC}}{\delta \varrho_\sigma(\vec{r})\delta \varrho_\sigma^\prime(\vec{r}^\prime)},
\end{align}
and the exchange-correlation hyper-kernel
\begin{align}
k_\text{XC}[\varrho_\sigma,\varrho_\sigma^\prime,\varrho_\sigma^{\prime\prime}](\vec{r}, \vec{r\prime}, \vec{r\prime\prime})  = \frac{\delta^3 E_\text{XC}}{\delta \varrho_\sigma(\vec{r})\delta \varrho_\sigma^\prime(\vec{r}^\prime)\delta \varrho_\sigma^{\prime\prime}(\vec{r}^{\prime\prime}) },
\end{align}
the first, second and third derivative of the second line of eq. 
(\ref{eq:ExYuk}) regarding to the density or using the chain rule 
and the definitions  given by 
refs. \cite{akinaga_range-separation_2008,iikura_long-range_2001} 
to $a_\sigma$ are needed. These are
\begin{align}
  \frac{\partial f(a_\sigma)}{\partial a_\sigma} &= \frac{4}{3} \left ( \left ( 2 a_\sigma^3 + 3 a_\sigma \right ) \ln \left ( 1 + \frac{1}{a_\sigma^2} \right ) \right. \notag \\
& \quad \quad \quad \left. - 2 a_\sigma  - 2 \arctan \left ( \frac{1}{a_\sigma} \right ) \right ), \\
\frac{\partial^2 f(a_\sigma)}{\partial a_\sigma^2} &= 4 \left ( \left ( (2 a_\sigma^2 + 1 \right ) \ln \left ( 1 + \frac{1}{a_\sigma^2} \right ) -  2\right),\\
\frac{\partial^3 f(a_\sigma)}{\partial a_\sigma^3} &= \frac{8 \left ( \left (2 a_\sigma^4 + 2 a_\sigma^2 \right ) \ln \left ( 1 + \frac{1}{a_\sigma^2} \right ) - 2 a_\sigma^2 -1 \right )}{a_\sigma^3 + a_\sigma } .
\end{align}

\section{Solution of the integral for a Gaussian shaped density times the Yukawa potential}\label{sec:AppendixScreenedPoissonGausDens}

A Gaussian distributed charge is subtracted in case of charged
systems in order to work with the (modified) Poisson equation for
involving a neutral charge density.  
This strategy involves
the integral of a parametrized Gaussian shaped 
density $n(\vec{r}, \vec{r_0}, \sigma)$ times the 
Yukawa potential $f(\vec{r}, \vec{r^\prime}, \gamma)$, which reads
\begin{align}
I &= \int n(\vec{r}, \vec{r_0}, \sigma) f(\vec{r}, \vec{r}^\prime, \gamma) \text{d}^3 \vec{r} \\
&= \int  \underbrace{\exp \left ( - \frac{ \left ( \vec{r} - \vec{r_0} \right )^2 }{2 \sigma^2} \right )}_{= n(\vec{r}, \vec{r_0}, \sigma)} \underbrace{\frac{ e^{- \gamma | \vec{r}^\prime -  \vec{r} |}}{| \vec{r}^\prime -  \vec{r} |}}_{=  f(\vec{r}, \vec{r}^\prime, \gamma)} \text{d}^3 \vec{r} \label{eq:YukDensity} \; .\\
\intertext{This can be rewritten as}
I &= \int  \exp \left ( - \frac{ \left ( \vec{\varrho_0} - \vec{\varrho}^\prime \right )^2 }{2 \sigma^2} \right ) \frac{ e^{- \gamma | \vec{\varrho}^\prime |}}{| \vec{\varrho}^\prime |} \text{d}^3 \varrho^\prime
\end{align}
by substituting $\vec{\varrho}^\prime = \vec{r}^\prime - \vec{r}$ and $\vec{\varrho_0} = \vec{r}^\prime - \vec{r_0}$. Using spherical coordinates and completing the square in the exponent, the integral can be written as
\begin{align}
I &= \int_0^{2\pi} \text{d}\varphi \int_0^\pi \sin \theta \text{d} \theta \notag \\
&\quad \times \int_0^R \varrho^2 \frac{e^{-\gamma \left | \vec{\varrho}\right|}}{\left | \vec{\varrho}\right|} \exp \left( - \frac{\varrho^2 + \varrho_0^2 - 2 \varrho\varrho_0 \cos \theta}{2\sigma^2} \right )\text{d}\varrho \; .
\end{align}
After solving the angular integrals, the remaining integral reads 
(for infinite $R$)
\begin{align}
I &=2 \pi \frac{\sigma^2}{\varrho_0}e^{-\frac{\varrho_0^2}{2 \sigma^2}} \notag \\
& \quad \times \int_0^\infty \left (e^\frac{\varrho_0 \varrho}{\sigma^2} - e^{-\frac{\varrho_0 \varrho}{\sigma^2}} \right) e^{-\gamma \varrho} e^{-\frac{\varrho^2}{2 \sigma^2}} \text{d}\varrho \; ,
\intertext{which can be divided into two parts}
I_1 &= 2 \pi \frac{\sigma^2}{\varrho_0}e^{-\frac{\varrho_0^2}{2 \sigma^2}} \int_0^\infty e^\frac{\varrho_0 \varrho}{\sigma^2}  e^{-\gamma \varrho} e^{-\frac{\varrho^2}{2 \sigma^2}} \text{d}\varrho \\
&= 2 \pi \frac{\sigma^2}{\varrho_0}e^{-\frac{\varrho_0^2}{2 \sigma^2}}  \notag \\
& \quad \times  \exp \left ( \frac{\left ( \sigma^2 \gamma - \varrho_0 \right )^2}{2 \sigma^2} \right ) \sigma \sqrt{\frac{\pi}{2} } \erfc \left ( \frac {\sigma^2 \gamma - \varrho_0}{\sqrt{2} \sigma} \right),
 \\
 \intertext{and}
I_2 &= - 2 \pi \frac{\sigma^2}{\varrho_0}e^{-\frac{\varrho_0^2}{2 \sigma^2}}  \int_0^\infty e^{-\frac{\varrho_0 \varrho}{\sigma^2}}  e^{-\gamma \varrho} e^{-\frac{\varrho^2}{2 \sigma^2}} \text{d}\varrho \\
&= - 2 \pi \frac{\sigma^2}{\varrho_0}e^{-\frac{\varrho_0^2}{2 \sigma^2}} \notag \\
& \quad \times  \exp \left ( \frac{\left ( \sigma^2 \gamma + \varrho_0 \right )^2}{2 \sigma^2} \right ) \sigma \sqrt{\frac{\pi}{2} } \erfc \left ( \frac {\sigma^2 \gamma + \varrho_0}{\sqrt{2} \sigma} \right).
\end{align}
Combining these results leads to
\begin{align}
I_1 + I_2 &= \frac{\sigma^3 \pi^{3/2} \sqrt{2}}{\varrho_0} e^{\frac{\sigma^2 \gamma^2}{2}}  \notag \\
 & \quad \times \left [ \exp \left( - \gamma \varrho_0 \right) \erfc \left( \frac{\sigma^2 \gamma - \varrho_0}{\sqrt{2} \sigma} \right) \right. \notag\\
 & \left. \quad \quad - \exp \left( \gamma \varrho_0 \right) \erfc \left( \frac{\sigma^2 \gamma + \varrho_0}{\sqrt{2} \sigma} \right) \right] \; .
\end{align}
By comparison of the density $n(\vec{r}, \vec{r_0}, \sigma)$ from eq. (\ref{eq:YukDensity}) with a normalized Gaussian
\begin{align}
n(\vec{r}, \vec{r_0}, \sigma) &= \frac{1}{\left ( 2 \pi \right)^{3/2} \sigma^3}  \exp \left( - \frac{ \left( \vec{r} - \vec{r_0} \right)^2}{2 \sigma^2} \right),\label{eq:GaussYuk}
\end{align}
the potential for a screened normalized Gaussian can be written as
\begin{align}
\varphi (\varrho_0, \gamma, \sigma) &= \frac{1}{2 \varrho_0} e^{\frac{\sigma^2 \gamma^2}{2}} \notag \\
 & \quad \times \left [ \exp \left( - \gamma \varrho_0 \right) \erfc \left( \frac{\sigma^2 \gamma - \varrho_0}{\sqrt{2} \sigma} \right) \right. \notag\\
 & \left. \quad \quad - \exp \left( \gamma \varrho_0 \right) \erfc \left( \frac{\sigma^2 \gamma + \varrho_0}{\sqrt{2} \sigma} \right) \right].\label{eq:YukScrGaus}
\end{align}
According to the computer algebra system 
{\tt maxima}\cite{maxima_maxima_2014} this converges 
to the error-function divided by $\varrho_0$ for $\gamma \to 0$
\begin{align}
\lim_{\gamma \to 0} \varphi (\varrho_0, \gamma, \sigma) &= - \frac{\erfc \left ( \frac{\varrho_0}{\sqrt{2}\sigma} \right) -1}{\varrho_0} \\
&= \frac{\erf \left ( \frac{\varrho_0}{\sqrt{2}\sigma} \right) }{\varrho_0},
\end{align}
which is known as the Coulomb potential of a Gaussian shaped 
charge density\cite{jackson_classical_1998}.

\section{Comparsion of the eigenvalues for Be\textsubscript{2}}\label{sec:AppendixBe2uocc}

\begin{table}
\begin{center}
\begin{tabular}{l|l|D{.}{.}{2}|D{.}{.}{2}|D{.}{.}{2}}
\multicolumn{1}{c|}{$n$}&\multicolumn{1}{c|}{$f_n$}&\multicolumn{1}{c|}{$\unit[\epsilon_{\hat{F}^\text{IVO}_\text{Huz}}]{(eV)}$}&\multicolumn{1}{c|}{$\unit[\epsilon_{\hat{F}^\text{IVO}_\text{Kel}}]{(eV)}$}&\multicolumn{1}{c}{$\unit[\epsilon_{\hat{F}^\text{IVO}_\text{Kel}(n_\text{max} =15)}]{(eV)}$}\\
\hline
0 & 2 & -10.47 & -10.47 & -10.47 \\
1 & 2 &  -7.49 &  -7.49  &   -7.49 \\
2 & 0 &  -5.23 & -5.23 &  -5.23 \\
3 & 0 & -5.23 & -5.23 &   -5.23 \\
4 & 0 & -3.51 & -3.48 &  -3.49 \\
5 & 0 & -2.33 & -2.34 &  -2.34 \\
6 & 0 & -2.33 & -2.34 &  -2.34 \\
7 & 0 & -1.95 & -1.94 &  -1.95 
\end{tabular}
\caption{Comparison of the eigenvalues of the Be$_2$ molecule using the different modified Fock operators. $n$ band-index, $f_\text{n}$ occupation number. Gridspacing $h=\unit[0.25]{\text{\AA}}$, amount of vacuum around the atoms, $x_\text{vac}=\unit[6]{\text{\AA}}$.
}
\label{tab:Num_IVO_HuziDiff}
\end{center}
\end{table}

To verify the effects of dropping $\hat{P}$ in eq. (20) of the main text, a HFT calculation using the modified Fock operator from Huzinaga, $\epsilon_{\hat{F}^\text{IVO}_\text{Huz}}$, including the projection operator $\hat{P}$ is compared to two calculations which use the rotation operator $\Omega_k$, but drop the usage of $\hat{P}$ and blank the HFT exchange cross-elements between occupied and unoccupied states $\epsilon_{\hat{F}^\text{IVO}_\text{Kel}}$. 
The results are listed in tab. \ref{tab:Num_IVO_HuziDiff}.  The calculated eigenvalues are nearly identical, such that the procedure of dropping $\hat{P}$ and blanking the cross term is suitable for the calculation of excited states using eigenvalue-differences or lrTDDFT. 

\section{Including RSF in lrTDDFT}
\label{sec:AppendixOmegaK}

We consider the changes brought into generalized lrTDDFT through the terms from
range separated functionals.
The most general case is the CAM scheme, where the Coulomb interaction kernel
of the exchange integral, eq.  (3) in the main text, is split into two 
parts \cite{yanai_new_2004}
\begin{align}
  \frac{1}{r_{12}} =& \underbrace{\frac{1 - \left [ \alpha + \beta  \left ( 1 - \omega_\text{RSF}(\gamma, r_{12})\right )\right]}{r_{12}}}_\text{SR, DFT}
   + \underbrace{\frac{\alpha + \beta \left (1- \omega_\text{RSF}(\gamma, r_{12})\right )}{r_{12}}.}_\text{LR, HFT} \label{eq:APCAM}
\end{align}

In linear response TDDFT or TDHFT we have to solve the 
generalized eigenvalue problem\cite{dreuw_long-range_2003,casida_time-dependent_2009,akinaga_intramolecular_2009}
\begin{align}
  \begin{pmatrix}
    \mathbf{A} & \mathbf{B} \\
    \mathbf{B^\ast} & \mathbf{A^\ast}
  \end{pmatrix} \begin{pmatrix} \vec{X} \\ 
    \vec{Y} 
  \end{pmatrix} &= \omega 
  \begin{pmatrix}
    \mathbf{1} & \mathbf{0} \\
    \mathbf{0} & -\mathbf{1}
  \end{pmatrix} 
  \begin{pmatrix} \vec{X} \\ 
    \vec{Y} 
  \end{pmatrix},
\end{align}
with
\begin{align}  
  A_{ia\sigma,jb\tau} &= \delta_{ij}\delta_{ab} \delta_{\sigma\tau} \left( \epsilon_a - \epsilon_i \right) + K_{ia\sigma,jb\tau} \label{eq:lrASI} \\
B_{ia\sigma,jb\tau} &=  K_{ia\sigma,bj\tau}\label{eq:lrBSI}
\end{align}
and 
the element $K_{pq\sigma,rs\tau}$ of the lrTDDFT coupling matrix reads\cite{casida_time-dependent_2009}
\begin{align}
K_{pq\sigma,rs\tau}^\text{HFT} &= \left( pq | rs \right) - \delta_{\sigma \tau} \left( pr | sq \right)&\text{for HFT} \label{eq:APKHFT}\\
K_{pq\sigma,rs\tau}^\text{DFT} &= \left( pq | rs \right) + \left( pq | f_\text{XC} | rs \right)&\text{for DFT}, \label{eq:APKDFT}
\end{align}
 where $p,q,r,s$ denotes arbitrary orbital-indices, $\sigma$ and $\tau$  are spin-indices, which are left out for brevity if selected by the Kronecker delta, $\delta_{\sigma\tau}$. $f_\text{xc} =  \frac{\delta^2E_\text{xc}}{\delta\varrho\left(\vec{r_1}\right)\delta\varrho\left(\vec{r_2}\right)}$ denotes the exchange-correlation kernel of the functional. Mulliken like notations
\begin{align}
\left( p q | r s \right) &= \int \int \frac{p^\ast(\vec{r_1}) q(\vec{r_1}) r^\ast(\vec{r_2}) s(\vec{r_2})}{|\vec{r_1} - \vec{r_2}|} \text{d}\vec{r_1} \text{d}\vec{r_2} \\
\noalign{\text{and}}
\left( p q | \hat{x} | r s \right) &= \int \int p^\ast(\vec{r_1}) q(\vec{r_1}) \hat{x} (\vec{r_1}, \vec{r_2}, ...)  r^\ast(\vec{r_2}) s(\vec{r_2}) \text{d}\vec{r_1} \text{d}\vec{r_2} 
\end{align}
are used. The first terms in eqs. (\ref{eq:APKHFT}) and (\ref{eq:APKDFT}) are the so called RPA terms, exchange and correlation plugs into the second terms. 

Applying eq. (\ref{eq:APCAM}) into the exchange-dependent terms, 
and keeping in mind that the HFT-exchange is negative $E_\text{X}^\text{HFT} = - \frac{1}{2} \sum_{ij} (ij|ji)$, we get
\begin{align}
K_{pq\sigma,rs\tau} &= \left( p_\sigma q_\sigma|r_\tau s_\tau \right) + K_{pq\sigma,rs\tau}^\text{CAM-DFT} + K_{pq\sigma,rs\tau}^\text{CAM-HFT} \label{eq:supCamDFT}\\
\noalign{\text{with}}
K_{pq\sigma,rs\tau}^\text{CAM-DFT} &= \left( 1 - \alpha - \beta \right) \left(p_\sigma q_\sigma |f_\text{xc}|r_\tau s_\tau\right) + \beta \left(p_\sigma q_\sigma |f_\text{xc}^\text{RSF}|r_\tau s_\tau\right)\\
K_{pq\sigma,rs\tau}^\text{CAM-HFT} &= - \delta_{\sigma\tau} \left(pr|\frac{\alpha +  \left( 1 - \omega_\text{RSF}\right)}{r_{12}}|sq\right),\label{eq:supCamHFT}
\end{align}
where $f_\text{XC}^{RSF}$ denotes the dampened kernel of the RSF.

This resembles a global hybrid for $\alpha \ne 0$ and $\beta=0$\cite{dreuw_long-range_2003}
\begin{align}
K_{pq\sigma, rs\tau} &= \left( pq\sigma | rs\tau\right) + \left(1-\alpha\right) K^\text{DFT}_{pq\sigma, rs\tau} + \alpha K^\text{HFT}_{pq\sigma, rs\tau} \\
\noalign{\text{with}}
K^\text{DFT}_{pq\sigma, rs\tau} &= \left(pq\sigma| f_\text{XC}|rs\tau \right) \\
K^\text{HFT}_{pq\sigma, rs\tau} &= - \delta_{\sigma\tau}\left(pr|sq \right)
\end{align}
and the pure LC scheme for $\alpha = 0$, $\beta = 1$ \cite{KuritzChargeTransferLikeExcitationsTimeDependent2011}
\begin{align}
K_{pq\sigma, rs\tau} &= \left( pq\sigma | rs\tau\right) + \beta K^\text{LC-DFT}_{pq\sigma, rs\tau} + \beta K^\text{HFT}_{pq\sigma, rs\tau} \\
\noalign{\text{with}}
K^\text{DFT}_{pq\sigma, rs\tau} &= \left(pq\sigma| f_\text{XC}^\text{RSF}|rs\tau \right) \\
K^\text{HFT}_{pq\sigma, rs\tau} &= - \delta_{\sigma\tau}\left(pr|\frac{1-\omega_\text{RSF}}{r_{12}}|sq \right) \, .
\end{align}
Note, that the sign of the last (HFT) term in 
eqs. (\ref{eq:supCamDFT}) and (\ref{eq:supCamHFT})
is erroneously positive in refs. 
\cite{tawada_long-range-corrected_2004} and 
\cite{akinaga_intramolecular_2009}.

\section{Impact of different forms of the IVO operator on excitions in Na$_2$}
\label{sec:AppendixIVONa2}

\begin{table}
\begin{center}
\begin{tabular}{l|D{.}{.}{2}|D{.}{.}{2}@{}D{.}{.}{2}|D{.}{.}{2}@{}D{.}{.}{2}|D{.}{.}{2}@{}D{.}{.}{2}|D{.}{.}{2}@{}D{.}{.}{2}|D{.}{.}{2}@{}D{.}{.}{2}|D{.}{.}{2}@{}D{.}{.}{2}}
\multicolumn{1}{c|}{state}&\multicolumn{1}{c|}{Exp.}&\multicolumn{4}{c|}{$\Omega_k^S$}&\multicolumn{4}{c|}{$\Omega_k^A$}&\multicolumn{4}{c}{$\Omega_k^T$}\\
\multicolumn{1}{c|}{}&\multicolumn{1}{c|}{(eV)}&\multicolumn{2}{c|}{$e_\nu \unit{(eV)}$}&\multicolumn{2}{c|}{$f_\nu$}&\multicolumn{2}{c|}{$e_\nu \unit{(eV)}$}&\multicolumn{2}{c|}{$f_\nu$}&\multicolumn{2}{c|}{$e_\nu \unit{(eV)}$}&\multicolumn{2}{c}{$f_\nu$}\\
\hline
a $^3\Sigma_u^+$     & 0.71 & 0.66 &(0.68) & 0.   & (0.)  & 0.62 & (0.64) & 0.   & (0.)   & 0.57 &(0.58) & 0. & (0.)\\
b  $^3\Pi_u$       & 1.68 & 1.41 &(1.56) & 0.   & (0.)  & 1.38 & (1.45) & 0.   & (0.)   & 1.36 &(1.37) & 0. & (0.)\\
A $^1\Sigma_u^+$ & 1.81 & 2.09 &(2.10) & 0.83 & (0.84)& 2.09 & (2.10) & 0.64 & (0.65) & 2.08 &(2.12) & 0.44 & (0.45) \\
1 $^3\Sigma_g^+$ & 2.25 & 2.37 & (2.41) & 0. & (0.) & 2.36 & (2.47) & 0. & (0.) & 2.35 & (2.39) & 0. & (0.)\\
B $^1\Pi_u$ & 2.52 & 2.61 & (2.68) & 1.38 & (1.50) & 2.62 & (2.75) & 1.16 & (1.32)& 2.63 & (2.85) & 0.94 & (1.06)\\ 
\end{tabular}
\caption{ 
  Excitation-energies and oscillator-strengths of Na${}_2$ calculated 
  by lrTDDFT utilizing the combination of RSF and IVOs. 
  Experimental energies (Exp.) and state assignments from refs. 
  \cite{linstrom_constants_2016,LiTripletstatesNa21999}. 
  $e_\nu$: excitation energy, $f_\nu$: oscillator strength. 
  $\Omega_k^{S,A,T}$: values calculated using the according 
  rotation-operator. A large basis of 100 unoccupied states was used,
  while the values in brackets use a small basis of seven unoccupied states
  only. The gridspacing was set to $h=\unit[0.20]{\text{\AA}}$ and
  the simulation box contained at least $x_\text{vac}=\unit[11]{\text{\AA}}$
  around each atom.}
\label{tab:Num_IVO_OmegaDiff}
\end{center}
\end{table}

We now discuss the changes introduced by IVOs to the linear response
matrices in HFT\cite{berman_fast_1979}. The matrix $\mathcal{B}$ is 
the same as with canonical HFT unoccupied orbitals.
The changes on the $\mathcal{A}$ matrix in HFT
\begin{align}
\mathbf{A}^\text{HFT}_{ia\sigma,jb\tau} &= \delta_{ij}\delta_{ab}\delta_{\sigma\tau} \left( \epsilon_{a\sigma} - \epsilon_{i\tau} \right) + K_{ia\sigma,jb\tau} \\
&= \delta_{ij}\delta_{ab}\delta_{\sigma\tau} \left( \epsilon_{a\sigma} - \epsilon_{i\tau} \right)+  \left(i_\sigma a_\sigma | j_\tau b_\tau \right) - \delta_{\sigma\tau}  \left( i_\sigma j_\sigma | a_\tau b_\tau \right)\label{eq:AppHFTA}
\end{align}
by applying the operator of eq. (27) in the main text are
\begin{align}
\mathbf{A}^\text{IVO}_{ia\sigma,jb\tau} &= \delta_{ij}\delta_{ab}\delta_{\sigma\tau} \left( \epsilon_{a\sigma}^\text{IVO} - \epsilon_{i\tau} \right)+  \left(i_\sigma a_\sigma | j_\tau b_\tau \right) - \delta_{\sigma\tau}  \left( i_\sigma j_\sigma | a_\tau b_\tau \right) \notag \\
& \quad + \delta_{ab}\delta_{\sigma\tau} \left[\left(a_\sigma a_\sigma|k_\tau k_\tau \right) - \left(a_\sigma k_\sigma|k_\tau a_\tau \right) \mp \left(a_\sigma k_\sigma |k_\tau a_\tau \right)\right] \; ,
\end{align}
where the negative sign is to be used for singlets, the positive sign 
for triplets.
Disregarding changes in the wavefunctions, 
the IVOs change the eigenvalue $\epsilon_{a}^\text{IVO}$ 
as compared to the 
canonical unoccupied states value $\epsilon_{a}$ to
\begin{align}
\epsilon_{a}^\text{IVO} &= \epsilon_{a} - \left(aa|kk\right) + \left(ak|ka\right) \pm \left(ak|ka \right),
\end{align}
where the positive (negative) sign is for singlets (triplets). 
Therefore $\mathbf{A}^\text{IVO}_{ia\sigma,jb\tau}$ becomes
\begin{align}
\mathbf{A}^\text{IVO}_{ia\sigma,jb\tau} &= \delta_{\sigma\tau} \lbrace \delta_{ab} \left[ \delta_{ij} \left( \epsilon_{a_\sigma} - \left(a_\sigma a_\sigma|k_\tau k_\tau\right) + \left(a_\sigma k_\sigma |k_\tau a_\tau \right) \pm \left(a_\sigma k_\sigma |k_\tau a_\tau\right) - \epsilon_{i\tau} \right)  \right. \notag \\
 & \quad \quad \quad \quad   \left. + \left(a_\sigma a_\sigma |k_\tau k_\tau\right) - \left(a_\sigma k_\sigma|k_\tau a_\tau \right) \mp \left(a_\sigma k_\sigma |k_\tau a_\tau\right) \right] - \left( i_\sigma j_\sigma| a_\tau b_\tau \right) \rbrace\notag \\
 & \quad \quad + \left(i_\sigma a_\sigma | j_\tau b_\tau \right)\label{eq:AppIVOA}
\end{align}

If we discuss a single, isolated excitation (single pole approximation, SPA, $i=j$, $a=b$), eq. (\ref{eq:AppHFTA}) becomes:
\begin{align}
\mathbf{A}_{ia\sigma,ia\tau}^\text{HFT,SPA} &=\delta_{\sigma\tau} \left( \epsilon_{a\sigma} - \epsilon_{i\tau} \right)+  \left(i_\sigma a_\sigma | i_\tau a_\tau \right) - \delta_{\sigma\tau}  \left( i_\sigma i_\sigma | a_\tau a_\tau \right),\label{eq:AppHFTASP}
\end{align}
and eq. (\ref{eq:AppIVOA}) becomes (using $a^\prime$ for the IVO orbitals):
\begin{align}
\mathbf{A}_{ia^\prime \sigma, ia^\prime \tau}^\text{IVO,SPA} &= \delta_{\sigma\tau} \lbrace  \epsilon_{a\sigma} - \left(a_\sigma^\prime a_\sigma^\prime |k_\tau k_\tau \right) + \left(a_\sigma^\prime k_\sigma |k_\tau a_\tau^\prime \right) \pm \left(a_\sigma^\prime k_\sigma|k_\tau a_\tau^\prime \right) - \epsilon_{i_\tau}  \notag \\
 &  \quad  \quad \quad  \quad  + \left(a_\sigma^\prime a_\sigma^\prime |k_\tau k_\tau \right) - \left(a_\sigma^\prime k_\sigma |k_\tau a_\tau^\prime \right) \mp \left(a_\sigma^\prime k_\sigma|k_\tau a_\tau^\prime \right) - \left(i_\sigma i_\sigma| a_\tau^\prime a_\tau^\prime \right) \rbrace\notag \\
 & \quad \quad + \left(i_\sigma a^\prime_\sigma | i_\tau a^\prime_\tau \right)\\
 &= \delta_{\sigma\tau} \lbrace  \epsilon_{a\sigma} - \epsilon_{i\tau} - \left(i_\sigma i_\sigma| a_\tau^\prime a_\tau^\prime \right) \rbrace + \left(i_\sigma a^\prime_\sigma | i_\tau a^\prime_\tau \right) \label{eq:AppIVOAS}
\end{align}

Eqs. (\ref{eq:AppHFTASP}) are (\ref{eq:AppIVOAS}) equal for $a = a^\prime$, 
but $a$ and $a^\prime$ are eigenfunctions to different operators, 
thus eqs. (\ref{eq:AppHFTASP}) and (\ref{eq:AppIVOAS}) will 
lead to different results. 
Thus the excitation energies have to differ in SPA. This also holds true for the eigenfunction of the different operators $\Omega_k^{S,A,T}$ between each other.  We've calculated the  excitation energies and oscillator strengths for the isolated Na${}_2$ molecule using the different operators for a small, seven, and a larger number of unoccupied states. The calculated values are listed in tab. \ref{tab:Num_IVO_OmegaDiff}.  
The energies generally agree within $\unit[100]{meV}$, while 
singlet energies agree even within $\unit[20]{meV}$.

Tab. \ref{tab:Num_IVO_OmegaDiff} also reveals, that there are 
differences in the oscillator strengths obtained
with the three operators.
The oscillator strengths are calculated by\cite{walter_time-dependent_2008}
\begin{align}
  f_{I \alpha} &= \frac{2m_e}{\hbar e^2} \left | \sum_{ia\sigma}\left( \vec{\mu}_{ia\sigma} \right)_\alpha \sqrt{\epsilon_{ia\sigma}} \left( F_\text{I} \right)_{ia\sigma} \right|^2 \label{eq:App_IVO_OSC}
\end{align}
where $\epsilon_{ia\sigma} = \epsilon_{a\sigma} - \epsilon_{i\sigma}$ denotes the eigenvalue-differences of the individual occupied $i$ and unoccupied $a$ 
Kohn-Sham (KS) states. 
The $\vec{\mu}_{ia\sigma} = - e \langle i\sigma | \vec{r} | a\sigma \rangle$ 
are the KS transition dipoles and $F_I$ denotes the eigenvector of the 
$\Omega$ matrix. 

\begin{table}
\begin{center}
\begin{tabular}{l|D{.}{.}{2}|D{.}{.}{2}|D{.}{.}{2}}
\multicolumn{1}{c|}{}&\multicolumn{1}{c|}{$\Omega_k^S$}&\multicolumn{1}{c|}{$\Omega_k^A$}&\multicolumn{1}{c}{$\Omega_k^T$}\\
\hline
$f_{ia\sigma}$ & 0.43 & 0.33 & 0.23\\
$\epsilon_{ia\sigma} (\unit{eV})$ & 1.99 & 1.54 & 1.06 \\
$\nicefrac{f_{ia\sigma}}{\epsilon_{ia\sigma}} (\unitfrac{1}{eV})$ & 0.22 & 0.21 & 0.22 
\end{tabular}
\caption{ 
  Oscillator strengths $f_{ia\sigma}$ of eq. (\ref{eq:App_IVO_OSCKS}),
    for the HOMO-LUMO transition, where the LUMO is obtained
    using different forms of the IVO operator $\Omega_k^{S,A,T}$.
    The corresponding excitation energies $\epsilon_{ia\sigma} = 
    \epsilon_{a\sigma} - \epsilon_{i\sigma}$ and the ratios are also given.
    Numerical settings as in tab. \ref{tab:Num_IVO_OmegaDiff}.
  }
\label{tab:Num_IVO_OmegafosDiff}
\end{center}
\end{table}

The differences in the oscillator strengths arise from differences in the
single particle energy differences $\epsilon_{ia\sigma}$. In order to show this
the mean KS oscillator strengths
\begin{align}
  f_{ia\sigma} &= \frac{2m_e}{3\hbar e^2} \; \epsilon_{ia\sigma} \sum_{\alpha = x,y,z} \left | \left( \vec{\mu}_{ia\sigma} \right)_\alpha  \right|^2 \label{eq:App_IVO_OSCKS}
\end{align}
are given in tab. \ref{tab:Num_IVO_OmegafosDiff}
for a single KS transition from HOMO to LUMO, 
where $\alpha$ denotes the spatial 
direction of $\vec{\mu}$. 
The value of $\epsilon_{a\sigma}$ depends on $\Omega_k^{S,A,T}$
which is reflected in $\epsilon_{ia\sigma}$.
While excitation energies and the oscillator strengths differ by up to a 
factor of two, their ratio is practically constant.
The variance in KS transition energies is compensated
in $\mathbf{A}^\text{IVO}$, but not in the matrix elements.


 \end{widetext}

\end{document}